\documentstyle[eqsecnum,amsmath,amsfonts,aps]{revtex}

\title{
Equation of motion for relativistic compact binaries with 
the strong field point particle limit: 
Third post-Newtonian order
}

\author{Yousuke Itoh\footnote{yousuke.itoh@aei.mpg.de}}
\address{
Max-Planck-Institut f\"ur Gravitationsphysik, 
Albert-Einstein-Institut \\ 
Am M\"uhlenberg 1, Golm 14476, Germany
}

\begin{document}
\maketitle

\begin{abstract}
An equation of motion for relativistic compact binaries 
is derived through the third post-Newtonian (3PN) approximation 
of general relativity. The strong field point 
particle limit and multipole expansion of the 
stars are used to solve iteratively 
the harmonically relaxed Einstein equations.
We take into account 
the Lorentz contraction on  
the multipole moments defined in our previous works.  
We then derive a 3PN acceleration of the  
binary orbital motion of the two spherical compact stars 
based on a surface integral approach which is 
a direct consequence of local energy momentum conservation.  
Our resulting equation of motion admits a conserved energy 
(neglecting the 2.5PN radiation reaction effect), 
is Lorentz invariant, and is unambiguous: 
there exist no undetermined parameters reported in the previous 
works. 
We shall show that our 3PN equation of motion agrees 
physically with the Blanchet and Faye 3PN equation of motion 
if $\lambda = - 1987/3080$, where $\lambda$ is the 
parameter which is undetermined within their framework.  
This value of $\lambda$ is consistent with the result of 
Damour, Jaranowski, and Sch\"afer, who first completed  
a 3PN iteration of the ADM Hamiltonian in the ADMTT gauge 
using dimensional regularization.
\end{abstract}

\begin{flushleft}
PACS Number(s): 04.25.Nx
\end{flushleft}

\newcommand{\pa}{\partial}

\section{Introduction}

Tremendous effort  
has been made for direct detection of gravitational waves. 
Interferometric gravitational wave detectors 
such as GEO600 \cite{GEO}, the Laser Interferometric Gravitational 
Wave Observatory(LIGO) \cite{LIGO},    
and TAMA300 \cite{TAMA} have been operating  
successfully. They have 
been actively investigating the data and 
reported 
results 
\cite{Tagoshi01,Ando01,Takahashi03,LIGOGEOS1,LIGOGEOS1CWPaper,LIGOGEOS1IBPaper}.

One promising source of gravitational 
waves for those detectors is a relativistic compact 
binary system in an inspiraling phase. 
The detectability and quality of measurements 
of astrophysical information of such gravitational wave sources
 rely on 
the accuracy of our theoretical knowledge about the waveforms. 
A high-order, say, third- or fourth-order, post-Newtonian equation 
of motion for an inspiraling compact binary   
is one of the necessary ingredients 
to construct and study such 
waveforms \cite{BCV02,CutlerEtAL93,TN94,TTS96}.

In addition to the purpose of making accurate enough templates, 
the high-order post-Newtonian approximation 
for an inspiraling compact binary 
is a fruitful tool, for example, 
to construct astrophysically realistic initial data for 
numerical simulations \cite{Alvi00,TBCD02,Blanchet03b,FS03}  
and to estimate  innermost circular 
orbits \cite{Blanchet02a,Blanchet03a}. 

The post-Newtonian (PN) equation of motion for relativistic 
compact binaries, a fundamental  
tool employed for the above purposes, has 
been derived by various authors (see reviews, 
e.g., \cite{Damour83,Damour87,Blanchet02b}). 
The equation of motion for a two point-masses binary   
in a harmonic coordinate up to  2.5PN order, 
at which the radiation reaction effect first appears, 
was derived by Damour and Deruelle \cite{DD81a,Damour82}  
based on the post-Minkowskian approach \cite{BDDIM81}. 
These works used Dirac delta distributions to express 
the point-masses mathematically, therefore 
they inevitably resorted to a purely mathematical regularization   
procedure to deal with divergences arising due to
the nonlinearity of general relativity. 
Damour \cite{Damour83} gave a physical argument known as 
the ``dominant Schwarzschild condition'' which supports 
the use of Dirac delta distributions and their regularization 
up to 2.5PN order.

Direct validations of the Damour and Deruelle 
2.5PN equation of motion have been obtained by 
several works \cite{GK83,Kopejkin85,BFP98,IFA01,PW02}. 
Grishchuk and 
Kopeikin \cite{GK83} and Kopeikin \cite{Kopejkin85} worked 
on extended bodies with weak internal gravity. 
On the other hand, the 
present author, Futamase, and Asada derived the 2.5PN 
equation of motion \cite{IFA01}
using the surface integral approach \cite{EIH1938} 
and the strong field 
point-particle limit \cite{Futamase87}.
All the works quoted above agree with each other. 
Namely, our work \cite{IFA01}  
shows the applicability of the Damour and Deruelle 
2.5PN equation of motion to a relativistic 
compact binary consisting of regular stars with strong internal 
gravity. 
We mention here that 
stars consisting of relativistic compact binaries, 
for which we are searching as gravitational wave sources, 
have 
strong internal gravitational field, and that it 
is a nontrivial question whether a  
star follows the same orbit regardless of 
the strength of its 
internal gravity.

A 3PN iteration was first 
reported 
by Jaranowski 
and Sch\"afer \cite{JS98}. There a 3PN 
Arnowitt-Deser-Misner (ADM) Hamiltonian 
in the ADM Transverse Traceless (ADMTT) gauge for two point-masses 
expressed as two Dirac delta distributions 
was derived based on the ADM canonical 
approach \cite{Schafer85,OOKH73}. 
However, it was found in \cite{JS98} that the 
regularizations they had used caused one 
coefficient $\omega_{{\rm kinetic}}$ undetermined 
in their framework. Moreover, they later found another 
undetermined coefficient in their Hamiltonian, 
called $\omega_{{\rm static}}$ \cite{JS99}. 
Origins of these two coefficients were attributed  
to some unsatisfactory features of regularizations  
they had used, such as violation of Leibniz's rule.  
The former coefficient was then fixed as 
$\omega_{{\rm kinetic}} = 41/24$  
by {\it a posteriori} imposing Poincar\'e invariance on 
their 3PN Hamiltonian \cite{DJS00}. 
As for the latter coefficient, 
Damour {\it et al}. \cite{DJS01a} 
succeeded in fixing it as 
$\omega_{{\rm static}} =0$, adopting dimensional 
regularization. 
Moreover, with this method they found the same value 
of $\omega_{{\rm kinetic}}$ as in \cite{DJS00}, which 
ensures Lorentz invariance of their Hamiltonian.

On the other hand, Blanchet and Faye have succeeded in 
deriving a 3PN equation of motion for two point-masses 
expressed as two Dirac delta distributions 
in a harmonic coordinate \cite{BF00a,BF01a} 
based on their previous work \cite{BFP98}. 
In this approach, they have assumed that 
two point-masses follow regularized geodesic 
equations 
(more precisely, they have assumed that the dynamics of 
two point-masses is described by a regularized action, 
from which the regularized geodesic equations were shown 
to be derived).
The divergences due to their use of  
Dirac delta distributions were systematically 
regularized with the help of generalized Hadamard's 
partie finie regularization, which they have devised \cite{BF00b}  
and carefully elaborated \cite{BF01b} so that 
it respects Lorentz invariance. 
They found, however, that there exists one and only one 
undetermined coefficient (which they call $\lambda$).  
(But see \cite{BDEF03} for the recent achievement of 
Blanchet and his collaborators.)

Interestingly, 
the two groups independently constructed a 
transformation between the two gauges and found 
that these two results coincide with each other when 
there exists a relationship \cite{DJS01b,ABF01} 
\begin{eqnarray}
\omega_{{\rm static}} &=& -\frac{11}{3}\lambda - \frac{1987}{840}.
\label{omegalambdarelation}
\end{eqnarray}
Therefore, it is possible to fix the $\lambda$ parameter from 
the result of \cite{DJS01a}. 
However, applicability of mathematical 
regularizations to the current problem is not a trivial issue, 
but an assumption to be verified, or at least supported by 
convincing arguments. 
There is no 
argument such as the ``dominant Schwarzschild condition'' at 
the 3PN order. 
Thus, it seems crucial to achieve  
unambiguous 3PN 
iteration without introducing singular sources and to support   
(or give counterevidence against) 
the result of \cite{DJS01a}.

In this paper, we derive a third post-Newtonian equation 
of motion in a harmonic coordinate 
applicable to inspiraling binaries consisting 
of regular spherically symmetric 
compact stars with strong internal gravity, 
using yet another method which is based on our previous papers
(\cite{IFA00} and \cite{IFA01}, referred to as paper I and 
paper II, respectively, henceforth). 
Namely, to treat strong internal gravity of the stars, 
we have used the strong field point-particle limit 
and surface integral approach. 
This point-particle limit 
enables us to incorporate a notion of 
a self-gravitating point particle 
into general relativity without introducing singular sources. 

In \cite{IF03}, we made a short report on our results.  
In this paper, we shall present the full explanation of 
our method and show our results.

The organization of this paper is as follows.  
In Secs. \ref{sec:PNA} and \ref{chap:formulation}, 
we briefly explain the basics of our method (see papers I 
and II for more details). After a short explanation on  
the structure of the 3PN equation of motion, 
we then compute the 3PN gravitational field required 
to derive the 3PN mass-energy 
relation from Secs. \ref{chap:3PNEOM} to \ref{tpnhc}.
The 3PN mass-energy relation and the 3PN 
momentum-velocity 
 relation are then derived in Secs.  
\ref{MeaningOfPt} and \ref{sec:3PNMomVelRelation}. 
Section \ref{TPNGravitationalField} describes how we 
evaluate the 3PN gravitational field necessary to 
derive a 3PN equation of motion. 
In these sections, 
we write down the formulas we have used and show several   
examples of our computations only, 
since the intermediate results are generally too lengthy  
to write down.   
We then show a 3PN equation of motion we obtained in 
Secs.  \ref{sec:3PNEOMwithLog} and \ref{sec:3PNEOM} 
and we shall compare it with the Blanchet 
and Faye 3PN equation of motion in Sec. \ref{subsec:comparison}. 
There, we found that our result is physically consistent 
with their result with $\lambda = -1987/3080$. 
Section \ref{subsec:summary} is devoted to a summary of 
our formalism and discussion.
Some useful formulas and supplementary computations are 
shown in Appendixes.

Throughout, we use units where $c=1=G$.
A tensor having alphabetical indexes, such as $x^i$ or $\vec x$,  
denotes a Euclidean three-vector. We raise or lower its indexes with 
a Kronecker delta. An object having Greek indexes, such as $x^{\mu}$,  
is a four-vector. Its indexes are raised or lowered by 
a flat Minkowskian metric. 

\section{Strong Field Point-Particle Limit and Scalings}
\label{sec:PNA}

In this section, we briefly review the 
strong field point-particle limit and  
associated scalings of the matter on the initial 
hypersurface. See  
\cite{IFA01,Futamase87,IFA00,Schutz85,AsadaFutamase97,Schutz80,FS83,Futamase83} 
for more details.

We first introduce 
a nondimensional small parameter $\epsilon$ which represents  
 slowness of a star's typical 
orbital velocity $\tilde v^i_{{\rm orb}}$ and 
thus which is a post-Newtonian expansion parameter,
\begin{eqnarray*}
&&
\tilde{v}^i_{{\rm orb}} 
\equiv \frac{d x^i}{d t} \equiv \epsilon \frac{d x^i}{d \tau},  
\end{eqnarray*}
where we set $v^i \equiv d x^i/d \tau$ of order unity.  
We call $\tau$ the 
{\it Newtonian dynamical time} \cite{Futamase87}. 
$\tau$ is a natural dynamical time scale of the orbital motion 
\cite{Futamase87,AsadaFutamase97}.

Then the post-Newtonian scaling 
implies that the typical scale of the mass of the star 
scales as $\epsilon^2$.

Henceforth, we call the $(\tau,x^i)$
coordinate the near zone coordinate.

\subsection{Strong field point-particle limit}

One would think that 
a point-particle limit may be achieved by setting 
the radius of the star to zero. In general relativity, however, 
by this procedure 
the star cannot become a ``point-particle'', rather it becomes an 
extended body (black hole) whose radius is of order of its gravitational 
radius. One solution to avoid this conceptual problem was proposed 
by Futamase \cite{Futamase87}. Following Futamase, we achieve 
a point-particle limit 
by letting the radius of the star shrink 
at the same rate as the mass of the star. This limit is nicely fit 
into the post-Newtonian approximation, since from the post-Newtonian 
scaling of the mass $(O(\epsilon^2))$, 
the radius of the star is $O(\epsilon^2) \rightarrow 0$  
as we take the post-Newtonian limit ($\epsilon \rightarrow 0$). 
This 
point-particle has finite internal gravity since a typical 
scale 
of the self-gravitational field of the star, the mass over the radius, 
is finite. 
Thus we call this limit the {\it strong field point-particle limit}.

\subsection{Surface integral approach and body zone}
\label{sec:EIHApproach}
 
We construct a sphere for each star,  
called the {\it body zone} $B_A$ 
for the star $A$ $(A=1,2)$ \cite{Futamase87}, which   
surrounds each star and does not overlap the other. 
More specifically,  
the scalings of the radius and the mass of the star motivate
us to introduce the body zone of the star $A$,
$B_A \equiv\{x^i | |\vec{x} - \vec{z}_A(\tau)| < \epsilon R_A \}$
and a body zone coordinate of the star $A$, 
$\alpha_A^{\underline{i}} \equiv \epsilon^{-2} (x^i - z_A^i(\tau))$.
Here $z_A^i(\tau)$ is
a representative point of the star $A$, e.g., the center of mass
of the star $A$. $R_A$, called the body zone radius, is an arbitrary 
length scale (much smaller than the orbital separation 
while $\epsilon R_A$ is larger than the radius of the star for 
any $\epsilon$) 
and constant (i.e., $d R_A/d\tau =0)$. Using 
the body zone coordinate,
the star does not shrink when $\epsilon \rightarrow 0$, 
while the boundary of the body zone
goes to infinity.
Thus, it is appropriate to define the 
star's characteristic quantities such as a mass using 
the body zone coordinate. 

On the other hand, the body zone serves us  
as a surface $\partial B_A$, through which gravitational 
energy momentum flux 
flows and in turn it amounts to gravitational force exerting on the
star $A$.
Since the body zone boundary $\pa B_A$ is far away from the surface
of the star $A$, we can evaluate 
explicitly 
the gravitational energy momentum flux
on $\pa B_A$ 
with the post-Newtonian gravitational field. 
After computing the surface integrals, 
we make the body zone shrink to derive the equation of 
motion for the compact star.

Effects of the internal structures of the compact stars 
may be coded in, e.g.,  multipole moments of the stars. 
These moments in turn 
appear in the gravitational energy momentum flux in the surface 
integral approach and affect the orbital motion.

\subsection{Scalings on initial hypersurface}
\label{sec:Scaling}

Following the works   
\cite{Schutz80,FS83,Futamase83}, we take the initial value 
formulation to solve Einstein equations.  
As the initial data for matter variables and   
gravitational field, 
we take a set of nearly stationary solutions  
of the exact Einstein equations representing two widely separated 
fluid balls. We assume that these solutions 
are parametrized by $\epsilon$ 
and the matter and the field variables have the following 
scalings on the initial hypersurface. 

The matter density scales as $\epsilon^{-4}$ 
(in the $(t,x^i)$ coordinate), implied by the scalings 
of the mass and the radius of the star. 
The internal time scale of the star is assumed to be comparable to 
that of the binary orbital motion. 
We assume that the star is pressure-supported.

{}From these initial data we have the following scalings of the star $A$'s
stress energy tensor components in the body zone coordinate,
$T_A^{\mu\nu}$: 
$T_A^{\tau\tau} = O(\epsilon^{-2})$,
$T_A^{\tau \underline{i}} = O(\epsilon^{-4})$,    
$T_A^{\underline{i}\underline{j}} = O(\epsilon^{-8})$ 
on the initial hypersurface.
Here the underlined indexes mean that for any tensor $S^i$,
$S^{\underline i} = \epsilon^{-2} S^i$.
In paper I, we regard the star $A$'s body zone coordinate as a 
Fermi normal coordinate of the star and  
we have transformed $T_N^{\mu\nu}$,
the components of the stress energy tensor of the matter in the
near zone coordinate, to  $T_A^{\mu\nu}$ using a transformation
from the near zone coordinate to  
the Fermi normal coordinate at 1PN order.
It is difficult, however, to
construct the Fermi normal 
coordinate at a high post-Newtonian order. 
Therefore, we shall not use it.
We simply assume that for $T_N^{\mu\nu}$ (or rather for 
$\Lambda_N^{\mu\nu}$, the source term of the harmonically 
relaxed Einstein equations),
$
T_N^{\tau\tau} = O(\epsilon^{-2}),
T_N^{\tau \underline{i}} = O(\epsilon^{-4}),
$
and 
$
T_N^{\underline{i}\underline{j}} = O(\epsilon^{-8}),
$
as their leading scalings. 

As for the field variables on the initial hypersurface, 
we simply make a reasonable assumption that the field is of 
2.5PN order except for the field determined by the 
constraint equations. 
If we take random initial data for the field 
\cite{Schutz80} supposed to be of 1PN order, 
they are irrelevant to the dynamics of the binary system up to
the radiation reaction order 
\cite{Futamase83}. 
Thus, we expect that the 2.5PN order free data of the 
gravitational field on the initial hypersurface 
do not affect the orbital motion 
up to 3PN order.

\section{Formulation}
\label{chap:formulation}

\subsection{Field equation}
\label{sec:FieldEq}

As discussed in the previous section,
we shall express our equation of motion in terms of
surface integrals over the body zone boundary where it is assumed
that 
the metric slightly deviates from the 
flat (auxiliary) metric 
$\eta^{\mu\nu} = {\rm diag}{(-\epsilon^2,1,1,1)}$
(in the near zone coordinate $(\tau,x^i)$).
We define a deviation field $h^{\mu\nu}$ as
\begin{eqnarray}
&&
h^{\mu\nu} \equiv  \eta^{\mu\nu} - \sqrt{-g}g^{\mu\nu},
\end{eqnarray}
where $g$ is the determinant of the metric. 
Indexes of $h^{\mu\nu}$ 
are raised or lowered  by the flat metric.

Now we impose a harmonic coordinate condition on the metric
$
 h^{\mu\nu}{}_{,\nu}=0 , 
$
where the comma denotes a  partial derivative. 
In the harmonic gauge, we can recast Einstein 
equations into a relaxed form, 
\begin{equation}
\Box h^{\mu\nu} = -16\pi \Lambda^{\mu\nu} , 
\end{equation}
where $
\Box = \eta^{\mu\nu}\pa_{\mu}\pa_\nu$
is the flat d'Alembertian and   
\begin{eqnarray}
&&\Lambda^{\mu\nu} \equiv \Theta^{\mu\nu}
+\chi^{\mu\nu\alpha\beta}{}_{,\alpha\beta} ,
\label{DefOfLambda} \\ 
&&\Theta^{\mu\nu} \equiv (-g) (T^{\mu\nu}+t^{\mu\nu}_{LL}) , \\
&&\chi^{\mu\nu\alpha\beta} \equiv \frac{1}{16\pi} 
(h^{\alpha\nu}h^{\beta\mu}
-h^{\alpha\beta}h^{\mu\nu}) . 
\label{eq:DefOfChi} 
\end{eqnarray}
Here, $T^{\mu\nu}$ and $t^{\mu\nu}_{LL}$ denote the stress-energy tensor 
of the stars 
and the Landau-Lifshitz pseudotensor \cite{LL1975}. 
In consistency with the harmonic condition, a local   
energy momentum conservation law is expressed as
\begin{equation}
\Lambda^{\mu\nu}{}_{,\nu}=0 . 
\label{conservation}
\end{equation}
Note that $\chi^{\mu\nu\alpha\beta}\mbox{}_{,\alpha\beta}$ 
itself is divergence-free.

Now we rewrite the relaxed Einstein equations into an integral form,  
\begin{equation}
h^{\mu\nu}(\tau,x^i)=4 \int_{C(\tau, x^k)} d^3y 
\frac{\Lambda^{\mu\nu}(\tau-\epsilon|\vec x-\vec y|, y^k; \epsilon)} 
{|\vec x-\vec y|} + h^{\mu\nu}_H(\tau,x^i),  
\label{IntegratedREE}
\end{equation}
where $C(\tau, x^k)$ means the past light cone emanating 
{}from the event $(\tau, x^k)$. $C(\tau,x^k)$ is truncated on  
the 
$\tau =0$ initial hypersurface.  $h^{\mu\nu}_H$ is a homogeneous solution
of a homogeneous wave equation in flat spacetime. 
In this paper, we shall 
adopt the no-incoming radiation condition  
(see, e.g., \cite{Fock1959}) by taking sufficiently 
large $\tau$, i.e., $h^{\mu\nu}_H = 0$.

We solve the Einstein equations as follows. First we split the
integration region into two zones: the near zone and the far zone. 

The near zone is 
the neighborhood of the gravitational wave source where the wave 
character of the gravitational radiation is not manifest. 
In this paper, we define the near zone as a sphere centered at
some fixed point, enclosing both of the stars,  and having a radius 
${\cal R}/\epsilon$, where ${\cal R}$ is arbitrary
but larger than the binary separation and 
the gravitational wavelength.
The scaling of the near zone radius is derived from the 
$\epsilon$ dependence of 
the wavelength of the gravitational radiation emitted by the binary 
due to its orbital motion. 
The center of the near zone sphere would be 
determined, if necessary, for example, to be the center of 
mass of the near zone. The outside of 
the  near zone is the far zone where  the retardation effect of the 
field is crucial.

We evaluated  
the integrals over the far zone which contribute to
the near zone field where the stars reside 
using the {\it 
direct integration of the relaxed Einstein equations}  
(DIRE) method, which was initiated by 
Will and his collaborators 
\cite{PW02,WW96,PW00}. DIRE directly and nicely fits into our 
formalism since it utilizes the relaxed Einstein equations 
in the harmonic gauge.
Although we do not show our explicit computation in this paper,  
we have followed the DIRE method and 
checked that the far zone contribution does not affect the 
equation of motion through 3PN order.
In fact, Blanchet and Damour \cite{BD88}, and later 
Pati and Will \cite{PW00}, showed that 
the far zone contribution to the near zone field 
affects (physically) the orbital motion 
starting at 4PN order. 
Henceforth we shall focus our attention on the 
near zone contribution to the near zone field,  
\begin{equation}
h^{\mu\nu}(\tau,x^i)= 4 \int_{N} d^3y 
\frac{\Lambda^{\mu\nu}_N(\tau-\epsilon|\vec x-\vec y|, y^k; \epsilon)} 
{|\vec x-\vec y|},  
\label{IntegratedREEinNZ}
\end{equation}
where $N$ denotes the near zone and we attach the subscript 
$N$ to $\Lambda^{\mu\nu}$ to clarify that they are quantities 
in the near zone.

\subsection{Near zone contribution}

For the near zone contribution, we first 
make retardation expansion 
and change the integral region to 
a $\tau =$ const spatial hypersurface 
\begin{equation}
h^{\mu\nu}=4
\sum_{n=0} 
\frac{(-\epsilon)^n}{n!}\left(\frac{\pa}{\pa \tau}\right)^n
\int_{N} d^3y 
|\vec x-\vec y|^{n-1}
\Lambda_N^{\mu\nu}(\tau, y^k; \epsilon).
\label{RetardedEIREE}
\end{equation}
Note that the above integral depends on the arbitrary length
${\cal R}$ in general. The cancellation between the 
${\cal R}$-dependent 
terms in the far zone contribution and those in the near zone 
contribution was shown by Pati and Will \cite{PW00} through  
sufficient post-Newtonian order. Moreover, the findings in
\cite{PW00,BD88} make us expect that we can remove ${\cal R}$-dependent 
terms via a suitable gauge transformation and then take a formal limit of 
infinite ${\cal R}$ in an equation of motion up to the 3PN level.
With this
expectation in mind, in the following  we shall omit the terms 
with negative powers of ${\cal R}$ (${\cal R}^{-k}; k > 0$) which 
vanish in the ${\cal R}$ infinite limit, while we shall retain terms 
with positive powers of ${\cal R}$ (${\cal R}^{k}; k > 0$) 
and logarithmic terms  ($\ln {\cal R}$) and see as a good computational 
check that these ${\cal R}$-dependent terms can be gauged away from our 
final result. Another reason to keep 
logarithmic terms ($\ln {\cal R}$) is to make the arguments 
of all possible logarithmic terms nondimensional.

Second we split
the integral into two parts: contribution from the body zone
$B= B_1 \cup B_2$, 
and from elsewhere, $N/B$. Schematically we evaluate
the following two types of integrals (we omit the indexes):  
\begin{eqnarray}
&&
h = \sum_{n=0}
\left(h_{Bn} + h_{N/Bn}\right), 
\label{TotFieldRetExpandAbst} 
\\
&&
h_{Bn} = 
4
\frac{(-\epsilon)^n}{n!}\left(\frac{\pa}{\pa \tau}\right)^n
\epsilon^6 \sum_{A=1,2}\int_{B_A}
 d^3\alpha_A 
\frac{f(\tau,\vec z_A + \epsilon^2\vec{\alpha}_A)}
{|\vec r_A-\epsilon^2\vec{\alpha}_A|^{1-n}} ,  
\\
 &&
h_{N/Bn} = 
4
\frac{(-\epsilon)^n}{n!}\left(\frac{\pa}{\pa \tau}\right)^n
\int_{N/B}  d^3y 
\frac{f(\tau,\vec{y})}
{|\vec x-\vec y|^{1-n}},
\label{NBContributionSC}
 \end{eqnarray}
where $\vec r_A \equiv \vec x - \vec z_A$ and 
$f(\tau,x^k)$ is some function. 
We shall deal with these two contributions successively 
in the following.

\subsubsection{Body zone contribution}

As for the body zone contribution,  
we make  a multipole expansion being concerned with  
the scaling of the integrand, i.e., 
$\Lambda^{\mu\nu}$ in the body zone. 
For example, the $n=0$ terms in  Eq. (\ref{RetardedEIREE}),
$h_{B n=0}^{\mu\nu}$, give
\begin{eqnarray}
h^{\tau\tau}_{B n=0} &=& 4 \epsilon^4 \sum_{A=1,2}
\left(\frac{P_A^{\tau}}{r_A} + \epsilon^2 \frac{D_A^k r^k_A}{r_A^3} +
 \epsilon^4 \frac{3 I_A^{<kl>} r^k_A r^l_A}{2 r_A^5} +
 \epsilon^6 \frac{5 I_A^{<klm>} r^k_A r^l_A r_A^m}{2 r_A^7} 
\right)
\nonumber \\
\mbox{} &&+ O(\epsilon^{12}),
\label{hBtt} \\ 
h^{\tau i}_{B n=0} &=& 4 \epsilon^4 \sum_{A=1,2}
\left(\frac{P_A^{i}}{r_A} + \epsilon^2 \frac{J_A^{ki} r^k_A}{r_A^3} 
+ \epsilon^4 \frac{3 J_A^{<kl>i} r^k_A r^l_A}{2 r_A^5} 
 \right)
 + O(\epsilon^{10}),
\label{hBti} \\ 
h^{ij}_{B n=0} &=& 4 \epsilon^2 \sum_{A=1,2}
\left(\frac{Z_A^{ij}}{r_A} + \epsilon^2 \frac{Z_A^{kij} r^k_A}{r_A^3} +
 \epsilon^4 \frac{3 Z_A^{<kl>ij} r^k_A r^l_A}{2 r_A^5} 
+
 \epsilon^6 \frac{5 Z_A^{<klm>ij} r^k_A r^l_A r^m_A}{2 r_A^7} 
\right)
\nonumber \\
\mbox{} &&+ O(\epsilon^{10}), 
 \label{hBij}
\end{eqnarray}
where $ r_A \equiv |\vec{r}_A|$.
The quantity with $<>$ 
denotes symmetric and trace-free (STF) on the indexes 
between the brackets. 
To derive the 3PN equation of motion,  we need  
$h^{\tau\tau}$ up to $O(\epsilon^{10})$ and 
$h^{\tau i}$ and $h^{i j}$ up to $O(\epsilon^8)$.

In the above equations we defined multipole moments of the star $A$ 
as 
\begin{eqnarray}
&&
I_A^{K_l} \equiv \epsilon^2 \int_{B_A}
d^3\alpha_A \Lambda^{\tau \tau}_N \alpha_A^{\underline{K_l}}, 
\label{eq:NZCTTmoment}
\\   
&&
J_A^{K_li} \equiv \epsilon^4 \int_{B_A}
d^3\alpha_A \Lambda^{\tau \underline i}_N \alpha_A^{\underline{K_l}},  
\label{eq:NZCTImoment}
\\
&&
 Z_A^{K_lij} \equiv \epsilon^8 \int_{B_A}
d^3\alpha_A \Lambda^{\underline{i}\underline{j}}_N 
\alpha_A^{\underline{K_l}},  
\label{eq:NZCIJmoment}
\end{eqnarray}
where the capital index denotes a set of collective indexes, 
$I_l \equiv i_1 i_2 \cdot\cdot\cdot i_l$  and
$\alpha_A^{\underline{I_l}} \equiv
\alpha_A^{\underline i_1}\alpha_A^{\underline i_2}
\cdot\cdot\cdot\alpha_A^{\underline i_l}$.  
Then 
$P_A^{\tau} \equiv I^{I_0}_A$, $D_A^{i_1} \equiv I^{I_1}_A$,
and $P_A^{i_1} \equiv J^{I_1}_A $.
We simply call $P^{\mu}_A$ the four-momentum of the star $A$,
$P^{i}_A$ the (three-)momentum, and $P^{\tau}_A$ the energy. 
Also we call $D^{i}_A$ the dipole moment and
$I^{ij}_A$ the quadrupole moment.

Then we transform these moments into more convenient forms 
using the conservation law Eq. (\ref{conservation}). 
In the following, $v_A^i \equiv \dot{z}^i_A$, an overdot 
denotes $\tau$ time
derivative, and $\vec y_A \equiv \vec y - \vec z_A$. 
Noticing that the body zone remains unchanged
(in the near zone coordinate), i.e., $\dot R_A = 0$, we have 
\begin{eqnarray}
&&
P^{i}_A = P^{\tau}_A v^i_A + Q_A^i +
\epsilon^2 \frac{d D_A^i}{d\tau},    
\label{MomVelRelation} \\ 
&&
J^{ij}_A =
\frac{1}{2}
\left(M_A^{ij} + \epsilon^2 \frac{d I_A^{ij}}{d\tau}\right) 
+ v_A^{(i} D_A^{j)} + \frac{1}{2}\epsilon^{-2} Q_A^{ij},
\label{JijToMij}
\\
Z^{ij}_A &=& \epsilon^2 P^{\tau}_A v_A^i v_A^j + 
 \frac{1}{2}\epsilon^6\frac{d^2 I_A^{ij}}{d\tau^2} +
2\epsilon^4 v_A^{(i}\frac{d D_A^{j)}}{d\tau} +
\epsilon^4 \frac{d v_A^{(i}}{d\tau}D_A^{j)}  \nonumber \\
\mbox{} &+& 
\epsilon^2 Q_A^{(i}v_A^{j)} + \epsilon^2 R_A^{(ij)} +
\frac{1}{2}\epsilon^2\frac{d Q_A^{ij}}{d\tau},
\label{StrVelRelation}
\\
&&
Z^{kij}_A = \frac{3}{2}A_A^{kij} - A_A^{(ij)k},
\label{ZkijToAkij}
\end{eqnarray}
where 
\begin{eqnarray}
&&
M_A^{ij} \equiv 2\epsilon^4\int_{B_A}d^3\alpha_A
\alpha_A^{[\underline i}\Lambda_N^{\underline j]\tau} , \\
&&
Q_A^{K_li} \equiv \epsilon^{-4}
 \oint_{\pa B_A} dS_m
 \left(\Lambda^{\tau m}_N  - v_A^m 
\Lambda^{\tau\tau}_N \right) y_A^{K_l} y_A^i
\label{QL} , \\ 
&&
R_A^{K_lij} \equiv \epsilon^{-4}
 \oint_{\pa B_A} dS_m
 \left(\Lambda^{m j}_N  - v_A^m 
\Lambda^{\tau j}_N \right) y_A^{K_l} y_A^i,
\label{RL} 
\end{eqnarray}
and
\begin{eqnarray}
&&
A_A^{kij} \equiv \epsilon^2 J_A^{k(i}v_A^{j)} + \epsilon^2 v_A^k J_A^{(ij)}
+R_A^{k(ij)} + \epsilon^4 \frac{dJ_A^{k(ij)}}{d\tau}.
\end{eqnarray}
$[$ $]$ and $($ $)$ 
denote antisymmetrization and symmetrization on the indexes 
between the brackets.
$M_A^{ij}$ is the spin of the star $A$ and  Eq. (\ref{MomVelRelation})
gives a  momentum-velocity relation. Thus our momentum-velocity 
relation is a direct analogue of the Newtonian momentum-velocity
relation. In general, 
we have\footnote{The equation in paper II corresponding to
Eq. (\ref{ZLijToJLij}) has a misprint, though it does not 
affect the 2.5PN equation of motion.}
\begin{eqnarray}
&&
J_A^{K_li} = J_A^{(K_li)} +  \frac{2l}{l+1}J_A^{(K_{l-1}[k_l)i]}, \\
&&
Z_A^{K_lij} = \frac{1}{2}
\left[Z_A^{(K_li)j} + \frac{2 l}{l+1}Z_A^{(K_{l-1}[k_l)i]j} +
Z_A^{(K_lj)i} + \frac{2 l}{l+1}Z_A^{(K_{l-1}[k_l)j]i} \right],
\end{eqnarray}
and 
\begin{eqnarray}
&&
J_A^{(K_li)} = \frac{1}{l+1}\epsilon^2\frac{d I_A^{K_li}}{d\tau}
+ v_A^{(i}I_A^{K_l)} + \frac{1}{l+1}\epsilon^{-2l}Q_A^{K_li}, 
 \label{JLiToILi}
\\
&&
Z_A^{(K_li)j} + Z_A^{(K_lj)i} =
\epsilon^2 v_A^{(i}J_A^{K_l)j} +
\epsilon^2 v_A^{(j}J_A^{K_l)i} +
\frac{2}{l+1}
\epsilon^4\frac{d J_A^{K_l(ij)}}{d\tau}
+ 
\frac{2}{l+1}
\epsilon^{-2l + 2}R_A^{K_l(ij)}.
\label{ZLijToJLij} 
\end{eqnarray}
The surface integrals $Q_A^{K_li}$ and $R_A^{K_lij}$ 
will be computed in Appendix \ref{RLijandQLi} and 
do contribute 
to the field and the equation of motion starting at  3PN 
order.

\subsubsection{$N/B$ contribution}

About the $N/B$ contribution, 
since the integrand 
$\Lambda_N^{\mu\nu} = -gt_{LL}^{\mu\nu} +
\chi^{\mu\nu\alpha\beta}\mbox{}_{,\alpha\beta}$ is at least 
quadratic in the small deviation field $h^{\mu\nu}$,
we make the post-Newtonian expansion in the integrand. 
Then, basically, with the help of (super)potentials $g(\vec{x})$
which satisfy $\Delta g(\vec x) = f(\vec{x})$,
$\Delta$ denoting a Laplacian, we have for each integral 
(e.g., the $n=0$ term in Eq. (\ref{NBContributionSC}))    
\begin{eqnarray}
\int_{N/B} d^3y \frac{f(\vec{y})}{|\vec{x}-\vec{y}|}  &=&
- 4 \pi g(\vec{x}) + \oint_{\pa(N/B)}dS_k
\left[\frac{1}{|\vec{x}-\vec{y}|}
 \frac{\pa g(\vec{y})}{\pa y^k} - g(\vec{y})\frac{\pa}{\pa y^k}
 \left(\frac{1}{|\vec{x} - \vec{y}|}\right) \right].  
\label{NBcontribution}
\end{eqnarray} 
We show a derivation of Eq. (\ref{NBcontribution}) in Appendix 
\ref{sec:proofeq331}.
For $n \ge 1$ terms in Eq. (\ref{NBContributionSC}), we use
(super)potentials many times to convert all the volume integrals into surface
integrals and the bulk terms (``$- 4 \pi g(\vec x)$'').   
\footnote{Notice that when solving a Poisson equation 
$\Delta g(\vec x) = f(\vec{x})$, a particular solution 
suffices for our purpose. By virtue of the surface integral 
term in Eq. (\ref{NBcontribution}), it is not necessary to be 
concerned about a homogeneous solution of the Poisson equation.}

Finding the superpotentials is one of the most formidable 
tasks, especially when we proceed to a high post-Newtonian order.   
Fortunately, up to 2.5PN order, all the required 
superpotentials are available \cite{BFP98,JS98}.  
At 3PN order, there appear  
many integrands  
for which we could not find the required superpotentials.  To 
obtain a 3PN equation of motion, we use an alternative 
method. The details of the method will be explained 
later.    

Finally, we note that 
$h^{\mu\nu} = O(\epsilon^4)$  
as shown in paper II.

\subsection{General form of the equation of motion}

{}From the definition of the four-momentum,  
\begin{eqnarray}
&&
P^{\mu}_A(\tau) \equiv \epsilon^2 \int_{B_A} d^3\alpha_A \Lambda^{\tau \mu}_N, 
\label{DefOfMomentum} 
\end{eqnarray}
and the conservation law, Eq. (\ref{conservation}), we have an 
evolution equation for the four-momentum,  
\begin{equation}
\frac{dP_A^{\mu}}{d\tau} = -\epsilon^{-4}
 \oint_{\pa B_A} dS_k \Lambda^{k\mu}_N
+\epsilon^{-4}
v_A^k \oint_{\pa B_A} dS_k \Lambda^{\tau\mu}_N. 
\label{EvolOfFourMom}
\end{equation}
Here we used the fact that the size and the shape of the
body zone are defined to be time-independent 
(in the near zone coordinate), while
the center of the body zone moves at the velocity of the star's 
representative point.

Substituting  the momentum-velocity relation, Eq. (\ref{MomVelRelation}), 
into the spatial components of  Eq. (\ref{EvolOfFourMom}), we obtain
the general form of the equation of motion for the star $A$,    
\begin{eqnarray}
P_A^{\tau}\frac{dv_A^i}{d\tau} &=&
 -\epsilon^{-4}
 \oint_{\pa B_A} dS_k \Lambda^{ki}_N
+ \epsilon^{-4}
v_A^k \oint_{\pa B_A} dS_k \Lambda^{\tau i}_N 
\nonumber\\
&&
 +\epsilon^{-4}
 v_A^i \left( \oint_{\pa B_A} dS_k \Lambda^{k\tau}_N
-v_A^k \oint_{\pa B_A} dS_k \Lambda^{\tau\tau}_N \right)
\nonumber\\
&&-\frac{dQ_A^i}{d\tau}  - \epsilon^2 \frac{d^2 D_A^i}{d \tau^2}.  
\label{generaleom}
\end{eqnarray} 

All the right-hand side terms in Eq. (\ref{generaleom}) except for the
dipole moment are expressed as surface integrals. 
We can specify the
value of $D_A^i$ freely to determine the representative point
$z_A^i(\tau)$ of the star $A$. For example, we may call 
$z_A^i(\tau)$ corresponding to $D_A^i = 0$ the center of 
mass of the star $A$ from an analogy of the Newtonian dynamics.

In Eq. (\ref{generaleom}), $P_A^{\tau}$
rather than the mass of the star $A$ appears. Hence
we have 
to derive a relation between the mass and $P_A^{\tau}$.
We shall derive the relation by solving the temporal component of
the evolution equation (\ref{EvolOfFourMom}) 
functionally. In fact, 
at lowest order, we have shown in paper II that  
\begin{eqnarray}
\frac{dP_A^{\tau}}{d\tau} &=& O(\epsilon^2).
\label{Eq:LowestPtaudot}
\end{eqnarray}
Then we define the mass of the star $A$ as  
the integrating constant of this equation,    
\begin{eqnarray}
&& 
m_A \equiv \lim_{\epsilon \to 0}P_A^{\tau}.  
\label{DefOfMass}
\end{eqnarray}
$m_A$ is the ADM mass that the star $A$ would have if it
 were isolated.  
We took the $\epsilon$ zero limit in Eq. (\ref{DefOfMass}) 
to ensure that the mass defined above  does not include the effect of the
companion star and the orbital motion of the star itself. 
Some subtleties
about this definition are discussed in paper II.
By definition, $m_A$ is constant. The procedure that we use to solve the 
evolution equation of $P_A^{\tau}$ and obtain the mass 
energy relation is achieved up to 3PN order successfully 
and the result will be shown later.

Since the general form of the equation of motion is  
expressed in terms of surface integrals 
over the body zone boundary,  
we can derive an 
equation of motion
for a strongly self-gravitating star
using the post-Newtonian approximation.   
Effects of the star's internal structure 
on the orbital motion  
such as tidally induced multipole moments  
appear through the field and hence the integrand 
$\Lambda_N^{\mu\nu}$ of the surface integrals.

\subsection{Lorentz contraction and multipole moments}
\label{LorentzContraction}

In this paper, we are concerned with a binary consisting of 
two spherically symmetric compact stars. In other words,
all the multipole moments of the star defined in an  
appropriate reference coordinate 
where effects of its orbital motion and 
the companion star are removed (modulo, namely,
the tidal effect) vanish.  
We adopt the generalized Fermi coordinate (GFC) 
\cite{AB86} as a star's reference coordinate for this purpose 

Then a question specific to our formalism is whether   
the differences between the multipole moments 
defined in Eqs. (\ref{eq:NZCTTmoment}), (\ref{eq:NZCTImoment}), 
and (\ref{eq:NZCIJmoment}) and 
the multipole moments in GFC give purely monopole terms.  
This problem is addressed in 
Appendix \ref{Appendix:CorrectionToMoments} and  
the differences are mainly attributed to the shape of 
the body zone. The body zone $B_A$ which is spherical 
in the near zone coordinate (NZC) is not spherical in the GFC 
mainly because of a kinematic effect (Lorentz contraction). 
To derive a 3PN equation of 
motion, it is sufficient to compute the difference 
in the STF quadrupole moment up to 1PN order.    
The result is 
\begin{eqnarray}
\delta I_A^{<ij>} 
&\equiv&  
I_{A,{\rm NZC}}^{<ij>} 
- 
I_{A,{\rm GFC}}^{<ij>}  
= - 
\epsilon^2
\frac{4 m_A^3}{5}v_A^{<i}v_A^{j>}
+ O(\epsilon^3), 
\label{LC:CorrectionToSTFQuadrupole}
\end{eqnarray}
where $I_{A,{\rm NZC}}^{ij} \equiv I_{A}^{ij}$.  
$I_{A,{\rm GFC}}^{ij}$ is the quadrupole moments defined 
in the generalized Fermi coordinate.
  
As is obvious from Eq. (\ref{LC:CorrectionToSTFQuadrupole}), 
this difference appears even if 
the companion star does not exist. 
We note that we could derive the 
3PN metric 
for an isolated star $A$ moving at a constant velocity 
using our method explained in this section 
by simply letting the mass of the companion star be zero.  
Actually, 
$\delta I_A^{<ij>}$ above is a necessary 
term which makes the so-obtained 3PN 
metric the same as the Schwarzschild metric 
boosted at the constant velocity $\vec v_A$ 
in the harmonic coordinate.

\subsection{On the arbitrary constant $R_A$}
\label{Arbitrariness}

Our final remark in this section is on the two  
arbitrary constants $R_A$. 
Since we introduce the body zones by hand,
the arbitrary body zone radii $R_A$
seem to appear in the metric, the multipole moments of the stars,  
and the equation of motion. As was discussed in detail in 
paper II,  we proved that the surface integrals 
in Eq. (\ref{generaleom}) 
that we evaluate to derive the equation of motion do  
not depend on $R_A$ through any order of the post-Newtonian 
iteration. As for the 
field and the multipole moments 
appearing in the field 
(mass, spin, quadrupoles, etc.), 
we reasonably expect that the $R_A$-dependent terms in the 
body zone contribution $h^{\mu\nu}_B$ and $N/B$ contribution 
$h^{\mu\nu}_{N/B}$ cancel each other out, since 
the total field $h^{\mu\nu} = h_{B}^{\mu\nu} + h_{N/B}^{\mu\nu}$ 
is obviously  independent of $R_A$.
In the cancellation, the multipole moments should 
be ``renormalized'' 
so that those moments do not depend on $R_A$. 
Such  
cancellation among $R_A$-dependent terms and 
the ``renormalization''  
was demonstrated explicitly 
in paper I up to 1PN order. 

Practically, the above observation enables us to discard 
safely all the $\epsilon R_A$-dependent terms except for 
logarithms of $\epsilon R_A$. We retain 
$\ln \epsilon R_A$-dependent 
terms to nondimensionalize the arguments of the logarithms.
Thus, instead of performing renormalization of 
the multipole moments, we simply discard the $\epsilon R_A$ 
dependences other than the $\ln \epsilon R_A$ dependences.
We emphasize here that we discard the  $\epsilon R_A$-dependent 
terms in the field first 
and  then evaluate the surface integrals in the 
general form of the equation of motion using the field which is 
independent of 
$\epsilon R_A$. We then 
discard the $\epsilon R_A$-dependent terms  
arising in the 
computation of the surface integrals. 

Appendix \ref{sec:renommultipole} is devoted to an explanation 
on the renormalization of the stars' multipole moments. There, we also 
give a justification for our procedure of discarding through  
our computation of  the field all the $\epsilon R_A$-dependent 
terms other than the $\ln \epsilon R_A$-dependent terms.

\section{Structure of the 3PN Equation of Motion}
\label{chap:3PNEOM}

In the following sections, we shall 
derive an acceleration for two spherical compact stars 
through third post-Newtonian accuracy. 

First, we split the four-momentum, the dipole moment, and 
the $Q_A^i$ integral into two parts, namely the $\Theta$ part and 
the $\chi$ part. 
\begin{eqnarray}
&&
P_{A\Theta}^{\mu} \equiv \epsilon^2 \int_{B_A}
d^3\alpha_A \Theta^{\mu \tau}_N, 
\\
&&
P_{A\chi}^{\mu} \equiv \epsilon^2 \int_{B_A}
d^3\alpha_A
\chi_N^{\mu\tau\alpha\beta}\mbox{}_{,\alpha\beta},   
\\
&&
D_{A\Theta}^i \equiv 
\epsilon^2 \int_{B_A}d^3\alpha_A 
\alpha_A^{\underline{i}}
\Theta_N^{\tau\tau},
\\  
&&
D_{A\chi}^i \equiv 
\epsilon^2 \int_{B_A}d^3\alpha_A 
\alpha_A^{\underline{i}}
\chi_N^{\tau\tau\alpha\beta}\mbox{}_{,\alpha\beta},  
\\
&&
Q_{A\Theta}^i \equiv   \epsilon^{-4}
 \oint_{\pa B_A} dS_k
 \left(\Theta^{\tau k}_N  - v_A^k 
\Theta^{\tau\tau}_N \right) y_A^i, 
\\ 
&&
Q_{A\chi}^i \equiv   \epsilon^{-4}
 \oint_{\pa B_A} dS_k
 \left(\chi^{\tau k\alpha\beta}_N\mbox{}_{,\alpha\beta}  - v_A^k 
\chi^{\tau\tau\alpha\beta}_N\mbox{}_{,\alpha\beta} \right) y_A^i. 
\end{eqnarray}
Correspondingly, we split the momentum-velocity relation 
Eq. (\ref{MomVelRelation}) and 
the evolution equation for the four-momentum 
Eq. (\ref{EvolOfFourMom}) into the $\Theta$ part 
and the $\chi$ part.

Now, $P_{A\chi}^{\mu}$ and $D_{A\chi}^i$ as well as $Q_{A\chi}^i$ 
can be expressed completely as surface integrals, and 
can be computed explicitly into functions of $m_A$, $\vec v_A$,  
and $\vec r_{12}$. It is straightforward to compute them 
up to 3PN order, since we only need the 2PN field 
to perform the surface integrals. 
The results are shown in Appendix \ref{chipart}.
There, we also compute the $\chi$ part of the 
momentum-velocity relation and 
the evolution equation for $P_{A\chi}^{\mu}$. Then 
comparing these 
equations with $P_{A\chi}^{\mu}$, $D_{A\chi}^i$, and 
$Q_{A\chi}^i$, we found that the $\chi$ part of these equations 
is an  identity up to 3PN order. 
This observation then means that the nontrivial 
momentum-velocity relation and the mass-energy relation, an 
equation of motion, are obtained from the $\Theta$ part 
of Eqs. (\ref{MomVelRelation}) and (\ref{EvolOfFourMom}).

Therefore, the equations that 
we have to evaluate to derive
an evolution equation for the energy and an  
equation of motion are actually 
\begin{eqnarray}  
\left(\frac{dP_{1\Theta}^{\tau}}{d\tau}\right)_{\le 3{\rm PN}} &=& 
\left(\frac{dP_{1\Theta}^{\tau}}{d\tau}\right)_{\le 2.5{\rm PN}} + 
\epsilon^6 \left[- 
\oint_{\pa B_1}dS_k\mbox{}_{10}\Theta_{N}^{\tau k} 
+ v_1^k 
\oint_{\pa B_1}dS_k\mbox{}_{10}\Theta_{N}^{\tau \tau}
\right], 
\label{EvolOfFourMom3PN}
\\ 
m_1 \left(\frac{d v_1^i}{d\tau}\right)_{\le 3{\rm PN}} 
&=& m_1\left(\frac{dv_1^i}{d\tau}\right)_{\le 2.5{\rm PN}} + 
\epsilon^6 \left[- 
\oint_{\pa B_1}dS_k\mbox{}_{10}\Theta_{N}^{ki} 
+ v_1^k 
\oint_{\pa B_1}dS_k\mbox{}_{10}\Theta_{N}^{\tau i}
\right] 
\nonumber \\
\mbox{} &&
+ \epsilon^6 
\left(\frac{dP_{1\Theta}^{\tau}}{d\tau}\right)_{3{\rm PN}} v_1^i
+ \epsilon^6 \left(
(m_1 - P_{1\Theta}^{\tau})\frac{d v_1^i}{d\tau}\right)_{3{\rm PN}} 
\nonumber \\
\mbox{} &&
- \epsilon^6 \frac{d \mbox{}_6 Q_{1\Theta}^i}{d\tau} 
- \epsilon^6 \frac{d^2 \mbox{}_4 D_{1\Theta}^i}{d\tau^2},     
\label{generaleom3PN}
\end{eqnarray}
where for an equation or a quantity $f$, 
$(f)_{\le n{\rm PN}}$ 
and $(f)_{n{\rm PN}}$ denote   
$f$ up to $n$PN order and $f$ at $n$PN order, 
respectively. 
$\mbox{}_{\le n}f$ and  
$\mbox{}_n f$, on the other hand, denote an equation or a 
quantity $f$ up to $O(\epsilon^n)$ and at $O(\epsilon^n)$. 
In paper II, we found $Q_{A\Theta}^i = O(\epsilon^6)$. 
For later convenience, in Eq. (\ref{generaleom3PN}) 
we retain $D_{A\Theta}^i$ of order $\epsilon^4$, 
which appears at a 3PN equation of motion
(in other words, we set $\mbox{}_{\le 3}D_{A\Theta}^i = 0$). 
It should be understood that 
in the second line of Eq. (\ref{generaleom3PN}), the acceleration 
$dv_1^i/d\tau$ should be replaced by the acceleration of 
an appropriate order lower than 2.5PN.  
We note that the $\chi$ parts of 
$Q_{A}^{K_li}$ and $R_{A}^{K_lij}$ integrals and 
the multipole moments including 
$P_{A\chi}^{\mu}$ and $D_{A\chi}^{i}$ 
affect an equation of motion through the field and hence the 
integrands of Eqs. (\ref{EvolOfFourMom3PN}) and (\ref{generaleom3PN}).
Henceforth, we call Eq. (\ref{generaleom3PN}) the general form 
of the 3PN equation of motion.

The explicit forms of the integrands 
$\mbox{}_{10}\Theta_N^{\mu\nu} = 
\mbox{}_{10}[(-g)t_{LL}^{\mu\nu}]$ (on $\pa B_A$) 
are (see Appendix \ref{pNELLPT}),  
\begin{eqnarray}
16\pi\mbox{}_{10}\Theta_{N}^{\tau\tau} &=&     
- \frac{7}{4}
\mbox{}_4h^{\tau\tau}\mbox{}_{,k}
\mbox{}_8h^{\tau\tau,k}  + \cdots, \\
16\pi\mbox{}_{10}\Theta_{N}^{\tau i} &=&     
2 \mbox{}_4h^{\tau\tau}\mbox{}_{,k}
\mbox{}_8h^{\tau [k,i]} + \cdots, \\ 
16\pi\mbox{}_{10}\Theta_{N}^{ij} &=&     
\frac{1}{4}
(\delta^i\mbox{}_k\delta^j\mbox{}_l + \delta^i\mbox{}_l\delta^j\mbox{}_k - 
\delta^{ij}\delta_{kl})
\left\{
\mbox{}_4h^{\tau\tau,k}
(\mbox{}_{10}h^{\tau\tau,l}
+ \mbox{}_8h^m\mbox{}_m\mbox{}^{,l} 
+ 4\mbox{}_8h^{\tau l} \mbox{}_{,\tau}
) +
8 \mbox{}_4h^{\tau}\mbox{}_m\mbox{}^{,k}\mbox{}_8h^{\tau [l,m]}
\right\}
\nonumber \\
\mbox{} &+&  2\mbox{}_4h^{\tau i}\mbox{}_{,k}\mbox{}_8h^{\tau [k,j]}
+ 2\mbox{}_4h^{\tau j}\mbox{}_{,k}\mbox{}_8h^{\tau [k,i]}
+ \cdots.  
\end{eqnarray}
The fields up to 2.5PN order, 
$\mbox{}_{\le 9}h^{\tau\tau}$, $\mbox{}_{\le 7}h^{\tau i}$, 
and $\mbox{}_{\le 7}h^{ij}$, are listed in paper II. 
Thus, to derive the 3PN mass-energy relation and 
the 3PN momentum-velocity relation, we have to derive 
$\mbox{}_8h^{\tau i}$. To derive the 3PN equation of motion, 
we further need $\mbox{}_{10}h^{\tau\tau} + \mbox{}_8h^k\mbox{}_k$.

Up to 2.5PN order, the superpotentials required  
to compute the field could be found \cite{BFP98,JS98}. 
However, at 3PN order, it is quite 
difficult to complete the required superpotentials. We take 
another method to overcome this problem. We shall explain our 
method in detail in Secs. \ref{sec:derivatoinOfh8ti} and 
\ref{TPNGravitationalField}. Below, 
we begin our calculation by deriving $\mbox{}_8h^{\tau i}$ and 
check (a part of) the 3PN harmonic condition 
$\mbox{}_{\le 8}h^{\tau \mu}\mbox{}_{,\mu} =0$.

\section{$\mbox{}_8h^{\tau i}$: $N/B$ integrals}
\label{sec:derivatoinOfh8ti}

In this section, we derive $\mbox{}_8h^{\tau i}$,   
\begin{eqnarray}
\mbox{}_{\le 8}h^{\tau i} &=& \mbox{}_{\le 8}h^{\tau i}_B 
+ 4 \epsilon^8 \int_{N/B}\frac{d^3y
\mbox{}_8\Lambda_N^{\tau i} 
}{|\vec x - \vec y|}
+ 
2 \epsilon^8 \frac{\pa^2}{\pa \tau^2}\left[
\int_{N/B} d^3y |\vec x - \vec y|
\mbox{}_6\Lambda_N^{\tau i}\right], 
\label{eq:h8tiGenkei}
\end{eqnarray}
where $\mbox{}_{\le 8}h^{\tau i}_B$ is the body zone contribution 
up to $O(\epsilon^8)$ and shown in Appendix \ref{RLijandQLi}  
as Eq. (\ref{hBtiInAp}).
The $Q_A^{kli}$ integral contained in $\mbox{}_{\le 8}h^{\tau i}_B$ 
is found in Appendix \ref{RLijandQLi} to vanish. 
We are concerned with the equation of motion for two 
spherical compact stars and we shall only retain monopole terms 
in $\mbox{}_{\le 8}h^{\tau i}_B$.

The second time derivative term in the retardation expansion 
(the last term in Eq. (\ref{eq:h8tiGenkei})) 
can be integrated explicitly via  
super-superpotentials (i.e., a particular solution 
of the Poisson equation with a superpotential as a source). 
The result is 
\begin{eqnarray}
\int_{N/B}d^3y |\vec x - \vec y|\mbox{}_6\Lambda^{\tau i}_N
&=&  \frac{11 P_1^{\tau}P_1^i}{3}
\ln\left(\frac{r_1}{{\cal R}/\epsilon}\right) + 
\frac{11P_1^{\tau}P_1^i}{6} - \frac{P_1^{\tau}P_1^k}{12}n_1^{<ik>} 
+ \frac{11P_1^{\tau}P_2^i}{6}  
\nonumber \\
\mbox{} &&+ 
8P_1^{\tau}P_2^{k}\left(\delta^{i}\mbox{}_k\Delta_{12} - 
\pa_{z_1^k}\pa_{z_2^i} + \frac{3}{4}\pa_{z_1^i}\pa_{z_2^k} 
\right)f^{(\ln S)} 
\nonumber \\
\mbox{}  &&- 
 \frac{11P_1^{\tau}P_2^i}{3}
\ln\left(\frac{2{\cal R}}{\epsilon}\right) + (1 \leftrightarrow 2), 
\end{eqnarray}
where $\Delta_{AA'} \equiv \delta^{ij}\pa_{z_A^i}\pa_{z_{A'}^j}$ 
and $\pa_{z_{A}^i} = \pa/\pa z_A^i$.      
The symbol $(1 \leftrightarrow 2)$ denotes  
the same terms but with the star's label 1 exchanged for 2. 
$f^{(\ln S)}$ is a superpotential satisfying 
$\Delta f^{(\ln S)} = \ln S$,  
where $S \equiv r_1 + r_2 + r_{12}$,  
and its explicit expression
is given in \cite{JS98} as 
\begin{eqnarray}
f^{(\ln S)} =&& \frac{1}{36}(-r_{1}^2 + 3 r_{1} r_{12} 
+ r_{12}^2 - 3 r_{1}r_{2} + 3 r_{12}r_{2} - r_{2}^2) 
+\frac{1}{12}(r_{1}^2 - r_{12}^2 + r_{2}^2) \ln S.
\end{eqnarray}

Now let us devote ourselves to an evaluation of the 
Poisson integral with the integrand 
$\mbox{}_8\Lambda_N^{\tau i}$.    
For this integrand, $\mbox{}_8\Lambda_N^{\tau i}$, 
it is difficult to find all the required superpotentials. 
We proceed as follows.  
First, we split the integrand $\mbox{}_8\Lambda_N^{\tau i}$ 
into two groups, one whose members depend on the negative power or 
logarithms of  $S$ or both, 
and the other whose members do not. 

The S-dependent integrands are 
\begin{eqnarray*}  
&& {\rm S-dependent~parts~of}~\{
- \mbox{}_4h^{kl,i}\mbox{}_4h^{\tau}\mbox{}_{k,l} 
+ 2 \mbox{}_4h^{i (k,l)}\mbox{}_4h^{\tau}\mbox{}_{k,l}
+ \mbox{}_4h^{ik}\mbox{}_{,\tau}
\mbox{}_4h^{\tau\tau}\mbox{}_{,k}
+ 2 \mbox{}_4h^{\tau\tau}\mbox{}_{,k}
\mbox{}_6h^{\tau[k,i]}\}.   
\end{eqnarray*}
For the remaining integrands in $\mbox{}_8\Lambda_N^{\tau i}$, 
we found all the required superpotentials except for 
essentially two Poisson equations 
whose source terms are $1/r_1^2/r_2$ and 
$(\ln r_1)/r_2^3$. 
Then in the following, we split the integrands into three groups:
\begin{description}
\item[(a)] Direct-integration part $=$ S-dependent group:\\
 the integrands which depend on inverse powers of S  
($S^k$, $k:$ negative integer), or have logarithms of S ($\ln S$), 
or both. 
\item[(b)] Superpotential part:\\   
the integrands which do {\it not} depend on inverse powers of S, 
 nor have logarithms of S,  
and for which the required superpotentials are available,      
\item[(c)] Superpotential-in-series part:\\  
the integrands which do {\it not} depend on  inverse powers of S, 
 nor have logarithms of S,  
and for which the required superpotentials are unavailable.     
\end{description}

We mention here that splitting the integrands into the S-dependent 
and the S-independent group is a rather rough technique; there may be 
some terms in the S-dependent group for which we could find  
superpotentials. We have made such
a classification since for the S-dependent group it seemed difficult to
find superpotentials.

For example, let us take 
$ - \mbox{}_4h^{kl,i}\mbox{}_4h^{\tau}\mbox{}_{k,l}$. Then  
\begin{eqnarray}
\lefteqn{
\left[
- \mbox{}_4h^{kl,i}\mbox{}_4h^{\tau}\mbox{}_{k,l}
\right]_{{\rm DIP}} 
\equiv 
{\rm direct-integration~part~of~} \{
- \mbox{}_4h^{kl,i}\mbox{}_4h^{\tau}\mbox{}_{k,l}\} }  \nonumber \\ &=& 
\frac{8m_1^2m_2}{r_1^2S} v_1^i
\left(
- \frac{2}{r_1r_{12}}+\frac{2}{r_1r_2}+\frac{1}{r_{12}r_2}
-\frac{r_{12}}{r_1^2r_2}+\frac{r_2}{r_1^2r_{12}}
\right) 
\nonumber \\
\mbox{} &&
+ \frac{16m_1^2m_2}{r_1S^2}n_1^i
\left(
-\frac{1}{r_1r_2}(\vec n_1\cdot\vec v_1) 
+\left(
- \frac{1}{r_{12}^2}
- \frac{2}{r_1r_{12}}
-\frac{2}{r_1^2}
-\frac{r_{12}}{r_1^2r_2}
+\frac{2r_2}{r_1^2r_{12}}
+\frac{r_2^2}{r_1^2r_{12}^2}
\right)(\vec n_{12}\cdot\vec v_1)  
\right. 
\nonumber \\
\mbox{} &&
\left.
+ 
\left(
\frac{r_2}{r_1^2r_{12}}
-\frac{1}{r_1^2}
-\frac{1}{r_1r_{12}}
\right)(\vec n_{2}\cdot\vec v_1)  
\right)
\nonumber \\
\mbox{} &&
+  \frac{16m_1^2m_2}{S^2}n_2^i
\left(
-\frac{1}{r_1^3}
+\frac{1}{r_{12}^3}
+\frac{2}{r_1r_{12}^2}
+\frac{1}{r_1^2r_{12}}
-\frac{2r_2}{r_1^2r_{12}^2}
-\frac{r_2}{r_1^3r_{12}}
-\frac{r_2^2}{r_1^2r_{12}^3}
\right)(\vec n_1\cdot\vec v_1)
\nonumber \\
\mbox{} &&
+  \frac{16m_1^2m_2}{S^2}n_2^i
\left(
- \frac{2}{r_1^3} - \frac{1}{r_1^2r_2}
-\frac{r_{12}}{r_1^3r_2} 
-\frac{r_2}{r_1r_{12}^3}
-\frac{2r_2}{r_1^2r_{12}^2}
+\frac{2r_2^2}{r_1^3r_{12}^2}
+\frac{r_2^3}{r_1^3r_{12}^3}
\right)(\vec n_2\cdot\vec v_1)
\nonumber \\
\mbox{} &&
+ (1 \leftrightarrow 2), 
\label{eq5-11}
\end{eqnarray}
\begin{eqnarray}
\lefteqn{
\left[
- \mbox{}_4h^{kl,i}\mbox{}_4h^{\tau}\mbox{}_{k,l}
\right]_{{\rm SP}} 
\equiv 
{\rm superpotential~part~of~}\{
- \mbox{}_4h^{kl,i}\mbox{}_4h^{\tau}\mbox{}_{k,l}\}}  \nonumber \\ &=& 
\frac{4m_1^3}{r_1^5}\left(
v_1^i - 3 (\vec n_1\cdot\vec v_1)n_1^i
\right) 
-\frac{16m_1^2 v_1^2}{r_1^4}(\vec n_1\cdot\vec v_1)n_1^i
\nonumber \\
\mbox{} &&+
\frac{2m_1^2m_2}{r_1^2}\left(
\frac{4 r_{12}^2}{r_1^2r_2^3}(\vec n_1\cdot\vec v_2)n_1^i
+\left(
\frac{1}{r_2^2}-\frac{r_{12}^2}{r_1^2r_2^2}+\frac{1}{r_1^2}
\right)\frac{v_2^i}{r_2}
\right)
- \frac{16m_1m_2
(\vec n_2\cdot\vec v_1)(\vec v_1\cdot\vec v_2)}{r_1^2r_2^2}n_1^i
\nonumber \\
\mbox{} &&
+ (1 \leftrightarrow 2), 
\label{eq5-12}
\end{eqnarray}
\begin{eqnarray}
\lefteqn{
\left[
- \mbox{}_4h^{kl,i}\mbox{}_4h^{\tau}\mbox{}_{k,l}
\right]_{{\rm SSP}} 
\equiv 
{\rm superpotential-in-series~part~of~}\{
- \mbox{}_4h^{kl,i}\mbox{}_4h^{\tau}\mbox{}_{k,l}\}}  \nonumber \\ &=& 
\frac{4m_1^2m_2}{r_1^2}\left(
-\frac{2}{r_2^3}(\vec n_1\cdot\vec v_2)n_1^i
-\frac{2}{r_1^2r_2}(\vec n_1\cdot\vec v_2)n_1^i
+ \frac{1}{r_1r_2^2}(\vec n_1\cdot\vec v_2)n_2^i
\right)
\nonumber \\
\mbox{} &&
+ (1 \leftrightarrow 2). 
\label{eq5-13}
\end{eqnarray}

Below, we first consider the superpotential part of 
$\mbox{}_8\Lambda_N^{\tau i}$.

\subsection{Superpotential part}
\label{ExplanationforSPP} 

The integrands $\mbox{}_8\Lambda_N^{\tau i} - 
({\rm S-dependent~parts~of}~\{
- \mbox{}_4h^{kl,i}\mbox{}_4h^{\tau}\mbox{}_{k,l} 
+ 2 \mbox{}_4h^{i (k,l)}\mbox{}_4h^{\tau}\mbox{}_{k,l}
+ \mbox{}_4h^{ik}\mbox{}_{,\tau}
\mbox{}_4h^{\tau\tau}\mbox{}_{,k}
+ 2 \mbox{}_4h^{\tau\tau}\mbox{}_{,k}
\mbox{}_6h^{\tau[k,i]}\}) $ can be simplified into a form which is 
independent of $S$. For the so-obtained S-independent group,  we 
have to find particular solutions of Poisson equations 
whose 
source terms are,\footnote{What particular solutions are required 
depends on how we simplify the integrands.}  
\begin{eqnarray}
&&
\left\{
\frac{1}{r_1^5}, \frac{1}{r_1^4}, \frac{1}{r_1^2}, 
\frac{r_1^ir_1^j}{r_1^7},r_1^ir_1^j,
\frac{r_1^ir_1^j}{r_1^6}, \frac{r_1^ir_1^j}{r_1^4},
\frac{r_1^ir_1^j}{r_1^2}, 
\frac{r_1^ir_1^jr_1^kr_1^l}{r_1^8},
\right.
\nonumber \\
&&
\left.
\frac{1}{r_1^5 r_2^4},\frac{1}{r_1^5r_2^2},\frac{1}{r_1^5r_2},
\frac{r_2}{r_1^5},\frac{r_2^3}{r_1^5},\frac{1}{r_1^4r_2^3},
\frac{1}{r_1^4r_2},\frac{r_2^2}{r_1^4},\frac{1}{r_1^3 r_2^3},
\frac{1}{r_1^3r_2^2},
\frac{1}{r_1^3r_2},\frac{r_2}{r_1^3},
\frac{1}{r_1 r_2},
\right.
\nonumber \\
&&
\left.
\frac{r_1^ir_1^j}{r_1^6r_2^3},\frac{r_1^ir_1^j}{r_1^6r_2},
\frac{r_1^ir_1^j r_2^2}{r_1^6},\frac{r_1^ir_1^j}{r_1^5r_2^3},
\frac{r_1^ir_1^j}{r_1^5r_2},\frac{r_1^ir_1^j}{r_1^4r_2^3},
\frac{r_1^ir_1^j r_2^2}{r_1^4},\frac{r_1^ir_1^j}{r_1^3r_2^3},
\frac{r_1^ir_1^j}{r_1^3r_2},\frac{r_1^ir_1^jr_2}{r_1^3},
\frac{r_1^ir_1^j}{r_2^2},\frac{r_1^ir_1^j}{r_1r_2},
\right.
\nonumber \\
&&
\left.
\frac{r_1^ir_2^j}{r_1^4r_2^3},
\frac{r_1^ir_2^j}{r_1^4},
\frac{r_1^ir_2^jr_2^2}{r_1^4},
\frac{r_1^ir_2^j}{r_1^3r_2^3},
\frac{r_1^ir_2^j}{r_1^3r_2},
\frac{r_1^ir_2^j r_2}{r_1^3},
\frac{r_1^ir_2^j}{r_1^2},
\frac{r_1^ir_2^j}{r_1 r_2},
\frac{r_1^ir_1^jr_1^kr_2^l}{r_1^5r_2^3}
\right\}.
\label{h8tiSPSourceList}
\end{eqnarray}
It should be understood that there are Poisson equations with 
the same sources but 
with $(1 \leftrightarrow 2)$ to be solved.

Our method to derive the superpotentials is heuristic;  
there are few guidelines available to find the required 
superpotentials. We proceed as follows.
First, we convert all the tensorial sources into scalars   
with spatial derivatives. For example,   
\begin{eqnarray}
\frac{r_1^ir_1^jr_1^kr_2^l}{r_1^5r_2^3} &=& 
- \frac{1}{3}\pa_{z_1^i}\pa_{z_1^j}\pa_{z_1^k}\pa_{z_2^l}
\left(\frac{r_1}{r_2}\right) + \frac{1}{3}
(\delta^{ij}\pa_{z_1^k} + 
\delta^{ik}\pa_{z_1^j} + 
\delta^{jk}\pa_{z_1^i} )\pa_{z_2^l}
\left(\frac{1}{r_1r_2}\right) 
\label{eq:superpotentialexample1}.  
\end{eqnarray}
(Here and henceforth, it should be understood that 
in general, 
``scalars'' can have tensorial indexes carried 
by $\vec v_A$ and $\vec r_{12}$, but 
do not have those by $\vec r_A$.)

Second, we find the particular solutions for Poisson 
equations with the scalars as sources using a formula 
$\Delta (f(\vec x)g(\vec x)) 
= g(\vec x)\Delta f(\vec x) + 
2 \vec \nabla f(\vec x) \cdot \vec \nabla g(\vec x) 
+ f(\vec x)\Delta g(\vec x)$ 
valid in $N/B$.  
We also use   
{\it superpotential chains} such as  
\begin{center}
\begin{tabular}{ccc}
$f^{(-3,-2)}$ & 
$\stackrel{\Delta_{11}}{\longrightarrow}$ & 
$6 f^{(-5,-2)}$\\
$\downarrow {\scriptstyle \Delta_{22}}$ & 
  & 
$\downarrow {\scriptstyle \Delta_{22}}$\\ 
$2 f^{(-3,-4)}$ &
$\stackrel{\Delta_{11}}{\longrightarrow}$ & 
$12 f^{(-5,-4)}$,  
\end{tabular}
\end{center}
where 
$$f^{(-3,-2)} =\frac{1}{r_1r_{12}^2}\ln\left(\frac{r_2}{r_1}\right),$$
and $f^{(m,n)}$ satisfies $\Delta f^{(m,n)} = r_1^m r_2^n$.
For example, 
for Eq. (\ref{eq:superpotentialexample1}), 
it is easy to find a particular solution, 
\begin{eqnarray*}
\frac{r_1^ir_1^jr_1^kr_2^l}{r_1^5r_2^3} &=& 
\Delta\left[
- \frac{1}{3}
\frac{\pa^4}{\pa z_1^i\pa z_1^j\pa z_1^k\pa z_2^l} 
f^{(1,-1)} + \frac{1}{3}
(\delta^{ij}\pa_{z_1^k} + 
\delta^{ik}\pa_{z_1^j} + 
\delta^{jk}\pa_{z_1^i} )\pa_{z_2^l}\ln S
\right],  
\end{eqnarray*}
where $f^{(1,-1)}$ is given in \cite{JS98}.
Another example is  
\begin{eqnarray*}
\frac{1}{r_1^5r_2} &=& 
\Delta \Delta_{11}\Delta_{22}\frac{1}{12}f^{(-3,1)}.
\end{eqnarray*}

Following the method described above, 
we could find all the required particular solutions   
other than  
$$\left\{
\frac{r_1^ir_1^j}{r_1^6r_2}, 
\frac{r_1^ir_1^j}{r_1^4r_2^3}, 
\frac{r_1^ir_2^j}{r_1^4r_2^3} 
\right\}.$$

In Appendix \ref{splist}, 
we list some of the particular solutions 
that we extensively used to 
derive the 3PN gravitational field. 
Other useful superpotentials are given in 
\cite{BFP98,JS98,BF01a}. 
The necessary particular solutions can be obtained by 
taking derivatives of these superpotentials with respect to 
$x^i$ or $z_1^i$ or $z_2^i$ or some combinations of them.

For example, a Poisson integral of 
the superpotential part of 
$- \mbox{}_4h^{kl,i}\mbox{}_4h^{\tau}\mbox{}_{k,l}$ 
(Eq. (\ref{eq5-12})) can be evaluated as follows. 
We find in $N/B$ 
\begin{eqnarray}
\lefteqn{
\left[
- \mbox{}_4h^{kl,i}\mbox{}_4h^{\tau}\mbox{}_{k,l}
\right]_{{\rm SP}}  
} 
\nonumber \\ &=& 
\Delta \left[      
4m_1^3\left(
\frac{1}{15r_1^3}v_1^i + \frac{1}{5}v_1^k\pa_{ik}\frac{\ln r_1}{r_1} 
\right) 
-2m_1^2 v_1^2 v_1^k 
\left(
\pa_{ik}\ln r_1
+ \frac{\delta^{ik}}{r_1^2}
\right)
\right. 
\nonumber \\
\mbox{} &&\left.  + 
m_1^2m_2 r_{12}^2v_2^k\left(
\frac{\pa^2 f^{(-2,-3)} }{\pa z_1^i\pa z_1^k}
+ \frac{2}{r_{12}^2} \delta^{ik}
\left(f^{(-2,-3)}+f^{(-4,-1)}
\right)\right)
\right. 
\nonumber \\
\mbox{} &&\left.  
- 16m_1m_2(\vec v_1\cdot\vec v_2)v_1^k\frac{\pa^2 \ln S}{\pa z_1^iz_2^k}
+ (1 \leftrightarrow 2)
\right]
\nonumber \\
\mbox{} &&\equiv 
\Delta  
SP(\tau,\vec x).  
\end{eqnarray}
Then the surface integral in Eq. (\ref{NBcontribution}) 
gives  
\begin{eqnarray}
\lefteqn{\frac{1}{-4\pi}
\oint_{\pa(N/B)}dS_k
\left[\frac{1}{|\vec{x}-\vec{y}|}
 \frac{\pa }{\pa y^k}
SP(\tau,\vec y)
 - SP(\tau,\vec y)
\frac{\pa}{\pa y^k}
 \left(\frac{1}{|\vec{x} - \vec{y}|}\right) \right]}   
\nonumber \\ &=& 
\frac{4 m_1^3}{5r_1^3}v_1^kn_{1}^{<ik>}\left(
\frac{23}{5} - 3 \ln \epsilon R_1 \right)
- \frac{m_1^2m_2}{r_{12}^2r_2}
\left(
v_2^i + 2 (\vec n_{12}\cdot \vec v_2)n_{12}^i + 
8 (\vec n_{12}\cdot \vec v_2)n_{12}^i
\ln\left(\frac{r_{12}}{\epsilon R_2}\right) 
\right) 
\nonumber \\
\mbox{} &&
+ \frac{16m_1^2v_1^2v_1^i}{3\epsilon R_1r_1} 
- \frac{8m_1^2m_2v_2^i}{3\epsilon R_1r_1r_{12}} 
\nonumber \\
\mbox{} &&
+ (1 \leftrightarrow 2).
\end{eqnarray}
We shall ignore the last two terms in the second to last line of 
the above equation,
$16m_1^2v_1^2v_1^i/(3\epsilon R_1r_1) 
- 8m_1^2m_2v_2^i/(3\epsilon R_1r_1r_{12})$
 (and the same terms with 
the label of the star exchanged, hidden in 
$(1 \leftrightarrow 2)$),   
because they have negative powers   
of $R_A$. We combine the above results and evaluate 
the Poisson integral of $
\left[
- \mbox{}_4h^{kl,i}\mbox{}_4h^{\tau}\mbox{}_{k,l}
\right]_{{\rm SP}}
$ as 
\begin{eqnarray}
\lefteqn{
\int_{N/B}\frac{d^3y}{|\vec x - \vec y|}
 \left[
- \mbox{}_4h^{kl,i}\mbox{}_4h^{\tau}\mbox{}_{k,l}
\right]_{{\rm SP}}} \nonumber \\ 
 &=& 
\frac{12m_1^3}{5r_1^3}v_1^kn_{1}^{<ik>}
\left(\frac{1}{5} + \ln\left(\frac{r_1}{\epsilon R_1}\right) 
\right) 
+ \frac{4 m_1^2}{r_1^2}v_1^2
\left((\vec n_1\cdot\vec v_1)n_1^i - v_1^i\right)
\nonumber \\
\mbox{} && 
+ m_1^2m_2\left(
-  \frac{2}{r_1^2r_2}(\vec n_1\cdot\vec v_2)n_1^i
+  \frac{2}{r_1r_{12}r_2}(\vec n_{12}\cdot\vec v_2)n_1^i
+  \frac{2}{r_1r_{12}r_2}(\vec n_1\cdot\vec v_2)n_{12}^i
\right. 
\nonumber \\
\mbox{} &&+ \left. 
 \frac{2}{r_{12}^2r_2}
(\vec n_{12}\cdot\vec v_2)n_{12}^i 
\left(
4\ln\left(\frac{r_{1}}{r_2}\right) - 
4\ln\left(\frac{r_{12}}{\epsilon R_2}\right)
-  1  
\right)
+ v_2^i\left(
 \frac{1}{r_1^2r_2} - \frac{1}{r_{12}^2r_2} 
+ \frac{r_2}{r_1^2r_{12}^2}
\right)
\right)
\nonumber \\
\mbox{} && 
+ \frac{16m_1m_2}{S}(\vec v_{1}\cdot\vec v_2)
\left(
\frac{v_1^i}{r_{12}} + 
\frac{n_{1}^i}{S}\left(
(\vec n_{12}\cdot\vec v_1) 
+ (\vec n_{2}\cdot\vec v_1)
\right)
\right. 
  \nonumber \\
\mbox{} && +\left. 
n_{12}^i \left(
- \frac{1}{S}
(\vec n_{12}\cdot\vec v_1)
- \frac{1}{r_{12}}
(\vec n_{12}\cdot\vec v_1) 
- \frac{1}{S}
(\vec n_{2}\cdot\vec v_1)
\right)
\right)+ (1 \leftrightarrow 2).
\end{eqnarray}
Notice that in the above equation there appear 
$\ln (\epsilon R_1)$ and  $\ln (\epsilon R_2)$.

In general, with the help of the superpotentials  
and using Eq. (\ref{NBcontribution}), 
we can integrate the superpotential part of 
$\mbox{}_8\Lambda_{N}^{\tau i}$. 
The explicit result is too long
to write down here. 
\footnote{The number of terms are $\sim 10^3$, 
depending on how we simplify the result.}

\subsection{Superpotential-in-series part}
\label{ExplanationforSSPP}

We could not find particular solutions for the following sources; 
\begin{eqnarray*}
\frac{r_1^ir_1^j}{r_1^6r_2} &=& 
\frac{1}{8}\pa_{z_1^i}\pa_{z_1^j}
\left(
\frac{1}{r_1^2r_2}
\right)
+ \Delta \left[
\frac{1}{4}\delta^{ij}f^{(-4,-1)}
\right], \\
\frac{r_1^ir_1^j}{r_1^4r_2^3} &=& 
- \frac{1}{2}\pa_{z_1^i}\pa_{z_1^j}
\left(
\frac{\ln r_1}{r_2^3}
\right)
+ \Delta \left[
\frac{1}{2}\delta^{ij}f^{(-2,-3)}
\right], 
\nonumber \\  
\frac{r_1^ir_2^j}{r_1^4r_2^3} &=&
\frac{1}{2}\pa_{z_1^i}\pa_{z_2^j} 
\left(
\frac{1}{r_1^2r_2}  
\right).
\end{eqnarray*}
For these innocent looking  
sources, we did not find particular solutions valid 
throughout $N/B$ in closed forms. 
Instead we looked for those valid near the star, say the
star 1; the 
field we need when we  evaluate the evolution equation for 
$P_{1\Theta}^{\tau}$ 
and the equation of motion for the star 1 is the field around the star 1.

Now, all the integrands classified into the 
superpotential-in-series part 
at 3PN order are found to have the following form 
(neglecting the $m_A$, 
$\vec v_A$, and $\vec r_{12}$ dependence appearing in 
actual applications of the following formulas):  
\begin{eqnarray}
\pa_{z_A^i}\pa_{z_{A'}^j} g(\vec x) &\equiv& 
\pa_{z_A^i}\pa_{z_{A'}^j}
\left(
\frac{(\ln r_1)^p (\ln r_2)^q}{r_1^ar_2^b}
\right),
\end{eqnarray}
where $a$ and $b$ are integers and $p=0, 1$, $q=0, 1$.  
$A,A' = 1,2$.

Then, we take spatial derivatives out
of the Poisson integral,  
\begin{eqnarray}
\int_{N/B}\frac{d^3y}{|\vec x - \vec y|}
\pa_{z_A^i}\pa_{z_{A'}^j} g(\vec y) &=&  
\pa_{z_A^i}\pa_{z_{A'}^j}
\int_{N/B}d^3y\frac{g(\vec y)}{|\vec x - \vec y|} 
\nonumber \\
\mbox{} &+&
 \pa_{z_A^i}\oint_{\pa B_{A'}}
dS_j\frac{g(\vec y)}{|\vec x - \vec y|} 
+ \oint_{\pa B_{A}}
dS_i\frac{\pa_{z_{A'}^j}g(\vec y)}{|\vec x - \vec y|}.   
\label{eq5-5}
\end{eqnarray}
For the remaining volume integral, we change the integration variable 
$\vec y$ into $\vec y_1$, namely, $\vec y_2 = \vec r_{12} + \vec y_1$. 
We also change 
the integration region $N/B$ into $N_1/B$,  
where $N_1 \equiv \{\vec y | |\vec y - \vec z_1| \le {\cal
R}/\epsilon\}$,  
\begin{eqnarray}
\lefteqn{
\int_{N/B}d^3y\frac{g(\vec y)}{|\vec x - \vec y|}} \nonumber \\  
&=& 
\int_{N_1/B}d^3y_1\frac{g(\vec y_1 + \vec z_1)}{|\vec r_1 - \vec y_1|} 
- z_1^k\oint_{\pa N}dS_k\frac{g(\vec y)}{|\vec x - \vec y|}
- \frac{1}{2!}
z_1^k z_1^l\oint_{\pa N}dS_k
\pa_{y^l}
\left(
\frac{g(\vec y)}{|\vec x - \vec y|}\right)
\nonumber \\
&&-
 \frac{1}{3!}
z_1^k z_1^l z_1^m\oint_{\pa N}dS_k
\pa_{y^l}\pa_{y^m}
\left(
\frac{g(\vec y)}{|\vec x - \vec y|}\right)
\nonumber \\
&&-
 \frac{1}{4!}
z_1^k z_1^l z_1^mz_1^n\oint_{\pa N}dS_k
\pa_{y^l}\pa_{y^m}\pa_{y^n}
\left(
\frac{g(\vec y)}{|\vec x - \vec y|}\right)
+ \cdot\cdot\cdot, 
\label{eq5-6}
\end{eqnarray}
where $\pa_{y^i} = \pa/\pa y^i$.  
The surface integrals and 
terms expressed as $\cdot\cdot\cdot$ arise due to 
the difference between $N$ and $N_1$. See, e.g.,  \cite{WW96}.
Note that $\vec r_A = \vec x - \vec z_A$, where $\vec x$ is the 
field point, while 
$\vec y_A = \vec y - \vec z_A$, where $\vec y$ is the integral 
variable.

For the first volume integral on the right-hand side of the 
above equation, we expand $|\vec r_1 - \vec y_1|$ as  
\begin{eqnarray}
\frac{1}{|\vec r_1 - \vec y_1|} &=& 
\sum_{c=0}\frac{1}{r_{>}}\left(\frac{r_{<}}{r_{>}}\right)^c 
{\rm P}_{c}\left(\frac{\vec r_1\cdot\vec y_1}{r_1y_1}\right), 
\end{eqnarray}  
where $r_{>} = {\rm max}(r_1,y_1)$, 
$r_{<} = {\rm min}(r_1,y_1)$, and 
${\rm P}_c(x)$ is a Legendre function of order $c$.

Now we split the integration region into four parts according 
to where the radial variable $y_1$ is as follows: 
region I,  
$y_1 \in [\epsilon R_1,r_1]$;  
region II, 
$y_1 \in [r_1,r_{12}-\epsilon R_2]$;  
region III, 
$y_1 \in [r_{12}-\epsilon R_2,r_{12}+\epsilon R_2]$;  
and region IV, 
$y_1 \in [r_{12}+\epsilon R_2, {\cal R}/\epsilon]$.  
In the third integral region, angular 
integration is incomplete due to the body zone 2 ($B_2$), which 
the Poisson integration over $N/B$ does {\it not} cover.

Then first for region I, 
\begin{eqnarray}
\int_{I}\frac{d^3y_1}{|\vec r_1 - \vec y_1|}g(\vec y)   
&=& 
2\pi\sum_{c=0}\frac{1}{r_1^{c+1}}
\int^{r_1}_{\epsilon R_1}dy_1\frac{(\ln y_1)^p}{y_1^{a-c-2}} 
\nonumber \\
\mbox{} &&\times 
\int^{1}_{-1}d\cos\theta
\frac{{\rm P}_c(\cos\theta){\rm P}_c(\cos \gamma)} 
{2^q (r_{12}^2+y_1^2-2r_{12}y_1\cos\theta)^{b/2}}
(\ln (r_{12}^2+y_1^2-2r_{12}y_1\cos\theta))^q  
\nonumber \\
\mbox{}&=& 
2 \pi \sum_{c=0} 
\frac{{\rm P}_c(\cos \gamma)}
{r_1^{c+1}r_{12}^{a+b-c-3}}
\int_{\epsilon R_1/r_{12}}^{r_1/r_{12}}
d\zeta (\ln \zeta + \ln r_{12})^p\zeta^{c+2-a}
\nonumber \\
\mbox{} &&\times
\int_{-1}^{1}\frac{dt {\rm P}_c(t)
\left(\frac{1}{2}\ln (1-2\zeta t + \zeta^2) + 
\ln r_{12}\right)^q}{(1-2\zeta t + \zeta^2)^{b/2}}, 
\end{eqnarray}
where 
$\cos \gamma \equiv  - (\vec r_1 \cdot \vec r_{12})/r_1/r_{12}$, 
$t \equiv \cos\theta \equiv - (\vec y_1 \cdot \vec r_{12})/y_1/r_{12}$, 
and $\zeta \equiv y_1/r_{12}$.

Next for region II,   
\begin{eqnarray}
\int_{II}\frac{d^3y_1}{|\vec r_1 - \vec y_1|}g(\vec y)   
&=& 
2\pi\sum_{c=0}r_1^{c}
\int^{r_{12}-\epsilon R_2}_{r_1}dy_1\frac{(\ln y_1)^p}{y_1^{a+c-1}} 
\nonumber \\
\mbox{} &&\times 
\int^{1}_{-1}d\cos\theta
\frac{{\rm P}_c(\cos\theta){\rm P}_c(\cos \gamma)} 
{2^q (r_{12}^2+y_1^2-2r_{12}y_1\cos\theta)^{b/2}}
(\ln (r_{12}^2+y_1^2-2r_{12}y_1\cos\theta))^q  
\nonumber \\
\mbox{}&=& 
2 \pi \sum_{c=0} 
\frac{r_1^c {\rm P}_c(\cos \gamma)}{r_{12}^{a+b+c-2}}
\int_{r_1/r_{12}}^{1 - \epsilon R_2/r_{12}}
\frac{d\zeta (\ln \zeta + \ln r_{12})^p}{\zeta^{a+c-1}}
\nonumber \\
\mbox{} &&\times
\int_{-1}^{1}\frac{dt {\rm P}_c(t)
\left(\frac{1}{2}\ln (1-2\zeta t + \zeta^2) + 
\ln r_{12}\right)^q}
{(1-2\zeta t + \zeta^2)^{b/2}}.  
\end{eqnarray}

Third for region IV,   
\begin{eqnarray}
\int_{IV}\frac{d^3y_1}{|\vec r_1 - \vec y_1|}g(\vec y)   
&=& 
2\pi\sum_{c=0}r_1^{c}
\int^{{\cal R}/\epsilon}_{r_{12}+\epsilon R_2}
dy_1\frac{(\ln y_1)^p}{y_1^{a+c-1}} 
\nonumber \\
\mbox{} &&\times 
\int^{1}_{-1}d\cos\theta
\frac{{\rm P}_c(\cos\theta){\rm P}_c(\cos \gamma)} 
{2^q (r_{12}^2+y_1^2-2r_{12}y_1\cos\theta)^{b/2}}
(\ln (r_{12}^2+y_1^2-2r_{12}y_1\cos\theta))^q  
\nonumber \\
\mbox{}&=& 
2 \pi \sum_{c=0} 
\frac{r_1^c {\rm P}_c(\cos \gamma)}{r_{12}^{a+b+c-2}}
\int_{1 + \epsilon R_2 /r_{12}}^{{\cal R}/(\epsilon r_{12})}
\frac{d\zeta (\ln \zeta + \ln r_{12})^p}{\zeta^{a+c-1}}
\nonumber \\
&&\times
\int_{-1}^{1}\frac{dt {\rm P}_c(t)
\left(\frac{1}{2}\ln (1-2\zeta t + \zeta^2) + 
\ln r_{12}\right)^q}{(1-2\zeta t + \zeta^2)^{b/2}}. 
\end{eqnarray}

Now for region III, the angular deficit $\theta_0$ 
due to the body zone 2 is determined by 
\begin{eqnarray}
 (\epsilon R_2)^2 &=& y_1^2 + r_{12}^2 - 2 r_{12}y_1\cos\theta_0. 
\label{eq:angulardeficit} 
\end{eqnarray}
It is convenient to redefine $\zeta$ as  $\zeta \equiv y_1/r_{12} -1$. 
Then $\theta$ ranges from $-1 < \cos\theta < \cos\theta_0 
\equiv 1 - \alpha({\zeta})$,   
where $\alpha(\zeta) = 
(\epsilon R_2/r_{12} - \zeta)(\epsilon R_2/r_{12} + \zeta)/
2/(1 +\zeta)$. Thence 
\begin{eqnarray}
\int_{III}\frac{d^3y_1}{|\vec r_1 - \vec y_1|}g(\vec y)   
&=& 
2\pi\sum_{c=0}r_1^{c}
\int^{r_{12}+\epsilon R_2}_{r_{12}-\epsilon R_2}
dy_1\frac{(\ln y_1)^p}{y_1^{a+c-1}} 
\nonumber \\
\mbox{} &&\times 
\int^{\cos\theta_0}_{-1}d\cos\theta
\frac{{\rm P}_c(\cos\theta){\rm P}_c(\cos \gamma)} 
{2^q (r_{12}^2+y_1^2-2r_{12}y_1\cos\theta)^{b/2}}
(\ln (r_{12}^2+y_1^2-2r_{12}y_1\cos\theta))^q  
\nonumber \\
\mbox{}&=& 
2 \pi \sum_{c=0} 
\frac{r_1^c {\rm P}_c(\cos \gamma)}{r_{12}^{a+b+c-2}}
\int_{- \epsilon R_2/r_{12}}^{\epsilon R_2/r_{12}}
\frac{d\zeta (\ln (1 + \zeta) + \ln r_{12})^p}{(1+\zeta)^{a+c-1}}
\nonumber \\
&&\times
\int_{-1}^{1-\alpha(\zeta)}
\frac{dt {\rm P}_c(t)
\left(\frac{1}{2}\ln (2 + 2\zeta- 2(1 + \zeta) t + \zeta^2) + 
\ln r_{12}\right)^q}{(2 + 2\zeta-2(1+ \zeta) t + \zeta^2)^{b/2}}.   
\end{eqnarray}

Then summing up the above results, 
we obtain the following formula, by which we evaluate 
the first volume integral on the right-hand side of 
Eq. (\ref{eq5-6}):   
\begin{eqnarray}
\lefteqn{
\int_{N_1/B}d^3y_1\frac{g(\vec y_1 + \vec z_1)}{|\vec r_1 - \vec y_1|}} 
\nonumber \\
&=& 
2 \pi \sum_{c=0} 
\frac{{\rm P}_c(\cos \gamma)}
{r_1^{c+1}r_{12}^{a+b-c-3}}
\int_{\epsilon R_1/r_{12}}^{r_1/r_{12}}
d\zeta (\ln \zeta + \ln r_{12})^p \zeta^{c+2-a}
\nonumber \\
&&\times
\int_{-1}^{1}\frac{dt {\rm P}_c(t)
\left(\frac{1}{2}\ln (1-2\zeta t + \zeta^2) + 
\ln r_{12}\right)^q}{(1-2\zeta t + \zeta^2)^{b/2}}  \nonumber \\
&+& 
2 \pi \sum_{c=0} 
\frac{r_1^c {\rm P}_c(\cos \gamma)}{r_{12}^{a+b+c-2}}
\int_{r_1/r_{12}}^{1 - \epsilon R_2/r_{12}}
\frac{d\zeta (\ln \zeta + \ln r_{12})^p}{\zeta^{a+c-1}}
\int_{-1}^{1}\frac{dt {\rm P}_c(t)
\left(\frac{1}{2}\ln (1-2\zeta t + \zeta^2) + 
\ln r_{12}\right)^q}
{(1-2\zeta t + \zeta^2)^{b/2}} \nonumber  \\
&+& 
2 \pi \sum_{c=0} 
\frac{r_1^c {\rm P}_c(\cos \gamma)}{r_{12}^{a+b+c-2}}
\int_{- \epsilon R_2/r_{12}}^{\epsilon R_2/r_{12}}
\frac{d\zeta (\ln (1 + \zeta) + \ln r_{12})^p}{(1+\zeta)^{a+c-1}}
\nonumber \\
&&\times
\int_{-1}^{1-\alpha(\zeta)}
\frac{dt {\rm P}_c(t)
\left(\frac{1}{2}\ln (2 + 2\zeta- 2(1 + \zeta) t + \zeta^2) + 
\ln r_{12}\right)^q}{(2 + 2\zeta-2(1+ \zeta) t + \zeta^2)^{b/2}}  
\nonumber \\
&+& 
2 \pi \sum_{c=0} 
\frac{r_1^c {\rm P}_c(\cos \gamma)}{r_{12}^{a+b+c-2}}
\int_{1 + \epsilon R_2 /r_{12}}^{{\cal R}/(\epsilon r_{12})}
\frac{d\zeta (\ln \zeta + \ln r_{12})^p}{\zeta^{a+c-1}}
\nonumber \\
&&\times
\int_{-1}^{1}\frac{dt {\rm P}_c(t)
\left(\frac{1}{2}\ln (1-2\zeta t + \zeta^2) + 
\ln r_{12}\right)^q}{(1-2\zeta t + \zeta^2)^{b/2}}.
\label{eq5-7}
\end{eqnarray}

As an example, when the source term is $r_1^ir_2^j/r_1^4/r_2^3$, 
we have 
\begin{eqnarray}
\int_{N/B}\frac{d^3y}{- 4 \pi|\vec x - \vec y|}
\frac{r_1^ir_2^j}{r_1^4r_2^3} &=& 
\frac{1}{2}
\pa_{z_1^i}\pa_{z_{2}^j} F_{[1,2]}^{(-2,-1)} 
+ O(\epsilon R_A),
\end{eqnarray}
where 
\begin{eqnarray}
F_{[1,2]}^{(-2,-1)} &=&   
 \int_{N/B}\frac{d^3y}{-4 \pi |\vec x - \vec y|}\frac{1}{y_1^2y_2} =
 \int_{N_1/B}\frac{d^3y_1}{-4 \pi|\vec r_1 - \vec y_1|}\frac{1}{y_1^2y_2}  
+ O\left(\left(\frac{\epsilon}{{\cal R}}\right)^2\right)
\nonumber \\
&=& 
-\frac{2}{r_{12}} + 
\frac{1}{r_{12}}\ln\left(\frac{r_1}{r_{12}}\right) +
\frac{P_1(\cos \gamma)}{3 r_{12}^2}
\left(- \frac{2}{3} + 
\ln\left(\frac{r_1}{r_{12}}\right)
\right) r_1 
\nonumber \\
&+&
 \frac{P_2(\cos \gamma)}{5r_{12}^3}
\left(- \frac{2}{5} + \ln\left(\frac{r_1}{r_{12}}\right)
\right) r_1^2
+ \frac{P_3(\cos \gamma)}{7r_{12}^4}
\left(- \frac{2}{7} + \ln\left(\frac{r_1}{r_{12}}\right)
\right) r_1^3
\nonumber \\
&+&
 \frac{P_4(\cos \gamma)}{9r_{12}^5}
\left(- \frac{2}{9} + \ln\left(\frac{r_1}{r_{12}}\right)
\right) r_1^4 + O\left(\frac{r_1^5}{r_{12}^5}\right) 
+ O\left(\left(\frac{\epsilon}{{\cal R}}\right)^2,\epsilon R_A\right).  
\end{eqnarray}
$F_{[1,2]}^{(-2,-1)}$ satisfies 
\begin{eqnarray}
\Delta F_{[1,2]}^{(-2,-1)} - \frac{1}{r_1^2r_2} = 
O\left(\frac{r_1^3}{r_{12}^3}\right)   
{\rm ~~~as~~r_1~~} \rightarrow 0. 
\label{seriesedpotExample}
\end{eqnarray}
Thus, in the neighborhood of $\vec z_1$,   
$F_{[1,2]}^{(-2,-1)}$ is the required solution  
of the Poisson equation 
in the sense of Eq. (\ref{seriesedpotExample}). In general,  
$F_{[A,c]}^{(m,n)}$ denotes a function which satisfies 
$$\Delta F_{[A,c]}^{(m,n)} - r_1^mr_2^n = 
O\left(\frac{r_A^{c+1}}{r_{12}^{c+1}}\right)
{\rm ~~~as~~r_A~~} \rightarrow 0.$$ An appropriate 
value of the index $c$ depends on how many times we should 
take derivatives of $F^{(m,n)}_{[A,c]}$ to derive 
an equation of motion.

Now, to illustrate our method, 
let us evaluate the Poisson integral of 
$\left[- \mbox{}_4h^{kl,i}\mbox{}_4h^{\tau}\mbox{}_{k,l}
\right]_{{\rm SSP}}$ defined by Eq. (\ref{eq5-13}),  
\begin{eqnarray}
\left[- \mbox{}_4h^{kl,i}\mbox{}_4h^{\tau}\mbox{}_{k,l}
\right]_{{\rm SSP}} &=& 
m_1^2m_2v_2^k\left(
 4 \frac{\pa^2}{\pa z_1^iz_1^k} 
\left(
\frac{\ln r_1}{r_2^3}
\right)
- \frac{\pa^2}{\pa z_1^iz_1^k} 
\left(
\frac{1}{r_1^2r_2}
\right)
+ 2\frac{\pa^2}{\pa z_1^kz_2^i}
\left(
\frac{1}{r_1^2r_2}
\right)
\right)
\nonumber \\
\mbox{} &&
+ 4m_1^2m_2v_2^i  
\Delta \left[ - f^{(-2,-3)} - \frac{1}{2}f^{(-4,-1)}\right]
+ (1 \leftrightarrow 2).
\end{eqnarray}
Then we evaluate the Poisson integral around $\vec z_1$ as 
\begin{eqnarray}
\lefteqn{
\int_{N/B}\frac{d^3y}{-4\pi|\vec x - \vec y|} 
\left[- \mbox{}_4h^{kl,i}\mbox{}_4h^{\tau}\mbox{}_{k,l}
\right]_{{\rm SSP}}} \nonumber \\
&=& 
m_1^2m_2v_2^k\left(
 4 \frac{\pa^2}{\pa z_1^iz_1^k} F^{(\ln,-3)}_{[1,2]}
- \frac{\pa^2}{\pa z_1^iz_1^k} F^{(-2,-1)}_{[1,2]}
+ 2\frac{\pa^2}{\pa z_1^kz_2^i}F^{(-2,-1)}_{[1,2]}
\right) 
\nonumber \\
\mbox{} &&
+ m_1m_2^2v_1^k\left(
 4 \frac{\pa^2}{\pa z_2^iz_2^k} F^{(-3,\ln)}_{[1,2]}
- \frac{\pa^2}{\pa z_2^iz_2^k} F^{(-1,-2)}_{[1,2]}
+ 2\frac{\pa^2}{\pa z_1^iz_2^k}F^{(-1,-2)}_{[1,2]}
\right) 
\nonumber \\
\mbox{} &&
+ \frac{4m_1^2m_2}{r_{12}^2}v_2^i  
\left(
\frac{1}{r_2}\ln\left(\frac{r_2}{\epsilon R_2}\right) 
+ \frac{1}{r_2} 
- \frac{r_2}{4r_1^2}
\right)
\nonumber \\
\mbox{} &&
+ \frac{4m_1m_2^2}{r_{12}^2}v_1^i  
\left(
\frac{1}{r_1}\ln\left(\frac{r_1}{\epsilon R_1}\right) 
+ \frac{1}{r_1} 
- \frac{r_1}{4r_2^2}
\right)
\nonumber \\
\mbox{} &&
+ R(\vec x),  
\end{eqnarray}
where 
$R(\vec x)$ is the remainder.  
$F^{(\ln,-3)}_{[1,2]}$ satisfies 
$$
\Delta F^{(\ln,-3)}_{[1,2]} - \frac{\ln r_1}{r_2^3} = 
O\left(\frac{r_1^3}{r_{12}^3}\right).   
$$
Similar equations hold for 
$F^{(-3,\ln)}_{[1,2]}$ and $F^{(-1,-2)}_{[1,2]}$.
 
We note that for the superpotential-in-series part, 
there is no need to add terms corresponding to 
the surface integral terms in Eq. (\ref{NBcontribution}). 
We note also that the surface integrals in Eqs. 
(\ref{eq5-5}) and (\ref{eq5-6}) do contribute to the field in the 
neighborhood of  the star.

We could evaluate the Poisson integral of 
the superpotential-in-series part of 
$\mbox{}_8\Lambda_N^{\tau i}$ 
in the neighborhood of the star 1 by means of 
the method described in this subsection.

\subsection{Direct-integration part}
\label{ExplanationforCompStarInt}

For the direct-integration part 
(e.g., Eq. (\ref{eq5-11})), we evaluate the 
surface integral in Eq. (\ref{EvolOfFourMom3PN}) directly,  
while we give up deriving the corresponding contributions to the 
field valid throughout $N/B$ in a closed form. 
In this subsection, we consider only the effect of the 
direct-integration part of $\mbox{}_8h^{\tau i}$ on the evolution 
equation for $P_{A\Theta}^{\tau}$.

Let us define the ``DIP'' field $\mbox{}_8h^{\tau i}_{{\rm DIP}}$, 
\begin{eqnarray}
&&
\mbox{}_8h_{{\rm DIP}}^{\tau i} \equiv 
({\rm direct-integration~part~of~}\mbox{}_8h^{\tau i}) = 
4 \int_{N/B}\frac{d^3y}{|\vec x - \vec y|}
\mbox{}_8\Lambda_{S}^{\tau i}, 
\label{eq:h8tiDIPDefinition}
\end{eqnarray}
with 
$16 \pi\mbox{}_8\Lambda_S^{\tau i} \equiv ({\rm S-dependent~parts~of}~ 
\{- \mbox{}_4h^{kl,i}\mbox{}_4h^{\tau}\mbox{}_{k,l} 
+ 2 \mbox{}_4h^{i (k,l)}\mbox{}_4h^{\tau}\mbox{}_{k,l}
+ \mbox{}_4h^{ik}\mbox{}_{,\tau}
\mbox{}_4h^{\tau\tau}\mbox{}_{,k}
+ 2\mbox{}_4h^{\tau\tau}\mbox{}_{,k}
\mbox{}_6h^{\tau[k,i]}\})$. 
Then in the derivation of 
the evolution equation for 
$P_{1\Theta}^{\tau}$, the direct-integration part appears as 
(see Eq. (\ref{tLLti10})) 
\begin{eqnarray}
&&
\frac{d P_{1\Theta}^{\tau}}{d \tau} = 
- \oint_{\pa B_1}\frac{dS_k}{8 \pi}
\mbox{}_4h^{\tau\tau}\mbox{}_{,l}
\mbox{}_8h_{{\rm DIP}}^{\tau [l,k]} 
+ \cdots \nonumber \\
&&=
 \oint_{\pa B_1}\frac{dS_k}{2 \pi}
\frac{m_2r_2^l}{r_2^3}
\mbox{}_8h_{{\rm DIP}}^{\tau [l,k]}
+ \cdots. 
\label{eq-511}
\end{eqnarray}
Then it is sufficient to compute the following integral:  
\begin{eqnarray}
 \oint_{\pa B_1}\frac{dS_k}{2 \pi}
\frac{m_2r_2^l}{r_2^3}
\mbox{}_8h_{{\rm DIP}}^{\tau [l,k]} 
&=& \oint_{\pa B_1}dS_k\frac{2}{\pi}
\frac{m_2r_2^l}{r_2^3}
\left[
\int_{N/B}d^3y
\frac{\mbox{}_8\Lambda_S^{\tau[l,k]}}{|\vec x - \vec y|}
- \oint_{\pa (N/B)}
\frac{dS_{[k} \mbox{}_8\Lambda_S^{l]\tau}}{|\vec x - \vec y|}
\right].
\label{eq-512}
\end{eqnarray}
Straightforward calculation shows that 
$\mbox{}_8\Lambda_S^{\tau i}(\vec y) \sim 1/y^3$ as 
$y \rightarrow \infty$, and thus no contribution arises 
from the surface integral over $\pa N$.   
On the other hand, 
\begin{eqnarray}
&&
- \oint_{\pa B}
\frac{dS_{[k} \mbox{}_8\Lambda_S^{l]\tau}}{|\vec x - \vec y|} = 
\frac{4 m_1m_2}{r_{12}^3}\left(
\frac{m_1}{r_1}n_{12}^{[k}v_1^{l]}
-\frac{2m_1}{3r_1}n_{12}^{[k}v_2^{l]}
\right)
+ (1\leftrightarrow 2).
\label{h8tiSurfIntDoesNotContribute}
\end{eqnarray}
Therefore, the second surface integral in the square brackets 
in Eq. (\ref{eq-512})  
gives no contribution to the evolution equation for $P_{1\Theta}^{\tau}$. 
The first  
integral requires special treatment, which we shall explain 
below.

Now let us consider an integral which has a form 
\begin{equation}
\oint_{\pa B_1}dS_k\frac{r_2^l}{r_2^3}
\mathop{{\rm disc}}_{\epsilon R_A} 
\int_{N/B}d^3y_1
\frac{f(\vec y_1)}{|\vec r_1 - \vec y_1|}. 
\label{CSIDefinition}
\end{equation}
Here $f(\vec x)$ carries tensorial indexes in general, but 
we do not write them explicitly for notational simplicity.
We call this type of integral the {\it companion star integral}.   
The first integral in Eq. (\ref{eq-512}) is a 
companion star integral with   
$f(\vec y_1) = 
(2 m_2/\pi)\mbox{}_8\Lambda_S^{\tau [l,k]}(\vec y_1 + \vec z_1)$ 
($\vec y_2$ must be replaced by $\vec r_{12} + \vec y_1$ in $f(\vec y_1)$). 
In the above equation, we defined 
$\mathop{{\rm disc}}_{\epsilon R_A}$,   
which means to discard 
all the $\epsilon R_A$-dependent terms other than 
logarithms of $\epsilon R_A$. Thus, for example, 
$$ 
\mathop{{\rm disc}}_{\epsilon R_A} 
\left[
\frac{1}{\epsilon R_1}
\ln\left(\frac{r_{12}}{\epsilon R_2}\right) + 
\frac{1}{r_1}\ln\left(\frac{r_{12}}{\epsilon R_2}\right) + 
\frac{1}{\epsilon R_1} + 
\frac{1}{r_1}
\right] = \frac{1}{r_1} +  
\frac{1}{r_1}\ln\left(\frac{r_{12}}{\epsilon R_2}\right).  
$$
The symbol 
$\mathop{{\rm disc}}_{\epsilon R_A}$ 
is introduced 
in Eq. (\ref{CSIDefinition}) 
to clarify that we discard $\epsilon R_A$ dependence in the 
field {\it before} we evaluate the surface integrals in the 
general form of the 3PN equation of motion.
  
To evaluate a companion star integral, 
we first exchange the order of integration,  
\begin{eqnarray}
\lefteqn{
\oint_{\pa B_1}dS_k
\frac{r_2^l}{r_2^3}
\mathop{{\rm disc}}_{\epsilon R_A}
\int_{N/B}d^3y
\frac{f(\vec y_1)}{|\vec r_1 - \vec y_1|}} \nonumber \\
&=& \lim_{r_1' \rightarrow \epsilon R_1}
\mathop{{\rm disc}}_{\epsilon R_A}
\int_{N/B}d^3y_1 f(\vec y_1)
\oint_{\pa B_1'}dS_k\frac{1}{|\vec y_1 - r_1' \vec n_1|}
\pa_{z_2^l}\frac{1}{|\vec r_{12} - r_1' \vec n_1|}, 
\label{eq513}
\end{eqnarray}
where we defined a sphere $B_1'$ 
whose center is $\vec z_1$ and radius is $r_1'$ which is 
a constant slightly larger than $\epsilon R_1$ for any 
(small) $\epsilon$ 
($\epsilon R_1 < r_1' << r_{12}$).  

We mention here that the procedure of exchanging 
the order of integrations here is motivated by the works 
of Blanchet and Faye \cite{BF00a,BF01a,BF00b}.

The reason we introduced $r_1'$ is as follows. 
Suppose that we treat an integrand for which 
the superpotential is available. By calculating 
the Poisson integral, we have a piece of field 
corresponding to the integrand. The piece generally 
depends on $\epsilon R_A$, however we reasonably 
discard such $R_A$-dependent terms 
(other than logarithmic dependence) as explained 
in Sec. \ref{Arbitrariness}.  Using so-obtained 
$R_A$-independent field, we evaluate the surface integrals  
in the general form of the 3PN equation of motion by discarding the 
$\epsilon R_A$ dependence emerging from the surface 
integrals, and obtain an equation of motion.
Thus discarding-$\epsilon R_A$ procedure 
must be employed at each time, when the field is derived and 
then when an equation of motion is derived, not in one time. 
Thus $r_1'$ was introduced to distinguish two species of 
$\epsilon R_A$ dependence and to discard $\epsilon R_A$ 
dependence in the right order. We show here a 
simple example.  Let us consider the following integral:  
\begin{equation}
\oint_{\pa B_1}dS_k \frac{r_1^k}{r_1^3}
\int_{N/B}\frac{d^3y}{|\vec x - \vec y|}\frac{1}{y_1^2}. 
\label{SimpExaForCSI}
\end{equation}
Using $\Delta \ln r_1 = 1/r_1^2$, we can integrate the Poisson 
integral and obtain the ``field'',  
$$
\int_{N/B}\frac{d^3y}{|\vec x - \vec y|}\frac{1}{y_1^2} = 
-4\pi \ln \left(\frac{r_1}{{\cal R}/\epsilon}\right) 
+ 4 \pi - 4\pi \frac{\epsilon R_1}{r_1} 
+ O\left(\left(\epsilon R_A\right)^2\right). 
$$
Since the ``body zone contribution'' must have an $\epsilon R_1$ 
dependence hidden in the ``moments'' as $4 \pi \epsilon R_1/r_1 
(+ O((\epsilon R_A)^2))$ 
(see Sec. \ref{Arbitrariness}),   
the terms $ - 4\pi \epsilon R_1/r_1  
+ O((\epsilon R_A)^2)$ 
should be discarded before we evaluate the 
``equation of motion''(the surface integral in 
Eq. (\ref{SimpExaForCSI})). The surface integral gives 
the ``equation of motion'', 
\begin{equation}
16\pi^2 \left(
\ln\left(\frac{{\cal R}/\epsilon}{\epsilon R_1}\right) + 1\right).
\label{SimpExaForCSICorrectOrder}
\end{equation}
On the other hand, we can derive the ``equation of motion'' by 
first evaluating the surface integral over $\pa B_1'$,   
\begin{eqnarray}
\oint_{\pa B_1}dS_k \frac{r_1^k}{r_1^3}
\int_{N/B}\frac{d^3y}{|\vec x - \vec y|}\frac{1}{y_1^2} 
&=& 
\int_{N/B}\frac{d^3y}{y_1^2}
\oint_{\pa B_1'}dS_k \frac{r_1^k}{r_1^3}\frac{1}{|\vec r_1 - \vec y_1|} 
\nonumber \\
\mbox{} &=&
16 \pi^2 \left[
\int_{\epsilon R_1}^{r_1'}\frac{dy}{r_1'} 
+ \int_{r_1'}^{{\cal R}/\epsilon}\frac{dy}{y_1} 
\right]
\nonumber \\
\mbox{} &=&
16\pi^2 \left(
\ln\left(\frac{{\cal R}/\epsilon}{r_1'}\right) + 1 
- \frac{\epsilon R_1}{r_1'}\right).
\label{SimpExaForCSItest} 
\end{eqnarray}
Thus, if we take $\pa B_1$ 
as the integral region instead of 
$\pa B_1'$ in the first equality in Eq. (\ref{SimpExaForCSItest}), 
or if we take $\lim_{r_1' \rightarrow \epsilon R_1}$ 
without employing $\mathop{{\rm disc}}_{\epsilon R_A}$ 
beforehand, we will obtain an incorrect result,  
$$
16\pi^2 
\ln\left(\frac{{\cal R}/\epsilon}{\epsilon R_1}\right),
$$
which disagrees with Eq. (\ref{SimpExaForCSICorrectOrder}).

Now let us return to Eq. (\ref{eq513}).  
To evaluate the 
surface integral, we expand the integrand, supposing $r_1'$ 
is small,\footnote{$0!! = (-1)!! =1$.}   
\begin{eqnarray}
\lefteqn{
\oint_{\pa B_1'}dS_k\frac{1}{|r_1' \vec n_1 - \vec y_1|}
\pa_{z_2^l}
\frac{1}{|\vec r_{12} + r_1' \vec n_1|} 
} \nonumber \\
&=& \pa_{z_2^l} \sum_{\stackrel{{\scriptstyle a=0}}{b=0}}(-1)^a 
\frac{(2a-1)!!(2b-1)!!}{a!b!}
\oint d\Omega_{{\bf n_1}} 
n_1^k n_1^{M_a} n_1^{N_b} n_{12}^{<M_a>} N_1^{<N_b>}
\frac{r_1^{'a+2}}{r_{12}^{a+1}}
\left\{
\frac{r_1^{'b}}{y_1^{b+1}},\frac{y_1^b}{r_1^{'b+1}}
\right\}
\nonumber \\
\mbox{} &=& 4\pi \pa_{z_2^l} \sum_{a=0}(-1)^a 
\frac{(2a-1)!!}{(2a+3)a!}
\frac{1}{r_{12}^{a+1}}n_{12}^{<M_a>} N_1^{<kM_a>}
\left\{
\frac{r_1^{'2a+3}}{y_1^{a+2}},y_1^{a+1}
\right\}
\nonumber \\
\mbox{} &&- 
4\pi \pa_{z_2^l} \sum_{a=0}(-1)^a 
\frac{(2a-1)!!}{(2a+3)a!}
\frac{1}{r_{12}^{a+2}}n_{12}^{<kM_a>} N_1^{<M_a>}
\left\{
\frac{r_1^{'2a+3}}{y_1^{a+1}},r_1^{'2}y_1^{a}
\right\}, 
\label{eq515}
\end{eqnarray}
where $N_1^i \equiv y_1^i/y_1$, and in $\{f,g\}$ in the above 
equation, $f$ denotes the result for $r_1' < y_1$ 
and $g$ denotes the result for $r_1' > y_1$.  
In the last equality of Eq. (\ref{eq515}), we used the 
following formula \cite{Thorne80}:  
\begin{eqnarray}
\lefteqn{\frac{1}{4 \pi}\oint
d\Omega_{{\bf n_1}} 
n_1^k
n_1^{I_a}
n_1^{J_b}
n_{12}^{<I_a>}
N_1^{<J_b>} 
}
\nonumber \\
&=&
\frac{a!}{(2a+1)!!}N_1^{<I_{a-1}>}n_{12}^{<k I_{a-1}>}\delta_{a,b+1}
+ 
\frac{b!}{(2b+1)!!}N_1^{<kJ_{b-1}>}n_{12}^{<J_{b-1}>}\delta_{a+1,b}.  
\end{eqnarray}
Substituting Eq. (\ref{eq515}) into Eq. (\ref{eq513}), we 
establish a formula 
\begin{eqnarray}
\lefteqn{\oint_{\pa B_1}dS_k\frac{r_2^l}{r_2^3}
\mathop{{\rm disc}}_{\epsilon R_A}
\int_{N/B}d^3y
\frac{f(\vec y_1)}{|\vec r_1 - \vec y_1|}}
\nonumber \\
&=& \lim_{r_1' \rightarrow \epsilon R_A} 
\mathop{{\rm disc}}_{\epsilon R_A}
4 \pi \sum_{a=0}(-1)^a\frac{(2a-1)!!}{(2a+3)a!}
\nonumber \\
&&\times
\left[
\int_{N/B'}d^3y_1 f(\vec y_1)
\left\{
\frac{r_1^{'2a+3}}{y_1^{a+2}}N_1^{<kM_a>}\pa_{z_2^l}
\left(\frac{n_{12}^{<M_a>}}{r_{12}^{a+1}}\right) 
- \frac{r_1^{'2a+3}}{y_1^{a+1}}N_1^{<M_a>}\pa_{z_2^l}
\left(\frac{n_{12}^{<kM_a>}}{r_{12}^{a+2}}\right) 
\right\} 
\right.
\nonumber \\  
&&+ \left. 
\int_{B_1'/B_1}d^3y_1 f(\vec y_1)
\left\{
y_1^{a+1}N_1^{<kM_a>}\pa_{z_2^l}
\left(\frac{n_{12}^{<M_a>}}{r_{12}^{a+1}}\right) 
- r_1^{'2}y_1^{a}N_1^{<M_a>}\pa_{z_2^l}
\left(\frac{n_{12}^{<kM_a>}}{r_{12}^{a+2}}\right) 
\right\}
\right], 
\label{CompanionStarFormula}
\end{eqnarray}
where $B' \equiv  B_1' \cup B_2$.

Now since we expect from the discussion in Sec. \ref{Arbitrariness}  
that   
an equation of motion does not depend on $\epsilon R_A$ and 
hence $r_1'$, in the formally infinite series in Eq. 
(\ref{CompanionStarFormula}), only finite terms for which the volume 
integral makes $r_1'$ dependence disappear do contribute. 
The $r_1'$ dependence of the above integral can be examined by 
the behavior of the integrand near $\pa B_1'$. Thus, to examine 
$r_1'$ dependence, we expand $f(\vec y_1)$ in 
the small $y_1$,\footnote{$\mbox{}_8\Lambda_S^{\tau i}$ has 
no logarithmic dependence.}   
\begin{equation}
f(\vec y_1) = \sum_{p=p_0}\frac{1}{y_1^p}
\mathop{{\hat f}}_{p}(\vec N_1\cdot\vec n_{12}),  
\label{eq5-23}
\end{equation}
where $p_0$, the lowest bound in the sum, 
is a finite integer.  Then, $r_1'$ dependence of the four 
volume integrals in Eq. (\ref{CompanionStarFormula}) 
can be evaluated as 
{
\setcounter{enumi}{\value{equation}}
\addtocounter{enumi}{1}
\setcounter{equation}{0}
\renewcommand{\theequation}{5.\theenumi\alph{equation}}
\begin{eqnarray} 
&&
\mathop{{\rm disc}}_{\epsilon R_A}
\int_{r_1'}dy_1 f(\vec y_1) \frac{r_1^{'2a+3}}{y_1^{a}} = 
 \sum_{p=p_0}\mathop{{\hat f}}_{p}(\vec N_1\cdot\vec n_{12}) 
 \left(\frac{r_1^{'a-p+4}}{a+p-1}\delta_{1-p \neq a}-
 r_1^{'2a+3}\delta_{1-p,a}\ln r_1^{'} 
\right),
\label{eq317a}\\
&&
\mathop{{\rm disc}}_{\epsilon R_A} 
\int_{r_1'}
dy_1 f(\vec y_1) \frac{r_1^{'2a+3}}{y_1^{a-1}} = 
 \sum_{p=p_0}\mathop{{\hat f}}_{p}(\vec N_1\cdot\vec n_{12}) 
\left(\frac{r_1^{'a-p+5}}{a+p-2}\delta_{2-p \neq a}-  
r_1^{'2a+3}\delta_{2-p,a}\ln r_1^{'}
\right),
\label{eq317b}\\
&&
\mathop{{\rm disc}}_{\epsilon R_A} 
\int^{r_1'}_{\epsilon R_1}
dy_1 f(\vec y_1) y_1^{a+3} = 
 \sum_{p=p_0}\mathop{{\hat f}}_{p}(\vec N_1\cdot\vec n_{12}) 
\left(
 \frac{r_1^{'a-p+4}}{a-p+4}\delta_{p-4 \neq a}  
+ \delta_{p-4,a}\ln\left(\frac{r_1'}{\epsilon R_1}\right)
\right),
\label{eq317c}\\
&&
\mathop{{\rm disc}}_{\epsilon R_A} 
\int^{r_1'}_{\epsilon R_1}
dy_1 f(\vec y_1) r_1^{'2}y_1^{a+2} = 
 \sum_{p=p_0}\mathop{{\hat f}}_{p}(\vec N_1\cdot\vec n_{12}) 
\left(
\frac{r_1^{'a-p+5}}{a-p+3}\delta_{p-3 \neq a}  
+ r_1^{'2}\delta_{p-3,a}\ln\left(\frac{r_1'}{\epsilon R_1}\right)
\right), 
\label{eq317d}
\end{eqnarray}
where $\delta_{a\neq b} = 0$ for $a=b$ and $1$ otherwise. 
Note that $a\ge0$. Since we take $r_1^{'} \rightarrow \epsilon R_A$ 
and discard $\epsilon R_A$ dependence, in the above four equations 
the only terms which we should retain are   
\setcounter{equation}{\value{enumi}}
}
{
\setcounter{enumi}{\value{equation}}
\addtocounter{enumi}{1}
\setcounter{equation}{0}
\renewcommand{\theequation}{5.\theenumi\alph{equation}}
\begin{eqnarray} 
\mathop{{\rm disc}}_{\epsilon R_A}
\lim_{r_1' \rightarrow \epsilon R_1}
\mathop{{\rm disc}}_{\epsilon R_A}
\int_{r_1'}dy_1 f(\vec y_1) \frac{r_1^{'2a+3}}{y_1^{a}} &=& 
 \sum_{p=p_0}
\frac{1}{2p-5}\delta_{a,p-4}
\mathop{{\hat f}}_{p}(\vec N_1\cdot\vec n_{12}), 
\label{eq318a}\\
\mathop{{\rm disc}}_{\epsilon R_A}
\lim_{r_1' \rightarrow \epsilon R_1}
\mathop{{\rm disc}}_{\epsilon R_A} 
\int_{r_1'}dy_1 f(\vec y_1) \frac{r_1^{'2a+3}}{y_1^{a-1}} &=& 
 \sum_{p=p_0}
\frac{1}{2p-7}\delta_{a,p-5}
\mathop{{\hat f}}_{p}(\vec N_1\cdot\vec n_{12}), 
\label{eq318b}\\
\mathop{{\rm disc}}_{\epsilon R_A}
\lim_{r_1' \rightarrow \epsilon R_1}
\mathop{{\rm disc}}_{\epsilon R_A} 
\int^{r_1'}_{\epsilon R_1}
dy_1 f(\vec y_1) y_1^{a+3} &=& 0,
\label{eq318c}\\
\mathop{{\rm disc}}_{\epsilon R_A}
\lim_{r_1' \rightarrow \epsilon R_1}
\mathop{{\rm disc}}_{\epsilon R_A} 
\int^{r_1'}_{\epsilon R_1}
dy_1 f(\vec y_1) r_1^{'2}y_1^{a+2} &=& 
- \sum_{p=p_0}
\frac{1}{2}\delta_{a,p-5}
\mathop{{\hat f}}_{p}(\vec N_1\cdot\vec n_{12}).    
\label{eq318d}
\end{eqnarray}
\setcounter{equation}{\value{enumi}}
}
Substituting Eqs. (\ref{eq318a}), (\ref{eq318b}), 
(\ref{eq318c}),  and (\ref{eq318d}) into Eq.  
(\ref{CompanionStarFormula}), we obtain

\begin{eqnarray}
\lefteqn{
\mathop{{\rm disc}}_{\epsilon R_A}
\oint_{\pa B_1}dS_k\frac{r_2^l}{r_2^3}
\mathop{{\rm disc}}_{\epsilon R_A}
\int_{N/B}d^3y
\frac{f(\vec y_1)}{|\vec r_1 - \vec y_1|}}\nonumber \\
&=& 
4 \pi 
\sum_{a=p_0-4 \ge 0}(-1)^{a}\frac{(2a-1)!!}{(2a+3)^2 a!}
\oint d\Omega_{{\bf N_1}}
\mathop{{\hat f}}_{a+4}(\vec N_1\cdot\vec n_{12})N_1^{<kM_a>}
\pa_{z_2^{l}}\left(\frac{n_{12}^{<M_a>}}{r_{12}^{a+1}}\right)   
\nonumber \\
&+& 
4 \pi 
\sum_{a=p_0-5 \ge 0}(-1)^{a}\frac{(2a+1)!!}{2 (2a+3)^2 a!}
\oint d\Omega_{{\bf N_1}}
\mathop{{\hat f}}_{a+5}(\vec N_1\cdot\vec n_{12})N_1^{<M_a>}
\pa_{z_2^{l}}\left(\frac{n_{12}^{<kM_a>}}{r_{12}^{a+2}}\right),    
\label{CompanionStarIntegral}
\end{eqnarray}
where we applied $\mathop{{\rm disc}}_{\epsilon R_A}$ to the 
left-hand side of Eq. (\ref{CompanionStarFormula})  
to make it clear that we discard the 
$\epsilon R_A$-dependent terms 
(other than the $\ln \epsilon R_A$ dependence) 
when deriving an equation of motion. 

To illustrate our method, 
let us take $\left[
- \mbox{}_4h^{kl,i}\mbox{}_4h^{\tau}\mbox{}_{k,l}
\right]_{{\rm DIP}}$ 
defined by Eq. (\ref{eq5-11}) as 
an integrand, for instance, in Eq. (\ref{CompanionStarIntegral}),  
$$f(\vec y_1) = \frac{2 m_2}{\pi}
\pa_{[j}\left[
- \mbox{}_4h^{|kl|,i]}\mbox{}_4h^{\tau}\mbox{}_{k,l}
\right]_{{\rm DIP}}, 
$$ 
where the 
vertical strokes denote that indexes between the strokes are 
excluded from (anti)symmetrization. 
We expand this integrand around $z_1^i$ as 
\begin{eqnarray}
\frac{2 m_2}{\pi}
\pa_{[j}\left[
- \mbox{}_4h^{|kl|,i]}\mbox{}_4h^{\tau}\mbox{}_{k,l}
\right]_{{\rm DIP}}&=& 
\frac{32m_1^2m_2^2}{\pi r_1^4 r_{12}^2}
\left(
- n_{12}^{[i}v_1^{j]} 
- 2(\vec n_1\cdot\vec v_1)n_1^{[i}n_{12}^{j]}
- 2 (\vec n_1\cdot\vec n_{12})n_1^{[i}v_{1}^{j]}
\right) 
\nonumber \\
\mbox{} &&
+ O\left(\left(\frac{r_{12}}{r_1}\right)^3\right). 
\end{eqnarray}
Thus we see that the integrand $\pa_{[j}\left[
- \mbox{}_4h^{|kl|,i]}\mbox{}_4h^{\tau}\mbox{}_{k,l}
\right]_{{\rm DIP}}$, for which $p_0=4$, does not 
contribute to the 3PN evolution equation of $P_{1\Theta}^{\tau}$ 
since the  angular integration of the integrand in Eq. 
(\ref{CompanionStarIntegral}) gives zero.  

Returning to the evaluation of $\mbox{}_8h_{{\rm DIP}}^{\tau i}$, 
we apply Eq. (\ref{CompanionStarIntegral}) 
to  $f(\vec y_1) = 
(2 m_2/\pi)
\mbox{}_8\Lambda_S^{\tau [l,k]}(\vec y_1 + \vec z_1)$.  
Expanding $\mbox{}_8\Lambda_S^{\tau [l,k]}(\vec y_1 + \vec z_1)$ 
around $z_1^i$, we find that $p_0=4$. Evaluating 
the angular integral in Eq. 
(\ref{CompanionStarIntegral}) 
for $\mbox{}_8\Lambda_S^{\tau [l,k]}(\vec y_1 + \vec z_1)$, 
however, results in zero.

\section{Harmonic condition}
\label{tpnhc}

In this section, we check a part of the harmonic condition,   
\begin{equation}
\mbox{}_{\le 8}h^{\tau\tau}\mbox{}_{,\tau} 
+ 
\mbox{}_{\le 8}h^{\tau i}\mbox{}_{,i} 
= 0.
\label{HarmonicCondition3PNFull}
\end{equation}
As explained in the previous section, we split  
the nonretarded part of the $N/B$ contribution 
$\mbox{}_8h^{\tau i}_{N/B n=0}$ into three groups: 
$\mbox{}_8h^{\tau i}_{{\rm SP}}$ 
(the superpotential part),
$\mbox{}_8h^{\tau i}_{{\rm SSP}}$ 
(the superpotential-in-series part), 
and 
$\mbox{}_8h^{\tau i}_{{\rm DIP}}$
(the direct-integration part). 
We could obtain an explicit form of only  
$\mbox{}_8h^{\tau i}_{{\rm SP}}$, 
the body zone contribution 
$\mbox{}_8h^{\tau i}_{B}$, 
and the second time derivative term in the retarded field 
$\mbox{}_8h^{\tau i}_{N/B n=2}$
(the last 
term in Eq. (\ref{eq:h8tiGenkei})).
Our trick is to transform the left-hand side
of Eq. (\ref{HarmonicCondition3PNFull}) into the following form:  
\begin{eqnarray}
{\cal H}(\tau,\vec x) &\equiv& 
\mbox{}_{\le 8}h^{\tau\tau}\mbox{}_{,\tau} 
+ \mbox{}_{\le 7}h^{\tau i}\mbox{}_{,i} 
+ \mbox{}_{8}h^{\tau i}_{B}\mbox{}_{,i} 
+ \mbox{}_{8}h^{\tau i}_{{\rm SP}}\mbox{}_{,i}
+ \mbox{}_{8}h^{\tau i}_{N/Bn=2}\mbox{}_{,i}  
\nonumber \\
\mbox{} && 
+ 4 \int_{N/B}\frac{d^3y}{|\vec x - \vec y|}
\frac{\pa}{\pa y^i} 
(\mbox{}_8\Lambda_{{\rm SSP}}^{\tau i} + \mbox{}_8\Lambda_{S}^{\tau i})
- 4 \oint_{\pa (N/B)}\frac{dS_i}{|\vec x - \vec y|}
(\mbox{}_8\Lambda_{{\rm SSP}}^{\tau i} + \mbox{}_8\Lambda_{S}^{\tau i}), 
\label{HarmonicCondition3PN} 
\end{eqnarray}
where $\mbox{}_8\Lambda_{{\rm SSP}}^{\tau i}$ is the integrand corresponding 
to the superpotential-in-series part. Now not surprisingly, 
we could evaluate the integrals explicitly using superpotentials. 
For example,  
\begin{eqnarray}
\frac{\pa}{\pa y^i}
16 \pi \mbox{}_8\Lambda_{{\rm SSP}}^{\tau i} &=& 
\frac{\pa}{\pa y^i}\left(
\frac{24 m_1^2m_2y_1^i (\vec y_1\cdot\vec v_2)}{y_1^6y_2} + \cdots 
\right) \nonumber \\ 
\mbox{} &=& 
\Delta \left(24 m_1^2m_2v_2^k
\left(
- \frac{1}{4}\pa_{z_1^k} f^{(-2,-3)} 
+ \frac{r_{12}^2}{8}\pa_{z_1^k} f^{(-4,-3)}
- \frac{5}{8}\pa_{z_1^k} f^{(-4,-1)} \right)
\right)
\nonumber \\
\mbox{} &&
+ \cdots.    
\end{eqnarray}

In this manner, 
though we could not evaluate Poisson integrals explicitly 
and consequently the field $\mbox{}_8h^{\tau i}$ valid 
throughout $N/B$ is not available in a closed form, we checked  the 
harmonic condition in the sense that ${\cal H}(\tau,\vec x) =0$ 
throughout $N/B$.

\section{The 3PN Mass-Energy Relation}
\label{MeaningOfPt}

As explained in Sec. \ref{chap:3PNEOM}, the direct-integration 
part does not play any role in the evaluation of the 
evolution equation of $P_{A\Theta}^{\tau}$. Thus we can take  
the same method as in the evaluation of the 2.5PN 
equation of motion. We first express 
$\mbox{}_{10}\Theta_{N}^{\tau \mu}$ 
explicitly by substituting  
the field derived previously into Eqs. 
(\ref{tLLtt10}) and 
(\ref{tLLti10}).  Then evaluating the surface integrals 
in Eq. (\ref{EvolOfFourMom3PN}), we obtain the evolution equation of 
$P_{A\Theta}^{\tau}$,  
\begin{eqnarray}
\left(\frac{d P_{1 \Theta}^{\tau}}{d \tau}\right)_{\le 3{\rm PN}} &=&
- \epsilon^2 \frac{m_1 m_2}{r_{12}^2}\left[
4(\vec{n}_{12}\cdot\vec{v}_1) - 3(\vec{n}_{12}\cdot\vec{v}_2)
\right]
\nonumber \\
\mbox{} &+&
\epsilon^4 \frac{m_1 m_2}{r_{12}^2}\left[
 -\frac{9}{2} (\vec{n}_{12}\cdot\vec{v}_2)^3
+ \frac{1}{2}v_1^2 (\vec{n}_{12}\cdot\vec{v}_2)
+ 6 (\vec{n}_{12}\cdot\vec{v}_1)
(\vec{n}_{12}\cdot\vec{v}_2)^2
\right. \nonumber \\
\mbox{} &&- \left.
 2 v_1^2 (\vec{n}_{12}\cdot\vec{v}_1)
+ 4 (\vec{v}_1\cdot\vec{v}_{2})
(\vec{n}_{12}\cdot\vec{V})
+ 5 v_2^2 (\vec{n}_{12}\cdot\vec{v}_2)
-4 v_2^2 (\vec{n}_{12}\cdot\vec{v}_1)
\right. \nonumber \\
\mbox{} &&+ \left.
 \frac{m_1}{r_{12}}
(-4 (\vec{n}_{12}\cdot\vec{v}_2)
 + 6 (\vec{n}_{12}\cdot\vec{v}_1))
+ \frac{m_2}{r_{12}}
(-10 (\vec{n}_{12}\cdot\vec{v}_1) + 11 (\vec{n}_{12}\cdot\vec{v}_2))
\right]
\nonumber \\
\mbox{} &+&
\epsilon^6\frac{m_1m_2}{r_{12}^2}
\left[
- \left(\frac{3}{2}v_1^4
+2v_1^2v_2^2 + 4v_2^4\right)
(\vec n_{12}\cdot\vec v_1) +
\left(\frac{5}{8}v_1^4
+\frac{3}{2} v_1^2v_2^2
+ 7v_2^4\right)
(\vec n_{12}\cdot\vec v_2) 
\right.
\nonumber \\
\mbox{} &&+ \left.
\left(2v_1^2+4v_2^2\right)
(\vec n_{12}\cdot\vec v_1)(\vec v_1\cdot\vec v_2)
- \left(2v_1^2+8v_2^2\right)
(\vec n_{12}\cdot\vec v_2)(\vec v_1\cdot\vec v_2)
\right.
\nonumber \\
\mbox{} &&+ \left.
\left(3 v_1^2 + 12 v_2^2\right)(\vec n_{12}\cdot\vec v_1)
(\vec n_{12}\cdot\vec v_2)^2 
- \left(\frac{3}{4}v_1^2
+ 12v_2^2\right)
(\vec n_{12}\cdot\vec v_2)^3    
\right.
\nonumber \\
\mbox{} &&+ \left.
 2 (\vec n_{12}\cdot\vec v_2)(\vec v_1\cdot\vec v_2)^2
-6 (\vec n_{12}\cdot\vec v_1)(\vec n_{12}\cdot\vec v_2)^2
(\vec v_1\cdot\vec v_2)
+6 (\vec n_{12}\cdot\vec v_2)^3(\vec v_1\cdot\vec v_2)
\right.
\nonumber \\
\mbox{} &&- \left.
\frac{15}{2}
(\vec n_{12}\cdot\vec v_1)(\vec n_{12}\cdot\vec v_2)^4
+\frac{45}{8}(\vec n_{12}\cdot\vec v_2)^5
\right. 
\nonumber \\
\mbox{} &&+ \left. 
 \frac{m_1}{r_{12}}
\left(
\left(-42v_1^2 - \frac{117}{4}v_2^2\right)(\vec n_{12}\cdot\vec v_1)
+
60(\vec n_{12}\cdot\vec v_1)^3
\right.\right.
\nonumber \\
\mbox{} &&+ \left.\left.
\left(\frac{137}{4} v_1^2
+\frac{37}{2}v_2^2
\right)(\vec n_{12}\cdot\vec v_2)
\right.\right.
\nonumber \\
\mbox{} &&+ \left.\left. 
 \frac{297}{4}(\vec n_{12}\cdot\vec v_1)(\vec v_1 \cdot\vec v_2)
- \frac{219}{4}(\vec n_{12}\cdot\vec v_2)(\vec v_1 \cdot\vec v_2)
- 151(\vec n_{12}\cdot\vec v_1)^2(\vec n_{12}\cdot\vec v_2)
\right.\right.
\nonumber \\
\mbox{} &&+ \left.\left. 
 109(\vec n_{12}\cdot\vec v_1)(\vec n_{12}\cdot\vec v_2)^2
- 23(\vec n_{12}\cdot\vec v_2)^3
\right)
\right.
\nonumber \\
\mbox{} &&+ \left.
 \frac{m_2}{r_{12}}
\left(
- \left(13 v_1^2 + 18 v_2^2 \right)(\vec n_{12}\cdot\vec v_1) 
+ \left(\frac{17}{2} v_1^2 + 25 v_2^2 \right)(\vec n_{12}\cdot\vec v_2) 
\right.\right.
\nonumber \\
\mbox{} &&+ \left.\left.
26 (\vec n_{12}\cdot\vec v_1) (\vec v_1\cdot\vec v_2) 
- 28(\vec n_{12}\cdot\vec v_2) (\vec v_1\cdot\vec v_2) 
+2(\vec n_{12}\cdot\vec v_1)^2(\vec n_{12}\cdot\vec v_2)
\right.\right.
\nonumber \\
\mbox{} &&+ \left.\left.
16(\vec n_{12}\cdot\vec v_1)(\vec n_{12}\cdot\vec v_2)^2
-20(\vec n_{12}\cdot\vec v_2)^3
\right) 
\right.
\nonumber \\
\mbox{} &&+ \left.
\frac{m_1^2}{r_{12}^2}
\left(\
\frac{33}{4} (\vec n_{12}\cdot\vec v_1)
- \frac{13}{2}(\vec n_{12}\cdot\vec v_2)
\right)
-\frac{m_1m_2}{r_{12}^2}
\left(\frac{35}{4}(\vec n_{12}\cdot\vec v_1)
+ \frac{17}{4}(\vec n_{12}\cdot\vec v_2) 
\right)
\right.
\nonumber \\
\mbox{} &&+ \left.
\frac{m_2^2}{r_{12}^2}
\left(\
- 12(\vec n_{12}\cdot\vec v_1) 
+ \frac{23}{2}(\vec n_{12}\cdot\vec v_2)
\right)
\right],
\label{TPNEMSurfaceIntegralFormTimeResult}
\end{eqnarray}
where $\vec V \equiv \vec v_1 - \vec v_2$. 

Remarkably, we can integrate Eq. 
(\ref{TPNEMSurfaceIntegralFormTimeResult}) functionally, 
\begin{eqnarray}
P^{\tau}_{1 \Theta} &=& m_1 \left( 1 + \epsilon^2 \mbox{}_2\Gamma_1
+ \epsilon^4 \mbox{}_4\Gamma_1 + \epsilon^6\mbox{}_6\Gamma_1 \right) + 
O(\epsilon^7),
\label{3PNEnergy}
\end{eqnarray}
with  
\begin{eqnarray}
\mbox{}_2\Gamma_1 &=& 
 \frac{1}{2}v_1^2 + \frac{3 m_2}{r_{12}},
\label{DefOfGamma2}
\\
\mbox{}_4\Gamma_1 &=& - \frac{3 m_2}{2 r_{12}} (\vec{n}_{12}\cdot\vec{v}_2)^2
+ \frac{2 m_2}{r_{12}}v_2^2
+ \frac{7 m_2}{2 r_{12}}v_1^2
- \frac{4 m_2}{r_{12}} (\vec{v}_1\cdot\vec{v}_{2})
+ \frac{3}{8} v_1^4
+ \frac{7 m_2^2}{2 r_{12}^2}
- \frac{5 m_1 m_2}{2 r_{12}^2},
\label{DefOfGamma4}
\end{eqnarray}

\begin{eqnarray}
\mbox{}_6\Gamma_1 &=& 
\frac{m_1^2m_2}{2r_{12}^3} + \frac{21m_1m_2^2}{4r_{12}^3} + 
\frac{5m_2^3}{2r_{12}^3} 
+ \frac{5}{16} v_1^6
\nonumber \\
\mbox{} &&+ 
 \frac{m_2^2}{r_{12}^2}\left(
\frac{45}{4}v_1^2
+\frac{19}{2} v_2^2
+ \frac{1}{2}(\vec n_{12}\cdot\vec v_1)^2
-19 (\vec v_1\cdot\vec v_2)
- (\vec n_{12}\cdot\vec v_1)(\vec n_{12}\cdot\vec v_2)
-3 (\vec n_{12}\cdot\vec v_2)^2
\right) 
\nonumber \\
\mbox{} &&+ 
 \frac{m_1m_2}{r_{12}^2}
\left(\frac{43}{8} v_1^2
+ \frac{53}{8}v_2^2
- \frac{69}{8}(\vec n_{12}\cdot\vec v_1)^2
- \frac{53}{4}(\vec v_1\cdot\vec v_2)
+\frac{85}{4}(\vec n_{12}\cdot\vec v_1)(\vec n_{12}\cdot\vec v_2)
\right. 
\nonumber \\
\mbox{} &&- \left. 
\frac{69}{8}(\vec n_{12}\cdot\vec v_2)^2
\right)
+ \frac{m_2}{r_{12}}
\left(\frac{33}{8}v_1^4
+  3 v_1^2v_2^2
+  2 v_2^4 
- 6 v_1^2(\vec v_1\cdot\vec v_2)
- 4 v_2^2(\vec v_1\cdot\vec v_2)
\right.
\nonumber \\ 
\mbox{}&&- \left.
\frac{7}{4}v_1^2(\vec n_{12}\cdot\vec v_2)^2
-\frac{5}{2}v_2^2(\vec n_{12}\cdot\vec v_2)^2
+  2(\vec v_1\cdot\vec v_2)^2
\right.
\nonumber \\ 
\mbox{}&&+ \left.
 2(\vec n_{12}\cdot\vec v_2)^2(\vec v_1\cdot\vec v_2)
+
\frac{9}{8}(\vec n_{12}\cdot\vec v_2)^4
\right). 
\label{DefOfGamma6}
\end{eqnarray}  
The mass-energy relation for the $\chi$ part up to 3PN 
order is given in Appendix \ref{chipart}.
Equations  
(\ref{3PNEnergy}) 
and (\ref{Ptchi3PN}) 
give the 3PN order mass-energy relation.

\subsubsection{Meaning of $P^{\tau}_{A \Theta}$}

In this section, we suggest an interesting interpretation 
of the mass-energy relation.
First of all, we expand in an $\epsilon$ series 
a four-velocity of the star $A$ normalized
as $g_{\mu\nu}u_A^{\mu}u_A^{\nu} = -\epsilon^{-2}$, where
$u_A^i = u_A^{\tau}v_A^i$. 
We have 
\begin{eqnarray}
u_A^{\tau} &=& 1 +
\epsilon^2\left[\frac{1}{2}v_A^2 + \frac{1}{4}\mbox{}_4h^{\tau\tau}\right]
\nonumber \\
\mbox{} &+&
\epsilon^4\left[
\frac{1}{4}\mbox{}_6h^{\tau\tau} +
\frac{1}{4}\mbox{}_4h^{k}\mbox{}_{k} -
\frac{3}{32}(\mbox{}_4h^{\tau\tau})^2
+ \frac{5}{8}\mbox{}_4h^{\tau\tau}v_A^2
-\mbox{}_4h^{\tau}\mbox{}_{k}v_A^k + \frac{3}{8}v_A^4 \right]
\nonumber \\
\mbox{} &+&
 \epsilon^5\frac{1}{4}\left[\mbox{}_7h^{\tau\tau}
+ \mbox{}_5h^k\mbox{}_k
\right] 
\nonumber \\
\mbox{} &+& 
\epsilon^6
\left[
\frac{1}{4}\mbox{}_8h^{\tau\tau} 
+ \frac{1}{4}\mbox{}_6h^k\mbox{}_k 
+ \frac{1}{16}\mbox{}_4h^{\tau\tau}\mbox{}_4h^k\mbox{}_k 
-\frac{3}{16}\mbox{}_4h^{\tau\tau}\mbox{}_6h^{\tau\tau}
+\frac{7}{128}(\mbox{}_4h^{\tau\tau})^3
+ \frac{1}{4}\mbox{}_4h^{\tau k}\mbox{}_4h^{\tau}\mbox{}_k
\right.
\nonumber \\
\mbox{}  &&- \left.
 \mbox{}_6h^{\tau}\mbox{}_kv_A^k
- \frac{1}{4}\mbox{}_4h^{\tau\tau}\mbox{}_4h^{\tau}\mbox{}_kv_A^k
+ \frac{1}{2}\mbox{}_4h_{kl}v_A^kv_A^l
+ \frac{1}{8}\mbox{}_4h^k\mbox{}_kv_A^2
+ \frac{5}{64}(\mbox{}_4h^{\tau\tau})^2v_A^2
\right.
\nonumber \\
\mbox{}  &&+ \left.
 \frac{5}{8}\mbox{}_6h^{\tau\tau}v_A^2
-\frac{3}{2}\mbox{}_4h^{\tau}\mbox{}_kv_A^kv_A^2 
+ \frac{27}{32}\mbox{}_4h^{\tau\tau}v_A^4 
+ \frac{5}{16}v_A^6
\right] + O(\epsilon^7).
\label{ExpandedUt} 
\end{eqnarray}
This is a formal series since 
the field should be evaluated somehow at $\vec z_A$ while  
the metric derived 
via the point-particle
description diverges at $\vec z_A$. 

Now let us regularize this equation with Hadamard's partie finie 
regularization (see, e.g., \cite{BF00b} 
in the literature of the post-Newtonian approximation). 
Evaluating with this procedure 
Eq. (\ref{ExpandedUt}) and $\sqrt{-g}$ expanded in 
$\epsilon$ up to $O(\epsilon^6)$, 
then comparing the result with Eq. (\ref{3PNEnergy}),  
we find   
at least up to 3PN order
\begin{eqnarray}
P_{A \Theta}^{\tau} = m_A[\sqrt{-g}u^{\tau}_A]_{A}^{{\rm ext}}. 
\label{MassEnergyRelationFull}
\end{eqnarray}
In the above equation, $[f]^{{\rm ext}}_{A}$
means that we regularize  
the quantity $f$ at the star $A$  
by Hadamard's partie finie or whatever regularizations which 
give the same result.

We emphasize that 
we have never assumed this ``natural''  
relation in advance. This relation Eq. (\ref{3PNEnergy}) 
has been
derived by solving the evolution equation for
$P_{A \Theta}^{\tau}$ functionally.

\section{The 3PN Momentum-Velocity Relation}
\label{sec:3PNMomVelRelation}

We now derive the 3PN momentum-velocity relation by 
calculating the $Q_{A}^i$ integral at 3PN order.  
From the definition of the $Q_A^i$ integral, Eq. (\ref{QL}), 
$$
Q_A^{i} = \epsilon^{6}
 \oint_{\pa B_A} dS_k
 \left(\mbox{}_{10}\Lambda^{\tau k}_N  - v_A^k 
\mbox{}_{10}\Lambda^{\tau\tau}_N \right) y_A^i + O(\epsilon^7), 
$$
we find that the calculation required is almost the same 
as that in the equation for $dP_{A\Theta}^{\tau}/d\tau$. Namely, 
$\mbox{}_8h^{\tau i}_{{\rm DIP}}$ does not contribute to     
the $Q_A^i$ integral due to (i) the antisymmetry of 
the direct-integration part (see Eq. (\ref{eq-511})), 
(ii) the behavior of the direct-integration part at infinity 
(see Eq. (\ref{eq-512})),  
and (iii) its behavior around $\vec z_A$ 
(see Eq. (\ref{CompanionStarIntegral}). Note  
that  $\mbox{}_8\Lambda_{S}^{\tau k}(\tau,\vec z_A + \vec y_A) y_A^i 
\sim 1/y_A^3$ in the neighborhood of the star $A$).  
Therefore, we need to compute 
the surface integrals 
in the definition of $Q_A^i$ 
using the field up to 2PN order, 
$\mbox{}_8h_B^{\tau i}$, $\mbox{}_8h_{N/Bn=2}^{\tau i}$,  
$\mbox{}_8h_{{\rm SP}}^{\tau i}$, 
and $\mbox{}_8h_{{\rm SSP}}^{\tau i}$. 
We show only $Q_{A\Theta}^i$ here,   
\begin{eqnarray}
\mbox{}_{\le 6}Q_{1\Theta}^i &=& 
- \epsilon^6 \frac{m_1^3m_2 n_{12}^{<ij>}v_{12}^j}{2r_{12}^3} 
= \epsilon^6 
\frac{d}{d\tau}\left(\frac{m_1^3m_2}{6r_{12}^3}r_{12}^i\right)
=- \epsilon^6 
\frac{d}{d\tau}\left(\frac{1}{6}m_1^3a_1^i\right), 
\label{eq:Q1ThetaIntegralOfOrdere6}
\end{eqnarray}
where it should be  understood that $a_A^i$ in the last expression  
is evaluated with the Newtonian acceleration.
We show $Q^i_{A\chi}$ in Appendix \ref{chipart}.

In the field, 
$Q_A^{i}$ of $O(\epsilon^6)$ appears at 4PN or higher 
order. Thus up to 3PN order, 
$\mbox{}_6Q_A^i$ 
affects the equation of motion only  
through the 3PN momentum-velocity relation. 
As stated in Sec. \ref{chap:3PNEOM},  
the nontrivial momentum-velocity relation 
which affects the equation of motion is 
that of the $\Theta$ part. 
This in turn motivates us to   
define the representative point of the star $A$,  $z_A^i$, 
by choosing the value of $D_{A\Theta}^i$.
We do not take into account $D^i_{A\chi}$ in the 
definition of $z_A^i$. 

Now with $Q_{A\Theta}^i$ in hand, we obtain the 
momentum-velocity relation,    
\begin{eqnarray}
&&
P_{1\Theta}^i 
= P_{1\Theta}^{\tau} v_1^i  
- \epsilon^6 \frac{d}{d\tau}\left(\frac{1}{6}m_1^3a_1^i\right) 
+ \epsilon^2 \frac{d D_{1\Theta}^i}{d\tau}. 
\label{eq:3PMMomuntumVelocityRelation}
\end{eqnarray}
This equation suggests that we choose  
\begin{eqnarray}
&&
D_{A\Theta}^i(\tau) =  \epsilon^4 \frac{1}{6} m_A^3a_A^i   
= \epsilon^4 \delta_{A\Theta}^i(\tau). 
\label{3PNDefOfCM}
\end{eqnarray}
For a while henceforth, we shall define $z_A^i$ by this equation.
We shall later give a more convenient definition of $z_A^i$.

Finally, we note that the nonzero 
dipole moment $D_{A\Theta}^i$ 
of order $\epsilon^4$ affects  
the 3PN field and the 3PN equation of motion in essentially 
the same manner as the Newtonian dipole moment affects  
the Newtonian field and the Newtonian equation of motion. 
From Eqs. (\ref{hBtt}), (\ref{hBti}), 
and (\ref{hBij}) (or Eqs. (\ref{hBttInAp}), (\ref{hBtiInAp}), and 
(\ref{hBijInAp}) for more explicit expressions), we see that   
$\delta_{A\Theta}^i$ appears only at 
$\mbox{}_{10}h^{\tau\tau}$ as  
\begin{eqnarray}
&&
h^{\tau\tau}|_{\delta_{A\Theta}} = 
4 \epsilon^{10}\sum_{A=1,2}\frac{\delta_{A\Theta}^kr_A^k}{r_A^3} 
+ O(\epsilon^{11}). 
\end{eqnarray}
Then the corresponding acceleration becomes 
\begin{eqnarray}
&&
m_1 a_1^i|_{\delta_{A\Theta}} = 
- \epsilon^6 \frac{3m_1\delta_{2\Theta}^k}{r_{12}^3}n_{12}^{<ik>} 
+ \epsilon^6\frac{3m_2\delta_{1\Theta}^k}{r_{12}^3}n_{12}^{<ik>} 
- \epsilon^6 \frac{d^2 \delta_{1\Theta}^i}{d\tau^2}. 
\label{EffectOf3PNDtoEOMviaMVR}
\end{eqnarray}
The last term comes from the momentum-velocity relation 
Eq. (\ref{eq:3PMMomuntumVelocityRelation}) 
and  
compensates the $Q_{A\Theta}^i$ integral contribution 
also appearing through 
Eq. (\ref{eq:3PMMomuntumVelocityRelation}).

Note that this change of the acceleration does not affect 
the conservation of the binary orbital energy,  
\begin{eqnarray}
m_1 a_1^i|_{\delta_{A\Theta}} v_1^i + 
m_2 a_2^i|_{\delta_{A\Theta}}v_2^i  &=& 
\epsilon^6 \frac{d}{d\tau}\sum_{A=1,2}\left[
\delta_{A\Theta}^k\frac{dv_A^k}{d\tau} - 
v_A^k \frac{d \delta_{A\Theta}^k}{d\tau}
\right]. 
\label{eq5-45}
\end{eqnarray}

\section{The 3PN Gravitational Field: $N/B$ Integrals}
\label{TPNGravitationalField}

To derive a 3PN equation of motion, we have to have 
$\mbox{}_{10}h^{\tau\tau} + \mbox{}_8h^{k}\mbox{}_k$ besides  
$\mbox{}_8h^{\tau i}$  and the 2.5PN field.  
The 2.5PN field is well known \cite{BFP98}. 
(See paper II for $\mbox{}_6h^{ij}$.)
At 3PN order, while the body zone contribution 
$\mbox{}_{\le 10}h_B^{\tau\tau} + 
\mbox{}_{\le 8}h_B^{k}\mbox{}_k$ is easily found as Eqs.  
(\ref{hBttInAp}) and (\ref{hBijInAp}) with the help 
of the 3PN mass-energy relation, Eqs. (\ref{3PNEnergy}) and 
(\ref{Ptchi3PN}), it is quite difficult to 
evaluate the Poisson-type integrals over the $N/B$ region.
The problem is again to find superpotentials required in the 
derivation of the field. Thus    
we have not evaluated all the Poisson integrals and we have 
applied the method used in the evaluation of the 
$\mbox{}_8h^{\tau i}$ contribution to $dP_A^{\tau}/d\tau$.

We shall deal with the nonretarded field from the next subsection,  
\begin{eqnarray}
&&
4 \int_{N/B}\frac{d^3y}{|\vec x - \vec y|}
\left(
\mbox{}_{10}\Lambda_{N}^{\tau\tau}(\tau,\vec y) + 
\mbox{}_8\Lambda_{N}^{k}\mbox{}_k(\tau,\vec y)
\right).  
\label{3pnnon-retardedfield}
\end{eqnarray}
In the last subsection, we consider the 
second and the fourth time derivative terms in the 
retarded field.

\subsection{Superpotential part}
\label{ExplanationforSPPAgain}

Using the 2PN field, 
we first write down 
$\mbox{}_{10}\Lambda_{N}^{\tau\tau} + 
\mbox{}_8\Lambda_{N}^{k}\mbox{}_k$ 
explicitly, and simplify it to remove 
its S-dependence as much as possible. Then    
as in Sec. \ref{sec:derivatoinOfh8ti}, 
we  split the integrand 
into two groups: the S-independent group and the 
S-dependent group. 
For the S-independent group, we could find the 
superpotentials required other than those which have 
the following sources in the Poisson equations: 
\begin{eqnarray}
&&
\left\{
\frac{1}{r_1^6r_2^6},
\frac{1}{r_1^6r_2^4},
\frac{1}{r_1^6r_2^2},
\frac{r_2^3}{r_1^6},
\frac{1}{r_1^4r_2^4},
\frac{1}{r_1^4r_2^2},
\frac{r_2}{r_1^4},
\frac{1}{r_1^2r_2},
\frac{r_1^ir_1^j}{r_1^6 r_2},
\frac{r_1^ir_1^j}{r_1^5 r_2^2},
\frac{r_2^5r_1^ir_1^j}{r_1^5},
\frac{r_1^ir_1^j}{r_1^4 r_2^3},
\frac{r_1^ir_1^j}{r_1^3 r_2^4},
\frac{r_2^3r_1^ir_1^j}{r_1^3},
\frac{r_1^ir_1^j}{r_1^2 r_2^5},
\right.
\nonumber \\
&&\left.
\frac{r_2r_1^ir_1^j}{r_1},
\frac{r_1r_1^ir_1^j}{r_2},
\frac{r_1^3r_1^ir_1^j}{r_2^3},
\frac{r_1^5r_1^ir_1^j}{r_2^5},
\frac{r_1^ir_2^j}{r_1^6 r_2},
\frac{r_1^ir_2^j}{r_1^5 r_2^2},
\frac{r_2^5r_1^ir_2^j}{r_1^5},
\frac{r_1^ir_2^j}{r_1^4 r_2^3},
\frac{r_2^3r_1^ir_2^j}{r_1^3},
\frac{r_2r_1^ir_2^j}{r_1}
\right\}.
\label{h10ttSSPList}
\end{eqnarray}
It should be understood that there are Poisson equations with 
the same sources but 
with $(1 \leftrightarrow 2)$. 
We shall treat the integrand corresponding to 
the list (\ref{h10ttSSPList}), 
the superpotential-in-series part of 
$\mbox{}_{10}\Lambda_{N}^{\tau\tau} + 
\mbox{}_8\Lambda_{N}^{k}\mbox{}_k$, 
in the next subsection.  

For the remaining integrand of 
$\mbox{}_{10}\Lambda_{N}^{\tau\tau} + 
\mbox{}_8\Lambda_{N}^{k}\mbox{}_k$ 
(the superpotential part of 
$\mbox{}_{10}\Lambda_{N}^{\tau\tau} + 
\mbox{}_8\Lambda_{N}^{k}\mbox{}_k$), 
we need to find the superpotentials whose sources are   
\begin{eqnarray}
&&
\left\{
\frac{1}{r_1^6},\frac{1}{r_1^3},
\frac{1}{r_1^6r_2^3},
\frac{1}{r_1^6r_2},
\frac{r_2}{r_1^6},
\frac{r_2^2}{r_1^6},
\frac{r_2^4}{r_1^6},
\frac{r_2^6}{r_1^6},
\frac{1}{r_1^5r_2^5},\frac{1}{r_1^5r_2^3},
\frac{r_2^2}{r_1^5},
\right. 
\nonumber \\
&&
\left. 
\frac{r_1^ir_1^j}{r_1^7r_2^4},
\frac{r_1^ir_1^j}{r_1^7r_2^2},
\frac{r_1^ir_1^j}{r_1^7r_2},
\frac{r_2r_1^ir_1^j}{r_1^7},
\frac{r_2^3r_1^ir_1^j}{r_1^7},
\frac{r_1^ir_1^j}{r_1^6r_2^5},
\frac{r_1^ir_1^j}{r_1^5r_2^4},
\frac{r_1^ir_1^j}{r_1^5r_2},
\frac{r_2r_1^ir_1^j}{r_1^5},
\frac{r_2^3r_1^ir_1^j}{r_1^5},
\frac{r_1^ir_1^j}{r_1^4r_2^5},
\frac{r_1^ir_1^j}{r_1^3r_2^5},
\frac{r_1^ir_1^j}{r_1r_2^5},
\frac{r_1^ir_1^j}{r_1r_2^3},
\right. 
\nonumber \\
&&
\left. 
\frac{r_1^ir_1^j}{r_2^4},
\frac{r_1r_1^ir_1^j}{r_2^5},
\frac{r_1r_1^ir_1^j}{r_2^3},
\frac{r_1^2r_1^ir_1^j}{r_2^2},
\frac{r_1^3r_1^ir_1^j}{r_2^5},
\frac{r_1^ir_2^j}{r_1^6r_2^3},
\frac{r_1^ir_2^j}{r_1^6},
\frac{r_2^2r_1^ir_2^j}{r_1^6},
\frac{r_2^4r_1^ir_2^j}{r_1^6},
\frac{r_1^ir_2^j}{r_1^5r_2^4},
\frac{r_1^ir_2^j}{r_1^5r_2^3},
\frac{r_1^ir_2^j}{r_1^5r_2},
\frac{r_2r_1^ir_2^j}{r_1^5},
\right.
\nonumber \\
&&
\left. 
\frac{r_2^3r_1^ir_2^j}{r_1^5},
\frac{r_1^ir_2^j}{r_1^4},
\frac{r_2^2r_1^ir_2^j}{r_1^4},
\frac{r_2^4r_1^ir_2^j}{r_1^4},
\frac{r_1^ir_2^j}{r_1^3r_2},
\frac{r_2r_1^ir_2^j}{r_1^3},
\frac{r_1^ir_2^j}{r_1^2},
\frac{r_2^2r_1^ir_2^j}{r_1^2},
r_1^ir_2^j,
\right.
\nonumber \\
&&
\left. 
\frac{r_1^ir_1^jr_1^kr_1^l}{r_1^7r_2^3},
\frac{r_1^ir_1^jr_1^kr_1^l}{r_1^7r_2},
\frac{r_1^ir_1^jr_1^kr_1^l}{r_1^5r_2^3},
\frac{r_1^ir_1^jr_2^kr_2^l}{r_1^5r_2^5},
\frac{r_1^ir_1^jr_2^kr_2^l}{r_1^5r_2^3}
\right\}.
\label{h10ttSPList}
\end{eqnarray}
Here we did not list the sources for which we already 
find the superpotentials 
(see the list (\ref{h8tiSPSourceList})).

We could derive 
the superpotentials corresponding to the list 
(\ref{h10ttSPList}) using the procedure described 
in Sec. \ref{ExplanationforSPP}. 
Useful particular solutions are given in 
\cite{BFP98,JS98,BF01a} and in Appendix \ref{splist}. 
Those superpotentials enable us to calculate the Poisson  
integral with the superpotential part as the integrand. 
We cannot write the result down here because of its 
enormous length.

\subsection{Superpotential-in-series part}
\label{ExplanationforSSPPAgain}

The superpotentials having the sources listed in 
(\ref{h10ttSSPList}) could not be found. Thus we 
employed the method explained in Sec. 
\ref{ExplanationforSSPP}. First we transform tensorial sources into
scalars.  For example, 
\begin{eqnarray}
\frac{r_1^5r_1^ir_1^j}{r_2^5} &=& \frac{1}{3}
\frac{\pa^2}{\pa z_1^iz_1^j}\frac{r_1^5}{r_2}
+ \Delta \left[
\frac{1}{63}\frac{\pa^2}{\pa z_1^iz_1^j}
\left(-\frac{r_1^9}{21r_2^3}+\frac{9}{7}r_{12}^2f^{(7,-5)}
- \frac{3r_1^7}{7r_2}+ 3r_{12}^2f^{(5,-3)}
\right) - \frac{\delta^{ij}}{7}f^{(7,-5)}
\right].\nonumber 
\end{eqnarray}
We apply  
Eqs. (\ref{eq5-5}), (\ref{eq5-6}), and (\ref{eq5-7}) to 
the sources in (\ref{h10ttSSPList}). 
The (scalar)sources derived by making the sources in the 
list (\ref{h10ttSSPList}) scalars  
to which we apply Eq. (\ref{eq5-7}) are; 
\begin{eqnarray}
&&
\left\{
\frac{1}{r_1^6r_2^6},
\frac{1}{r_1^6r_2^4},
\frac{1}{r_1^6r_2^2},
\frac{r_2^3}{r_1^6},
\frac{1}{r_1^4r_2^4},
\frac{1}{r_1^4r_2^2},
\frac{r_2}{r_1^4},
\frac{1}{r_1^2r_2},
\frac{r_1^5}{r_2},
\frac{r_1}{r_2^4},
\frac{r_1^2\ln r_1}{r_2^5},
\frac{1}{r_1^2 r_2},
\frac{\ln r_1}{r_2^3}
\right\},
\label{h10ttSSPListRefined}
\end{eqnarray}
and the same functions with their labels 1 and 2 exchanged. 
For these sources, we could evaluate the Poisson integrals in 
a similar sense to Eq. (\ref{seriesedpotExample}), and 
as a result we obtain 
in the neighborhood of the star 1   
the field  
corresponding to the superpotential-in-series part.

\subsection{Direct-integration part}
\label{ExplanationforDIPAgain}

We now consider the DIP field contribution to an  
equation of motion. At 3PN order, the DIP field 
appears in the integrands of the surface 
integrals of the general form of the 3PN equation of motion as  
(see Eqs. (\ref{tLLti10}) and (\ref{tLLij10}))
\begin{eqnarray}
\lefteqn{
\mbox{}_{10}[16\pi(-g)t_{LL}^{\tau i}]_{{\rm DIP}} = 
2\mbox{}_4h^{\tau\tau}\mbox{}_{,k}
\mbox{}_8h^{\tau[k,i]}_{{\rm DIP}},}  
\\
\lefteqn{
\mbox{}_{10}[16\pi(-g)t_{LL}^{ij}]_{{\rm DIP}}
} \nonumber \\
&=& 
\frac{1}{4}
(\delta^i\mbox{}_k\delta^j\mbox{}_l 
+ \delta^i\mbox{}_l\delta^j\mbox{}_k - 
\delta^{ij}\delta_{kl})
\left\{
\mbox{}_4h^{\tau\tau,k}
(\mbox{}_{10}h^{\tau\tau,l}_{{\rm DIP}} 
+ \mbox{}_8h^m_{{\rm DIP}}\mbox{}_{m}\mbox{}^{,l} 
+ 4\mbox{}_8h^{\tau l}_{{\rm DIP}} \mbox{}_{,\tau}
) +
8 \mbox{}_4h^{\tau}\mbox{}_m\mbox{}^{,k}\mbox{}_8h^{\tau [l,m]}_{{\rm DIP}} 
\right\} 
\nonumber \\
\mbox{} &+&  2\mbox{}_4h^{\tau i}\mbox{}_{,k}
\mbox{}_8h^{\tau [k,j]}_{{\rm DIP}} 
+ 2\mbox{}_4h^{\tau j}\mbox{}_{,k}
\mbox{}_8h^{\tau [k,i]}_{{\rm DIP}}.  
\end{eqnarray}
Here we added the $\mbox{}_8h^{\tau i}_{{\rm DIP}}$ contribution. (Note 
that in Sec. \ref{ExplanationforCompStarInt}, we 
evaluated the $\mbox{}_8h^{\tau i}_{{\rm DIP}}$ contribution to 
the evolution equation for $P_{A\Theta}^{\tau}$, but not 
to an equation of motion.) 
Then the DIP field contribution to a 3PN acceleration denoted 
by $a_{1{\rm DIP}}^i$ becomes 
\begin{eqnarray}
\lefteqn{m_1 a_{1{\rm DIP}}^i}  \nonumber \\  
&=&  
\frac{m_1}{4\pi}\oint_{\pa B_1}d\Omega\left[
\int_{N/B}\frac{d^3y}{|\vec x - \vec y|}
\left(
\mbox{}_{10}\Lambda_{S}^{\tau\tau,i} 
+\mbox{}_8\Lambda_{S}^k\mbox{}_{k}\mbox{}^{,i}
+ 4\mbox{}_8\Lambda_{S}^{\tau i}\mbox{}_{,\tau}
+ 8v_1^k\mbox{}_8\Lambda_{S}^{\tau [i,k]}
\right)
\right]
\nonumber \\
&+&
\frac{m_2}{4\pi}(
\delta^i\mbox{}_k\delta^j\mbox{}_l
+\delta^i\mbox{}_l\delta^j\mbox{}_k
-\delta^{ij}\delta_{kl})
\oint_{\pa B_1}dS_j\frac{r_2^k}{r_2^3}
\nonumber \\
&&\times
\int_{N/B}\frac{d^3y}{|\vec x - \vec y|}
\left(
\mbox{}_{10}\Lambda_{S}^{\tau\tau,l} 
+\mbox{}_8\Lambda_{S}^m\mbox{}_{m}\mbox{}^{,l}
+ 4\mbox{}_8\Lambda_{S}^{\tau l}\mbox{}_{,\tau}
+ 8v_2^m\mbox{}_8\Lambda_{S}^{\tau [l,m]}
\right)
\nonumber \\
&-& \frac{2 m_2}{\pi}V^k
\oint_{\pa B_1}dS_k\frac{r_2^l}{r_2^3}
\int_{N/B}
\frac{d^3y \mbox{}_8\Lambda_S^{\tau[l,i]}}{|\vec x - \vec y|}
+\frac{2 m_2}{\pi}v_2^i
\oint_{\pa B_1}dS_j\frac{r_2^k}{r_2^3}
\int_{N/B}
\frac{d^3y\mbox{}_8\Lambda_S^{\tau[k,j]}}{|\vec x - \vec y|}
\nonumber \\
&-& \frac{m_1}{4\pi}\oint_{\pa B_1}d\Omega\left[
\oint_{\pa(N/B)}
\frac{dS^i}{|\vec x - \vec y|}
(\mbox{}_{10}\Lambda^{\tau\tau}_{S}+ \mbox{}_8\Lambda^k_{S}\mbox{}_k)
+ 4 \sum_{A=1,2}v_A^k
\oint_{\pa B_A}
\frac{dS_k\mbox{}_8\Lambda_{S}^{\tau i}}{|\vec x - \vec y|}
\right. 
\nonumber \\
\mbox{} &&+ \left. 
 8 v_1^k
\oint_{\pa (N/B)}
\frac{dS_{[k}\mbox{}_8\Lambda_S^{i]\tau}}{|\vec x - \vec y|}
\right]
\nonumber \\
&-& \frac{m_2}{4\pi}(
\delta^i\mbox{}_k\delta^j\mbox{}_l
+\delta^i\mbox{}_l\delta^j\mbox{}_k
-\delta^{ij}\delta_{kl})
\oint_{\pa B_1}dS_j\frac{r_2^k}{r_2^3}
\nonumber \\
&&\times
\left[
\oint_{\pa(N/B)}\frac{dS_l}{|\vec x - \vec y|}
\left(
\mbox{}_{10}\Lambda_{S}^{\tau\tau} 
+\mbox{}_8\Lambda_{S}^m\mbox{}_{m}
\right)
+
4 \sum_{A=1,2}v_A^m
\oint_{\pa B_A}\frac{dS_m\mbox{}_8\Lambda_{S}^{\tau l}}
{|\vec x - \vec y|}
+ 
8 v_2^m 
\oint_{\pa(N/B)}
\frac{dS_{[m}\mbox{}_8\Lambda_{S}^{l]\tau}}{|\vec x - \vec y|}
\right]
\nonumber \\
&+&
\frac{2 m_2}{\pi}V^k
\oint_{\pa B_1}dS_k\frac{r_2^l}{r_2^3}
\oint_{\pa (N/B)}
\frac{dS^{[i} \mbox{}_8\Lambda_S^{l]\tau}}{|\vec x - \vec y|}
-\frac{2 m_2}{\pi}v_2^i
\oint_{\pa B_1}dS_j\frac{r_2^k}{r_2^3}
\oint_{\pa (N/B)}
\frac{dS_{[j} \mbox{}_8\Lambda_S^{k]\tau}}{|\vec x - \vec y|}, 
\label{DIPIntCont}
\end{eqnarray}
where we used a relation 
$$
\frac{d}{d\tau}\int_{N/B}d^3y
\frac{\mbox{}_8\Lambda_{S}^{\tau i}(\tau,\vec y)}{|\vec x - \vec y|}
= \int_{N/B}
\frac{d^3y}{|\vec x - \vec y|}
\frac{d}{d\tau}
\mbox{}_8\Lambda_{S}^{\tau i}(\tau,\vec y)
- \sum_{A=1,2}v_A^k\oint_{\pa B_A}dS_k
\frac{\mbox{}_8\Lambda_{S}^{\tau i}(\tau,\vec y)}{|\vec x - \vec y|}.
$$
All the integrals in Eq. (\ref{DIPIntCont})   
other than the first four terms can be easily 
evaluated. We now explain how to evaluate the first four 
integrals.

\subsubsection{Main star integral}

For the first integral in Eq. (\ref{DIPIntCont}),  
we change the integration variable $\vec y$ to $\vec y_1$ and also 
change the integration region $N$ to $N_1$ 
using 
Eq. (\ref{eq5-6}). The surface integrals over $\pa N$ can be 
easily evaluated. 
Note that $\vec y_2$ in the integrands must be replaced by 
$\vec r_{12} + \vec y_1$. 
For the remaining volume integral, we  
use computationally the same method as the 
one employed by Blanchet and Faye in  
\cite{BF01a,BF00b}.  
Let us consider the following integral, 
which we call the {\it main star integral}:  
\begin{eqnarray}
&&
\frac{1}{4\pi}\oint_{\pa B_1}d\Omega 
\mathop{{\rm disc}}_{\epsilon R_A}
\int_{N_1/B}d^3y_1\frac{f(\vec y_1)}{4 \pi |\vec r_1 - \vec y_1|}. 
\label{eq:mainstarintegral}
\end{eqnarray}
(For the definition of $\mathop{{\rm disc}}_{\epsilon R_A}$, 
see Sec. \ref{ExplanationforCompStarInt}.)
With the notice given below Eq. (\ref{eq513}) in mind,   
we first exchange the order of integration,  
\begin{eqnarray}
\lefteqn{
\frac{1}{4\pi}\oint_{\pa B_1}d\Omega 
\mathop{{\rm disc}}_{\epsilon R_A}
\int_{N_1/B}d^3y_1\frac{f(\vec y_1)}{|\vec r_1 - \vec y_1|}
}\nonumber \\ 
&&=
 \lim_{r_1' \rightarrow \epsilon R_1}
\mathop{{\rm disc}}_{\epsilon R_A}\left[
\int_{N_1/B'}\frac{d^3y_1}{y_1} f(\vec y_1)
+ \int_{B_1'/B_1}\frac{d^3y_1}{r_1'} f(\vec y_1)\right], 
\label{MainStarIntegralEq1}
\end{eqnarray}
where we used  
\begin{eqnarray}
&&
\oint_{\pa B_1}\frac{d\Omega}{|\vec r_1 - \vec y_1|} =\left\{
\begin{array}{ll}
\frac{4\pi}{r_1} & {\rm for~} r_1 \ge y_1, \\
\frac{4\pi}{y_1} & {\rm for~} r_1 < y_1.
\end{array}
\right.
\end{eqnarray}

Next we make a symmetric trace-free decomposition (STF decomposition) 
of the integrand on the indexes of $\vec n_1$,  
\begin{eqnarray}  
&&
f(\vec y_1) = \sum_{l=0} g_{l}(\cos\theta,y_1)n_1^{<I_l>}, 
\end{eqnarray}
where $\cos \theta = - \vec n_{12}\cdot \vec n_1$. 
Note that $g_{l}(\cos\theta,y_1)$ is not necessarily a scalar. 
In general, $g_{l}(\cos\theta,y_1)$ is a tensor whose indexes 
are carried by $\vec v_A$, $\vec r_{12}$, or some 
combinations of them. 
Then substituting back the STF-decomposed integrand into 
Eq. (\ref{MainStarIntegralEq1}), we have 
\begin{eqnarray}
\lefteqn{ 
\frac{1}{4\pi}\oint_{\pa B_1}d\Omega 
\mathop{{\rm disc}}_{\epsilon R_A}
\int_{N_1/B}d^3y_1\frac{f(\vec y_1)}{4 \pi |\vec r_1 - \vec y_1|}
}
\nonumber \\ 
&=&
\frac{1}{2}\sum_{l=0}n_{12}^{<I_l>} 
\lim_{r_1' \rightarrow \epsilon R_1}
\mathop{{\rm disc}}_{\epsilon R_A}\left[
\int_{N_1/B'}dy_1y_1
\int dt {\rm P}_{l}(-t)
g_{l}(t,y_1) 
\right.
\nonumber \\   
\mbox{}&&~~~~~~~~~~~~~~~~ + \left. 
 \int_{\epsilon R_1}^{r_1'}
\frac{dy_1y_1^2}{r_1'} 
\oint_{-1}^{1}dt {\rm P}_{l}(-t)
g_{l}(t,y_1)
\right], 
\label{MainStarIntegralEq2}
\end{eqnarray}
where (below, $\vec n$ and $\vec N$ are unit vectors)
\begin{eqnarray}
\int \frac{d\Omega_{{\bf n}}}{4\pi} n^{<I_l>}f(\vec N\cdot\vec n) 
&=& N^{<I_l>}\int
\frac{d\Omega_{{\bf n}}}{4\pi} 
{\rm P}_l(\vec N\cdot\vec n)
f(\vec N\cdot\vec n)   
\end{eqnarray}
was used. 
Notice that in Eq. (\ref{MainStarIntegralEq2}), 
when $y_1\in[r_{12}-\epsilon R_2,r_{12}+\epsilon R_2]$
the body zone 2 prevents  
the angular integration region from being complete. 
The angular deficit is given by Eq. 
(\ref{eq:angulardeficit}).

Now let us give an example. Take $\pa_{z_1^i}(\ln S) \pa_j (1/r_1)$ 
as an integrand,   
\begin{eqnarray}
\lefteqn{
\oint_{\pa B_1}\frac{d\Omega}{4\pi}  
\mathop{{\rm disc}}_{\epsilon R_A}
\int_{N_1/B}\frac{d^3y_1}{4 \pi |\vec r_1 - \vec y_1|}\pa_{z_1^i}\ln S 
\pa_j \frac{1}{y_1}
}
\nonumber \\
&=& 
\oint_{\pa B_1}\frac{d\Omega}{4\pi} 
\mathop{{\rm disc}}_{\epsilon R_A}
\int_{N_1/B}\frac{d^3y_1}{4 \pi|\vec r_1 - \vec y_1|}
\left(
\frac{n_1^in_1^j}{y_1^2S} - \frac{n_{12}^in_1^j}{y_1^2S}
\right)
\nonumber \\
&=& 
\frac{1}{2}n_{12}^{<ij>} 
\lim_{r_1' \rightarrow \epsilon R_1}
\mathop{{\rm disc}}_{\epsilon R_A}\left[
\int_{N_1/B'}\frac{dy_1}{y_1}
\int  \frac{dt {\rm P}_{2}(-t)}
{r_{12} + y_1 + \sqrt{r_{12}^2 + y_1^2 - 2 y_1r_{12}t}}
\right.
\nonumber \\
\mbox{} &&~~~~~~~~~~~~~~~~~~~~~~~~~~~~~~~~~~~~~~~~~+\left.
 \int_{\epsilon R_1}^{r_1'}
\frac{dy_1}{r_1'} 
\oint_{-1}^{1} \frac{dt {\rm P}_{2}(-t)}
{r_{12} + y_1 + \sqrt{r_{12}^2 + y_1^2 - 2 y_1r_{12}t}}
\right]
\nonumber \\
\mbox{} &&- 
\frac{1}{2}n_{12}^{i}n_{12}^j 
\lim_{r_1' \rightarrow \epsilon R_1}
\mathop{{\rm disc}}_{\epsilon R_A}\left[
\int_{N_1/B'}\frac{dy_1}{y_1}
\int  \frac{dt {\rm P}_{1}(-t)}
{r_{12} + y_1 + \sqrt{r_{12}^2 + y_1^2 - 2 y_1r_{12}t}}
\right.
\nonumber \\
\mbox{} &&~~~~~~~~~~~~~~~~~~~~~~~~~~~~~~~~~~~~~~~~+\left.
 \int_{\epsilon R_1}^{r_1'}
\frac{dy_1}{r_1'} 
\oint_{-1}^{1} \frac{dt {\rm P}_{1}(-t)}
{r_{12} + y_1 + \sqrt{r_{12}^2 + y_1^2 - 2 y_1r_{12}t}}
\right]  
\nonumber \\
\mbox{} &&+ 
\frac{1}{6}\delta^{ij} 
\lim_{r_1' \rightarrow \epsilon R_1}
\mathop{{\rm disc}}_{\epsilon R_A}\left[
\int_{N_1/B'}\frac{dy_1}{y_1}
\int  \frac{dt}
{r_{12} + y_1 + \sqrt{r_{12}^2 + y_1^2 - 2 y_1r_{12}t}}
\right.
\nonumber \\
\mbox{} &&~~~~~~~~~~~~~~~~~~~~~~~~~~~~~~~~+\left.
 \int_{\epsilon R_1}^{r_1'}
\frac{dy_1}{r_1'} 
\oint_{-1}^{1} \frac{dt}
{r_{12} + y_1 + \sqrt{r_{12}^2 + y_1^2 - 2 y_1r_{12}t}}
\right].   
\end{eqnarray}
Here 
\begin{eqnarray}
\lefteqn{
\int_{N_1/B'}\frac{dy_1}{y_1}
\int  \frac{dt {\rm P}_{l}(-t)}
{r_{12} + y_1 + \sqrt{r_{12}^2 + y_1^2 - 2 y_1r_{12}t}} 
}\nonumber \\ 
&=& 
\left\{
\begin{array}{ll}
\frac{5}{3r_{12}}-\frac{\pi^2}{6r_{12}} 
+O(\epsilon R1,r_1')  & {\rm for~} l= 2,\\
\frac{3}{2r_{12}}-\frac{\pi^2}{6r_{12}}
+O(\epsilon R1,r_1')  & {\rm for~} l= 1,\\
\frac{2}{r_{12}}-\frac{\pi^2}{6r_{12}} + 
\frac{1}{r_{12}}\ln\left(\frac{r_{12}}{r_1'}\right)
+O(\epsilon R1,r_1')  & {\rm for~} l= 0,
\end{array} 
\right. 
\end{eqnarray}
and 
\begin{eqnarray}
 \int_{\epsilon R_1}^{r_1'}
\frac{dy_1}{r_1'} 
\oint_{-1}^{1} \frac{dt {\rm P}_{l}(-t)}
{r_{12} + y_1 + \sqrt{r_{12}^2 + y_1^2 - 2 y_1r_{12}t}}
&=& 
\left\{
\begin{array}{ll}
O(\epsilon R_1,r_1')  & l= 2,\\
O(\epsilon R_1,r_1')  & l= 1,\\
\frac{1}{r_{12}}+O(\epsilon R1,r_1')   
& l= 0. 
\end{array} 
\right. 
\end{eqnarray}
Thus we obtain 
\begin{eqnarray}
\oint_{\pa B_1}\frac{d\Omega}{4\pi}  
\mathop{{\rm disc}}_{\epsilon R_A}
\int_{N_1/B}\frac{d^3y_1}{4 \pi |\vec r_1 - \vec y_1|}\pa_{z_1^i}\ln S 
\pa_j \frac{1}{y_1}
&=& \frac{2\delta^{ij}}{9r_{12}} + 
\frac{n_{12}^in_{12}^j}{12r_{12}} +
\frac{\delta^{ij}}{6r_{12}}
\ln \left(\frac{r_{12}}{\epsilon R_1}\right). 
\end{eqnarray}
The integrand $\pa_{z_1^i}(\ln S) \pa_j (1/r_1)$ 
was taken in \cite{BF01a} as an example of 
 Hadamard's partie finie of Poisson integrals. 
The results are the same. 
A difference in the two results 
here is simply in definitions of the lower 
bounds of the integration. Here in our case we set 
$\epsilon R_A$ as the lower bounds.

Now we return to the evaluation of the 3PN gravitational 
field. Applying Eq. (\ref{MainStarIntegralEq2}) to the 
first integral in Eq. (\ref{DIPIntCont}), where 
$$f(\vec y_1) = 
\left. 
\left[
\frac{\pa}{\pa y^i} 
\mbox{}_{10}\Lambda_{S}^{\tau\tau}(\tau,\vec y) 
+\frac{\pa}{\pa y^i} 
\mbox{}_8\Lambda_{S}^k\mbox{}_{k}(\tau,\vec y) 
+ 4\frac{d}{d \tau} 
\mbox{}_8\Lambda_{S}^{\tau i}(\tau,\vec y) 
+ 
8v_1^k\frac{\pa}{\pa y^{[k}} 
\mbox{}_8\Lambda_{S}^{i]\tau }(\tau,\vec y)
\right]
\right|_{\vec y = \vec y_1 + \vec z_1},
$$
we could evaluate the main star integral contribution to a  
3PN acceleration.

\subsubsection{Companion star integral}

The second, the third, and the fourth terms in Eq. (\ref{DIPIntCont}) 
have a form of a companion star integral and thus can be evaluated 
by the method described in Sec. \ref{ExplanationforCompStarInt}.   
The integrands have no logarithmic dependence and thus admit 
an expansion in the form of Eq. (\ref{eq5-23}), where we found 
$p_0=5$. Then using Eq. (\ref{CompanionStarIntegral}), we obtain 
the companion star integral contribution,  
\begin{eqnarray} 
\lefteqn{
\frac{m_2}{4\pi}(
\delta^i\mbox{}_k\delta^j\mbox{}_l
+\delta^i\mbox{}_l\delta^j\mbox{}_k
-\delta^{ij}\delta_{kl})}
\nonumber \\
&&\times
\oint_{\pa B_1}dS_j\frac{r_2^k}{r_2^3}
\int_{N/B}\frac{d^3y}{|\vec x - \vec y|}
\left(
\mbox{}_{10}\Lambda_{S}^{\tau\tau,l} 
+\mbox{}_8\Lambda_{S}^m\mbox{}_{m}\mbox{}^{,l}
+ 4\mbox{}_8\Lambda_{S}^{\tau l}\mbox{}_{,\tau}
+ 8v_2^m\mbox{}_8\Lambda_{S}^{\tau [l,m]}
\right)
\nonumber \\
&-& \frac{2 m_2}{\pi}V^k
\oint_{\pa B_1}dS_k\frac{r_2^l}{r_2^3}
\int_{N/B}d^3y
\frac{\mbox{}_8\Lambda_S^{\tau[l,i]}}{|\vec x - \vec y|}
+\frac{2 m_2}{\pi}v_2^i
\oint_{\pa B_1}dS_j\frac{r_2^k}{r_2^3}
\int_{N/B}d^3y
\frac{\mbox{}_8\Lambda_S^{\tau[k,j]}}{|\vec x - \vec y|}
\nonumber \\ 
\mbox{} &=&
\frac{m_1^2m_2^2}{r_{12}^5}r_{12}^i 
\left[
\frac{46v_1^2}{9} + \frac{16v_2^2}{3}-
16(\vec n_{12}\cdot\vec v_1)^2
-92(\vec v_{1}\cdot\vec v_2)
\right]. 
\end{eqnarray}

\subsection{Retarded field}

At 3PN order, we have to evaluate 
the integral over $N/B$ in the  
second retardation expansion term,  
\begin{eqnarray*}
&&
2 \frac{\pa^2}{\pa \tau^2}\int_{N/B}d^3y|\vec x - \vec y|
\left(
\mbox{}_8\Lambda_N^{\tau\tau} + 
\mbox{}_6\Lambda_N^{k}\mbox{}_k  
\right). 
\end{eqnarray*}
This integral can be evaluated via the super-superpotential    
$f(\vec y)$ satisfying $\mbox{}_8\Lambda_N^{\tau\tau} + 
\mbox{}_6\Lambda_N^{k}\mbox{}_k  = \Delta^2 f(\vec y)$ as  
\begin{eqnarray*}
\lefteqn{\int_{N/B}d^3y|\vec x - \vec y| \Delta^2 f(\vec y)}
\nonumber \\
&=& - 8\pi f(\vec x) 
\nonumber \\
\mbox{} &&+ 
\oint_{\pa (N/B)}dS_k\left[
|\vec x - \vec y|\pa_k\Delta f(\vec y) 
- \frac{y^k-x^k}{|\vec x - \vec y|}\Delta f(\vec y) 
+ \frac{2}{|\vec x - \vec y|}\pa_k f(\vec y)  
+ \frac{2(y^k-x^k)}{|\vec x - \vec y|^3} f(\vec y) 
\right].  
\end{eqnarray*}
The superpotential of $\mbox{}_8\Lambda_N^{\tau\tau} + 
\mbox{}_6\Lambda_N^{k}\mbox{}_k$ is easily found. (In the 
derivation of $\mbox{}_8h^{\tau\tau}$ and 
$\mbox{}_6h^{ij}$ we found the superpotentials 
required here.) The super-superpotentials 
we could not find are only those with $1/(r_1^2r_2), 1/(r_1r_2^2), 
r_1/r_2^2$, and $r_2/r_1^2$ as sources. The corresponding 
integrands are  
\begin{eqnarray*}
&&
\frac{\pa^2}{\pa \tau^2}\int_{N/B}
\frac{d^3y}{4\pi|\vec x - \vec y|}
\left(
-\frac{21m_1^2m_2}{y_1^2y_2} 
- \frac{15m_1^2m_2y_2}{2r_{12}^2y_1^2}
-\frac{21m_1m_2^2}{y_1y_2^2} 
- \frac{15m_1m_2^2y_1}{2r_{12}^2y_2^2}
\right). 
\end{eqnarray*}
Thus we use the method explained in Sec. \ref{ExplanationforSSPP} 
and obtain the field near the star 1, which is  
sufficient to derive the equation of motion.

The $N/B$ integral appearing in the fourth 
retardation expansion term can be evaluated via 
the following super-super-superpotentials:   
$$
\frac{1}{r_1^4} = \Delta \frac{1}{2r_1^2} = 
\Delta^2\frac{1}{2}\ln r_1 = \Delta^3 \frac{1}{2}f^{(\ln,0)},
$$
$$
\frac{\vec r_1\cdot\vec r_2}{r_1^3r_2^3} = \Delta\frac{1}{2r_1r_2}
= \Delta^2\frac{1}{2}\ln S = \Delta^3\frac{1}{2}f^{(\ln S)},
$$
and a formula 
\begin{eqnarray*}
\lefteqn{\int_{N/B}d^3y|\vec x - \vec y|^3\Delta^3f(\vec y)} 
\nonumber \\ 
&=& - 96\pi f(\vec x) + 
\oint_{\pa (N/B)}dS_k\left[
|\vec x - \vec y|^3\pa_k\Delta^2f(\vec y) 
-3 |\vec x - \vec y|(y^k-x^k)\Delta^2f(\vec y)
\right.
\nonumber \\
\mbox{} &&+\left.
12|\vec x - \vec y|\pa_k\Delta f(\vec y)  
-12\frac{y^k-x^k}{|\vec x - \vec y|}\Delta f(\vec y)
+\frac{24}{|\vec x - \vec y|}\pa_kf(\vec y)
+  \frac{24(y^k-x^k)}{|\vec x - \vec y|^3} f(\vec y)
\right].  
\end{eqnarray*}
It is straightforward to evaluate the surface integrals.

\section{The 3PN Equation of Motion with Logarithmic Terms}
\label{sec:3PNEOMwithLog}

To obtain a 3PN equation of motion, 
we evaluate the surface integrals in the general form of the 
3PN equation of motion Eq. (\ref{generaleom3PN}) 
using the field $\mbox{}_8h^{\tau\tau}, 
\mbox{}_{\le 6}h^{\mu\nu}$, 
the 3PN body zone contributions,  
and the 3PN $N/B$ contributions corresponding to  
the superpotential part and the superpotential-in-series part.  
We then combine the result with the contribution from the 
direct-integration part. For a computational check, 
we have used the 
direct-integration method (the method with which we evaluate 
the direct-integration part) to  evaluate the 
contributions to 
the equation of motion from 
all of the 3PN $N/B$ nonretarded field,  
$\mbox{}_8h_{N/B n=0}^{\tau i}$ and 
$\mbox{}_{10}h_{N/B n=0}^{\tau\tau} 
+ \mbox{}_8h_{N/B n=0}^{l}\mbox{}_l$,   
by assuming that they had belonged completely to a direct-integration part.  
As expected, we obtained the same result from two methods: the  
direct-integration method and the direct-integration method plus 
the superpotential method plus the superpotential-in-series method.

In the evaluation of the body zone field $h_B^{\mu\nu}$ 
(shown as 
Eqs. (\ref{hBttInAp}),  (\ref{hBtiInAp}), and (\ref{hBijInAp})), 
besides the explicitly seen energy monopole terms 
in these fields which 
must be converted into mass monopole terms via 
the 3PN mass-energy relation,   
the effects of the $Q_A^{K_li}$ and $R_A^{K_lij}$ integrals 
appearing in the 3PN field 
are properly 
taken into account through Eq. (\ref{QREffecttoEOMinAp}). 
$\mbox{}_6Q_{A\Theta}^i$ given in Eq. 
(\ref{eq:Q1ThetaIntegralOfOrdere6}) affects a 3PN equation 
of motion through the 3PN momentum-velocity relation. 
Since we 
define the representative points of the
 stars via Eq. (\ref{3PNDefOfCM}), 
we add the corresponding acceleration given by Eq. 
(\ref{EffectOf3PNDtoEOMviaMVR}).  Furthermore, our choice 
of the representative points of the stars makes $D_{A\chi}^i$ appear 
independently of $D_{A\Theta}^i$ 
in the field, and hence 
$\mbox{}_4D_{A\chi}^i$ affects the 3PN equation of motion. 
In summary, 
$\mbox{}_{\le 4}Q_{A}^{K_li}, \mbox{}_{\le 4}R_{A}^{K_lij},  
\mbox{}_{6}Q_{A\Theta}^i,  
\delta_{A\Theta}^i$, and $\mbox{}_4D_{A\chi}^i$
contributions to the 3PN field can be written as 
\begin{eqnarray}
\mbox{}_{10}h^{\tau\tau} +   
\mbox{}_8h^{k}\mbox{}_{k} &=& 
4\sum_{A=1,2}
\frac{r^k_A}{r_A^3}\left(
\delta_{A\Theta}^k + \mbox{}_4D_{A\chi}^k +  
\mbox{}_4R_A^{kll} 
-\frac{1}{2}\mbox{}_4R_A^{llk}
 \right) + \cdots, 
\end{eqnarray}
where $\cdots$ are other contributions. 
The effect of this field on the equation of motion 
is properly taken into 
account via Eq. (\ref{QREffecttoEOMinAp}) with 
 $C_{QR}$ replaced by 
 $C_{QR} + C_{D_{\chi}} + C_{\delta_{A\Theta}}$, where 
$C_{\delta_{A\Theta}} = 2/3$ and 
$C_{D_{\chi}} = -350/9$. 
(See Eq. (\ref{3PNDefOfCM}) for  $C_{\delta_{A\Theta}}$ 
and 
Eq. (\ref{eq:Dichi3PNinApChi}) for $C_{D_{\chi}}$.) 
On the other hand, $\mbox{}_{6}Q_{A\Theta}^i$ and $\delta_{A\Theta}^i$ 
affect the equation of motion through the momentum-velocity relation  
in Eq. (\ref{generaleom3PN}),  
\begin{eqnarray}
m_1 \left(\frac{d v_1^i}{d\tau}\right)_{\le 3{\rm PN}} 
&=& - \epsilon^6 \frac{d \mbox{}_6 Q_{1\Theta}^i}{d\tau} 
- \epsilon^6 \frac{d^2 \delta_{1\Theta}^i}{d\tau^2} + \cdots,     
\end{eqnarray}
but cancel each other out, since we chose Eq. (\ref{3PNDefOfCM}). 
We note that 
there is no need to take into account an effect of 
$D_{A\chi}^i$ through the momentum-velocity 
relation of the $\chi$ part on the equation of motion 
since we evaluate the general form of the 3PN 
equation of motion Eq. (\ref{generaleom3PN}) to 
derive a 3PN equation of motion.
Then these 
contributions to a 3PN acceleration can be summarized into 
\begin{eqnarray}
\lefteqn{
({\rm the~contribution~to~} m_1 a_1^i {\rm  
~from~~} \mbox{}_{\le 4}Q_{A}^{K_li}, \mbox{}_{\le 4}R_{A}^{K_lij},  
\mbox{}_{6}Q_{A\Theta}^i,  
\delta_{A\Theta}^i,
\mbox{}_4D_{A\chi}^i)}
\nonumber \\ 
&=&
- \epsilon^6
\left(C_{QR} + C_{D_{A\chi}}
+ C_{\delta_{A\Theta}}
\right) 
\frac{m_1^3m_2^2}{2 r_{12}^6}r_{12}^i
-  
\epsilon^6
\left(C_{QR} + C_{D_{A\chi}}
+ C_{\delta_{A\Theta}}
\right) 
\frac{m_1^2m_2^3}{2 r_{12}^6}r_{12}^i  
\nonumber \\ 
&=& \epsilon^6\frac{118}{9}
\frac{m_1^3m_2^2}{r_{12}^6}r_{12}^i 
+ \epsilon^6
\frac{118}{9}\frac{m_1^2m_2^3}{r_{12}^6}r_{12}^i.  
\end{eqnarray}


Collecting these contributions mentioned from the beginning 
of this section, 
we  obtain a 3PN equation of motion. 
However, we found 
that logarithmic terms having the arbitrary constants 
$\epsilon R_A$ in their arguments survive,  
  \begin{eqnarray}
m_1 
\left(
\frac{dv_1^i}{d\tau} 
\right)^{{\rm with ~log}}
&=&
m_1 
\left(
\frac{dv_1^i}{d\tau}
\right)_{\le 2.5{\rm PN}}
\nonumber \\ 
\mbox{} &+& \epsilon^6 
\frac{m_1^2m_2}{r_{12}^3}
\left[
\frac{44m_1^2}{3r_{12}^2}n_{12}^i
\ln \left(\frac{r_{12}}{\epsilon R_1}\right)
-\frac{44m_2^2}{3r_{12}^2}n_{12}^i
\ln \left(\frac{r_{12}}{\epsilon R_2}\right)
\right. 
\nonumber \\ 
\mbox{} &&+ \left. \frac{22m_1}{r_{12}}
\left(
5 (\vec n_{12}\cdot\vec V)^2 n_{12}^i - V^2n_{12}^i 
- 2 (\vec n_{12}\cdot\vec V)V^i
\right)\ln \left(\frac{r_{12}}{\epsilon R_1}\right)  
\right]
\nonumber \\ 
\mbox{} &+& \cdot\cdot\cdot 
+ O(\epsilon^7), 
\label{3PNEOM}
\end{eqnarray}
where the acceleration through 2.5PN order, 
$(dv_1^i/d\tau)_{\le 2.5{\rm PN}}$ 
is the Damour and Deruelle 2.5PN acceleration. 
In our formalism, we have computed it in paper II. 
The ``$\cdot\cdot\cdot$'' stands for the terms that do not include 
any logarithms.

Since this equation contains two arbitrary constants, 
the body zone radii $R_A$, at first sight its predictability on the 
orbital motion of the binary seems to be reduced. In the 
next section, we shall show that by reasonable 
redefinition of the representative points  of the stars, 
we can remove $R_A$ from our equation of motion. There, we show 
the explicit form of the 3PN equation of motion we obtained.

\section{The 3PN Equation of Motion}
\label{sec:3PNEOM}

The following alternative choice of the representative point 
of the star $A$ removes the $\epsilon R_A$ dependence in Eq. (\ref{3PNEOM}):  
\begin{eqnarray}
D_{A\Theta,{\rm New}}^{ i}(\tau) = 
\epsilon^4 \delta_{A\Theta}^i(\tau) -  
 \epsilon^4\frac{22}{3} 
m_A^3 a_A^i \ln \left(\frac{r_{12}}{\epsilon R_A}\right)
\equiv  
\epsilon^4 \delta^i_{A\Theta}(\tau) + 
\epsilon^4 \delta^i_{A\ln}(\tau)           
\equiv \epsilon^4 \delta^i_{A}(\tau).       
\label{ReDefineCM}
\end{eqnarray}
Note that this redefinition of the representative points does not 
affect the existence of the energy conservation, as was shown   
by Eq. (\ref{eq5-45}). We can examine the effect of this redefinition 
onto the equation of motion using Eq. 
(\ref{EffectOf3PNDtoEOMviaMVR}) 
(use $\delta_{A\ln}^i$ instead of $\delta^i_{A\Theta}$). 
Thence we have 
\begin{eqnarray}
m_1 a_1^i|_{\delta_{A\ln}} &=&
- \epsilon^6 \frac{3m_1\delta_{2\ln}^k}{r_{12}^3}n_{12}^{<ik>} 
+ \epsilon^6\frac{3m_2\delta_{1\ln}^k}{r_{12}^3}n_{12}^{<ik>} 
- \epsilon^6 \frac{d^2 \delta_{1\ln}^i}{d\tau^2} 
\nonumber \\
\mbox{} &=& 
- \frac{44m_1^4m_2}{3r_{12}^5}n_{12}^i
\ln \left(\frac{r_{12}}{\epsilon R_1}\right)
+ \frac{44m_1^2m_2^3}{3r_{12}^5}n_{12}^i
\ln \left(\frac{r_{12}}{\epsilon R_2}\right)
\nonumber \\ 
\mbox{} &&- \frac{22m_1^3m_2}{r_{12}^4}
\left(
5 (\vec n_{12}\cdot\vec V)^2 n_{12}^i - V^2n_{12}^i 
- 2 (\vec n_{12}\cdot\vec V)V^i
\right)\ln \left(\frac{r_{12}}{\epsilon R_1}\right) 
\nonumber \\ 
\mbox{} &&+ 
\frac{22m_1^3m_2}{3r_{12}^4}
\left(
\frac{m_1}{r_{12}}n_{12}^i 
+\frac{m_2}{r_{12}}n_{12}^i 
- V^2 n_{12}^i
\right. 
\nonumber \\ 
\mbox{} &&+ \left. 
8 (\vec n_{12}\cdot\vec V)^2n_{12}^i 
- 2 (\vec n_{12}\cdot\vec V) V^i 
\right). 
\label{eq:f1dlog}
\end{eqnarray}
Comparing the above equation with Eq. (\ref{3PNEOM}),  
we easily conclude that 
the representative point $z_A^i$ of the star $A$
defined by 
\begin{eqnarray}
D_{A\Theta,{\rm New}}^{i}(\tau) &=& 
\epsilon^{-6} \int_{B_A}d^3y
(y^i-z_A^i(\tau))\Theta_{N}^{\tau\tau}
(\tau,y^k) 
= \epsilon^4 \delta_A^i(\tau)  
\label{redefOfz1}
\end{eqnarray}
obeys an equation of motion free from logarithms and 
hence free from any ambiguity up to 3PN order 
inclusively.

We mention here that Blanchet and Faye \cite{BF01a} 
have already noticed that in their 3PN equation of motion  a suitable 
coordinate transformation removes (parts of) 
logarithmic dependences of arbitrary 
parameters corresponding (roughly) to our body zone 
radii.\footnote{Unlike our case, 
their coordinate 
transformation does not remove the logarithmic dependences 
of their free parameters  
completely. The remaining logarithmic dependence was used 
to make their equation of motion conservative.}  
It is well known that choosing different values of 
dipole moments corresponds to the coordinate transformation.

By adding $m_1 a_1^i|_{\delta_{A\ln}}$ 
to Eq. (\ref{3PNEOM}), we obtain our 3PN 
equation of motion for two spherical compact stars whose 
representative points are defined by Eq. (\ref{redefOfz1}),    
\begin{eqnarray}
m_1 \frac{dv_1^i}{d\tau}
 &=&
 - \frac{m_1m_2}{r_{12}^2}n_{12}^i 
\nonumber\\
&+& \epsilon^2 \frac{m_1m_2}{r_{12}^2} n_{12}^i 
\left[ -v_1^2-2v_2^2 +4(\vec v_1\cdot\vec v_2)
+\frac32 (\vec n_{12}\cdot\vec v_2)^2 
+\frac{5m_1}{r_{12}}
+\frac{4m_2}{r_{12}} \right]
\nonumber\\
&+& \epsilon^2 \frac{m_1m_2}{r_{12}^2}
V^i \left[ 4(\vec n_{12}\cdot\vec v_1) 
-3(\vec n_{12}\cdot\vec v_2)  \right]
\nonumber \\
&+&\epsilon^4 \frac{m_1 m_2}{r_{12}^2} n_{12}^i
\left[
-2 v_2^4 + 4 v_2^2 (\vec{v}_1 \cdot \vec{v}_2 )
       - 2 (\vec{v}_1 \cdot \vec{v}_2)^2
       + \frac32 v_1^2 (\vec{n}_{12}\cdot\vec{v}_2)^2
       + \frac92 v_2^2 (\vec{n}_{12}\cdot\vec{v}_2)^2
\right.
\nonumber \\
&&- \left. 6 (\vec{v}_1 \cdot \vec{v}_{2})
           (\vec{n}_{12}\cdot\vec{v}_2)^2
       - \frac{15}{8}(\vec{n}_{12}\cdot\vec{v}_2)^4
-  \frac{57}{4}\frac{m_1^2}{r_{12}^2}
   - 9 \frac{m_2^2}{r_{12}^2}
   - \frac{69}{2} \frac{m_1 m_2}{r_{12}^2}
\right.
\nonumber \\
&&+ \left. \frac{m_1}{r_{12}}
\left(
       -\frac{15}{4}v_1^2 + \frac54 v_2^2
       - \frac52 (\vec{v}_1 \cdot \vec{v}_{2})
+ \frac{39}{2} (\vec{n}_{12}\cdot\vec{v}_1)^2
       - 39 (\vec{n}_{12}\cdot\vec{v}_1)(\vec{n}_{12} \cdot \vec{v}_{2})
\right.\right.
\nonumber \\
\mbox{} &&+ \left.\left.
        \frac{17}{2} (\vec{n}_{12}\cdot\vec{v}_2)^2
\right)
\right. 
\nonumber \\
&&+ \left. \frac{m_2}{r_{12}}
\left(
       4v_2^2  - 8 (\vec{v}_1 \cdot \vec{v}_{2})
+ 2 (\vec{n}_{12}\cdot\vec{v}_1)^2
       - 4 (\vec{n}_{12}\cdot\vec{v}_1)(\vec{n}_{12} \cdot \vec{v}_{2})
       - 6 (\vec{n}_{12}\cdot\vec{v}_2)^2
\right)
\right]
\nonumber \\
&+&\epsilon^4 \frac{m_1 m_2}{r_{12}^2} V^i
\left[
\frac{m_1}{r_{12}}
\left(
- \frac{63}{4}(\vec{n}_{12}\cdot\vec{v}_1)
+ \frac{55}{4}(\vec{n}_{12}\cdot\vec{v}_2)
\right)
+\frac{m_2}{r_{12}}
\left(
- 2(\vec{n}_{12}\cdot\vec{v}_1)
- 2(\vec{n}_{12}\cdot\vec{v}_2)
\right)
\right.
\nonumber \\
\mbox{} &&+ \left.
v_1^2 (\vec{n}_{12}\cdot\vec{v}_2)
+ 4 v_2^2 (\vec{n}_{12}\cdot\vec{v}_1)
- 5 v_2^2 (\vec{n}_{12}\cdot\vec{v}_2)
- 4 (\vec{v}_1 \cdot \vec{v}_{2})(\vec{n}_{12}\cdot\vec{V})
\right.
\nonumber \\
&&- \left.6 (\vec{n}_{12}\cdot\vec{v}_1)(\vec{n}_{12}\cdot\vec{v}_2)^2
+ \frac92 (\vec{n}_{12}\cdot\vec{v}_2)^3
\right]
\nonumber \\
&+&\epsilon^5 \frac{4 m_1^2 m_2}{5 r_{12}^3} 
\left[n_{12}^i
(\vec{n}_{12}\cdot\vec{V})
\left(
-6\frac{m_1}{r_{12}} + \frac{52}{3}\frac{m_2}{r_{12}} + 3 V^2
\right)
+  V^i
\left(
2 \frac{m_1}{r_{12}} - 8 \frac{m_2}{r_{12}} - V^2
\right)
\right]
\nonumber 
\end{eqnarray}
\begin{eqnarray}
\mbox{} &+& 
\epsilon^6\frac{m_1m_2}{r_{12}^2}n_{12}^i
\left[
\frac{35}{16}(\vec n_{12}\cdot\vec v_2)^6
- \frac{15}{8}(\vec n_{12}\cdot\vec v_2)^4v_1^2
+ \frac{15}{2}(\vec n_{12}\cdot\vec v_2)^4(\vec v_1\cdot\vec v_2)
\right.
\nonumber \\
\mbox{} &&+\left.
3(\vec n_{12}\cdot\vec v_2)^2(\vec v_1\cdot\vec v_2)^2
- \frac{15}{2}(\vec n_{12}\cdot\vec v_2)^4v_2^2
+\frac{3}{2} (\vec n_{12}\cdot\vec v_2)^2v_1^2v_2^2
-12(\vec n_{12}\cdot\vec v_2)^2(\vec v_1\cdot\vec v_2)v_2^2
\right.
\nonumber \\
\mbox{} &&-\left.
2(\vec v_1\cdot\vec v_2)^2v_2^2
+\frac{15}{2}(\vec n_{12}\cdot\vec v_2)^2v_2^4
+4(\vec v_1\cdot\vec v_2)v_2^4 - 2v_2^6
\right. 
\nonumber \\
\mbox{} &&+ \left. 
 \frac{m_1}{r_{12}}
\left(
-\frac{171}{8}(\vec n_{12}\cdot\vec v_1)^4 
+ \frac{171}{2}(\vec n_{12}\cdot\vec v_1)^3(\vec n_{12}\cdot\vec v_2)
\right.\right.
\nonumber \\
\mbox{} &&-\left.\left.
\frac{723}{4}(\vec n_{12}\cdot\vec v_1)^2(\vec n_{12}\cdot\vec v_2)^2
+\frac{383}{2}(\vec n_{12}\cdot\vec v_1)(\vec n_{12}\cdot\vec v_2)^3
-\frac{455}{8}(\vec n_{12}\cdot\vec v_2)^4
\right.\right.
\nonumber \\
\mbox{} &&+\left.\left.
\frac{229}{4}(\vec n_{12}\cdot\vec v_1)^2v_1^2
-\frac{205}{2}(\vec n_{12}\cdot\vec v_1)
(\vec n_{12}\cdot\vec v_2)v_1^2
+\frac{191}{4}(\vec n_{12}\cdot\vec v_2)^2v_1^2      
-\frac{91}{8}v_1^4
\right.\right.
\nonumber \\
\mbox{} &&-\left.\left.
\frac{229}{2}(\vec n_{12}\cdot\vec v_1)^2(\vec v_1\cdot\vec v_2)
+244(\vec n_{12}\cdot\vec v_1)(\vec n_{12}\cdot\vec v_2)
(\vec v_1\cdot\vec v_2)
- \frac{225}{2}(\vec n_{12}\cdot\vec v_2)^2(\vec v_1\cdot\vec v_2)
\right.\right.
\nonumber \\
\mbox{} &&+\left.\left.
\frac{91}{2}v_1^2(\vec v_1\cdot\vec v_2)
-\frac{177}{4}(\vec v_1\cdot\vec v_2)^2
+\frac{229}{4}(\vec n_{12}\cdot\vec v_1)^2v_2^2
-\frac{283}{2}(\vec n_{12}\cdot\vec v_1)
(\vec n_{12}\cdot\vec v_2)v_2^2
\right.\right.
\nonumber \\
\mbox{} &&+\left.\left.
\frac{259}{4}(\vec n_{12}\cdot\vec v_2)^2v_2^2
-\frac{91}{4}v_1^2v_2^2
+43(\vec v_1\cdot\vec v_2)v_2^2
-\frac{81}{8}v_2^4
\right)
\right.
\nonumber \\
\mbox{} &&+\left.
\frac{m_2}{r_{12}}\left(
-6(\vec n_{12}\cdot\vec v_1)^2(\vec n_{12}\cdot\vec v_2)^2
+12(\vec n_{12}\cdot\vec v_1)(\vec n_{12}\cdot\vec v_2)^3
\right. \right. 
\nonumber \\ 
\mbox{} &&+ \left.\left. 
 6(\vec n_{12}\cdot\vec v_2)^4 
+4(\vec n_{12}\cdot\vec v_1)(\vec n_{12}\cdot\vec v_2)
(\vec v_1\cdot\vec v_2)
+12(\vec n_{12}\cdot\vec v_2)^2(\vec v_1\cdot\vec v_2)
+4(\vec v_1\cdot\vec v_2)^2
\right.\right.
\nonumber \\
\mbox{} &&-\left.\left.
4(\vec n_{12}\cdot\vec v_1)(\vec n_{12}\cdot\vec v_2)v_2^2
-12(\vec n_{12}\cdot\vec v_2)^2v_2^2
-8(\vec v_1\cdot\vec v_2)v_2^2+4v_2^4
\right)
\right.
\nonumber \\
\mbox{} &&+\left.
\frac{m_2^2}{r_{12}^2}\left(
- (\vec n_{12}\cdot\vec v_1)^2
+2(\vec n_{12}\cdot\vec v_1)(\vec n_{12}\cdot\vec v_2)
+\frac{43}{2}(\vec n_{12}\cdot\vec v_2)^2
+18(\vec v_1\cdot\vec v_2)
-9v_2^2
\right)
\right.
\nonumber \\
\mbox{} &&+\left.
\frac{m_1m_2}{r_{12}^2}\left(
\frac{415}{8}(\vec n_{12}\cdot\vec v_1)^2
-\frac{375}{4}(\vec n_{12}\cdot\vec v_1)
(\vec n_{12}\cdot\vec v_2)
+\frac{1113}{8}(\vec n_{12}\cdot\vec v_2)^2
\right.\right.
\nonumber \\
\mbox{} &&-\left.\left.
\frac{615 \pi^2}{64}(\vec n_{12}\cdot\vec V)^2
+ 18 v_1^2 + \frac{123\pi^2}{64}V^2
+33 (\vec v_1\cdot\vec v_2)
-\frac{33}{2}v_2^2
\right)
\right. 
\nonumber \\
\mbox{} &&+ \left. 
\frac{m_1^2}{r_{12}^2}
\left(
- \frac{2069}{8}(\vec n_{12}\cdot\vec v_1)^2
+ 543(\vec n_{12}\cdot\vec v_1)
(\vec n_{12}\cdot\vec v_2)
-\frac{939}{4}(\vec n_{12}\cdot\vec v_2)^2
+\frac{471}{8}v_1^2 
\right.\right.
\nonumber \\
\mbox{} &&-\left.\left.
\frac{357}{4}(\vec v_1\cdot\vec v_2)
+\frac{357}{8}v_2^2\right)
\right.
\nonumber \\
\mbox{} &&+\left.
\frac{16m_2^3}{r_{12}^3}
+\frac{m_1^2m_2}{r_{12}^3}
\left(\frac{547}{3}-\frac{41\pi^2}{16}\right)
- \frac{13m_1^3}{12r_{12}^3} 
+\frac{m_1m_2^2}{r_{12}^3}\left(
\frac{545}{3} 
- \frac{41\pi^2}{16}
\right)  
\right]
\nonumber 
\end{eqnarray}

\begin{eqnarray}
\mbox{}&+&
\epsilon^6\frac{m_1m_2}{r_{12}^2}V^i\left[
\frac{15}{2}(\vec n_{12}\cdot\vec v_1)
(\vec n_{12}\cdot\vec v_2)^4
-\frac{45}{8}(\vec n_{12}\cdot\vec v_2)^5
-\frac{3}{2}(\vec n_{12}\cdot\vec v_2)^3v_1^2
\right.
\nonumber \\
\mbox{}&&+\left.
6(\vec n_{12}\cdot\vec v_1)
(\vec n_{12}\cdot\vec v_2)^2(\vec v_1\cdot\vec v_2)
-6(\vec n_{12}\cdot\vec v_2)^3(\vec v_1\cdot\vec v_2)
-2(\vec n_{12}\cdot\vec v_2)(\vec v_1\cdot\vec v_2)^2
\right.
\nonumber \\
\mbox{}&&-\left.
12(\vec n_{12}\cdot\vec v_1)(\vec n_{12}\cdot\vec v_2)^2
v_2^2
+12(\vec n_{12}\cdot\vec v_2)^3v_2^2
+(\vec n_{12}\cdot\vec v_2)v_1^2v_2^2
-4(\vec n_{12}\cdot\vec v_1)(\vec v_1\cdot\vec v_2)v_2^2
\right.
\nonumber \\
\mbox{}&&+\left.
8(\vec n_{12}\cdot\vec v_2)(\vec v_1\cdot\vec v_2)v_2^2
+4(\vec n_{12}\cdot\vec v_1)v_2^4
-7(\vec n_{12}\cdot\vec v_2)v_2^4
+ \frac{m_2}{r_{12}}
\left(
-2(\vec n_{12}\cdot\vec v_1)^2(\vec n_{12}\cdot\vec v_2)
\right.\right.
\nonumber \\
\mbox{}&&+\left.\left.
8(\vec n_{12}\cdot\vec v_1)(\vec n_{12}\cdot\vec v_2)^2
+2(\vec n_{12}\cdot\vec v_2)^3
+2(\vec n_{12}\cdot\vec v_1)(\vec v_1\cdot\vec v_2)
+4(\vec n_{12}\cdot\vec v_2)(\vec v_1\cdot\vec v_2)
\right.\right.
\nonumber \\
\mbox{}&&-\left.\left.
2(\vec n_{12}\cdot\vec v_1)v_2^2
-4(\vec n_{12}\cdot\vec v_2)v_2^2
\right)
+\frac{m_1}{r_{12}}
\left(
-\frac{243}{4}(\vec n_{12}\cdot\vec v_1)^3
+\frac{565}{4}(\vec n_{12}\cdot\vec v_1)^2
(\vec n_{12}\cdot\vec v_2)
\right.\right.
\nonumber \\
\mbox{}&&-\left.\left.
\frac{269}{4}(\vec n_{12}\cdot\vec v_1)
(\vec n_{12}\cdot\vec v_2)^2
-\frac{95}{12}(\vec n_{12}\cdot\vec v_2)^3
+\frac{207}{8}(\vec n_{12}\cdot\vec v_1)v_1^2
-\frac{137}{8}(\vec n_{12}\cdot\vec v_2)v_1^2
\right.\right.
\nonumber \\
\mbox{}&&-\left.\left.
36(\vec n_{12}\cdot\vec v_1)(\vec v_1\cdot\vec v_2)
+\frac{27}{4}(\vec n_{12}\cdot\vec v_2)
(\vec v_1\cdot\vec v_2)
+\frac{81}{8}(\vec n_{12}\cdot\vec v_1)v_2^2
+\frac{83}{8}(\vec n_{12}\cdot\vec v_2)v_2^2
\right)
\right.
\nonumber \\
\mbox{} &&+ \left.
\frac{m_2^2}{r_{12}^2}
\left(
4(\vec n_{12}\cdot\vec v_1)
+5(\vec n_{12}\cdot\vec v_2)
\right)
\right.
\nonumber \\
\mbox{} &&+ \left.
\frac{m_1m_2}{r_{12}^2}\left(
-\frac{307}{8}(\vec n_{12}\cdot\vec v_1)
+\frac{479}{8}(\vec n_{12}\cdot\vec v_2)
+\frac{123\pi^2}{32}(\vec n_{12}\cdot\vec V)
\right)
\right.
\nonumber \\
\mbox{} &&+ \left.
\frac{m_1^2}{r_{12}^2}
\left(
\frac{311}{4}(\vec n_{12}\cdot\vec v_1)
-\frac{357}{4}(\vec n_{12}\cdot\vec v_2)
\right)
\right] 
\nonumber \\ 
\mbox{} &&
+ O(\epsilon^7), 
\label{3PNEOMFinal}
\end{eqnarray}
in the harmonic gauge.

Now we list  some features of our 3PN equation of motion.
In the test-particle limit, 
our 3PN equation of motion coincides  
with a 
geodesic equation for a test-particle in the 
Schwarzschild metric in the harmonic coordinate 
(up to 3PN order). 
With the help of the formulas developed in \cite{BF01b},  
we have checked the Lorentz invariance of Eq. (\ref{3PNEOMFinal}) 
(in the post-Newtonian perturbative sense).
Also, we have checked that our 3PN acceleration admits 
a conserved energy of the binary orbital motion 
(modulo the 2.5PN radiation reaction effect).  
In fact, the energy of the binary $E$ associated with  
Eq. (\ref{3PNEOMFinal}) is
\begin{eqnarray}
E &=& 
\frac{1}{2}m_1v_1^2 -\frac{m_1 m_2}{2r_{12}} 
\nonumber 
 \\
\mbox{} &+& 
\epsilon^2
\left[
\frac{3}{8}m_1v_1^4 + \frac{m_1^2m_2}{2r_{12}^2} 
+ \frac{m_1m_2}{2r_{12}}
\left(
3v_1^2 - \frac{7}{2}(\vec v_1\cdot\vec v_2)
- \frac{1}{2}(\vec n_{12}\cdot\vec v_1) 
(\vec n_{12}\cdot\vec v_2)
\right)
\right] 
\nonumber 
\end{eqnarray}
\begin{eqnarray}
\mbox{} &+& 
\epsilon^4
\left[
\frac{5}{16}m_1v_1^6 - \frac{m_1^3m_2}{2r_{12}^3}
- \frac{19 m_1^2m_2^2}{8r_{12}^3}
\right.
\nonumber \\
\mbox{} &&+ \left.
\frac{m_1^2m_2}{2 r_{12}^2}
\left(
-3v_1^2 + \frac{7}{2}v_2^2 + \frac{29}{2}(\vec n_{12}\cdot\vec v_1)^2
- \frac{13}{2}(\vec n_{12}\cdot\vec v_1)(\vec n_{12}\cdot\vec v_2)
+ (\vec n_{12}\cdot\vec v_2)^2
\right)
\right.
\nonumber \\
\mbox{} &&+ \left.
\frac{m_1m_2}{4r_{12}}
\left(
\frac{3}{2}(\vec n_{12}\cdot\vec v_1)^3(\vec n_{12}\cdot\vec v_2)
+ \frac{3}{4}(\vec n_{12}\cdot\vec v_1)^2(\vec n_{12}\cdot\vec v_2)^2
- \frac{9}{2}(\vec n_{12}\cdot\vec v_1)(\vec n_{12}\cdot\vec v_2)v_1^2
\right. \right. 
\nonumber \\
\mbox{} &&- \left. \left.
 \frac{13}{2}(\vec n_{12}\cdot\vec v_2)^2v_1^2
+ \frac{21}{2}v_1^4
+ \frac{13}{2}(\vec n_{12}\cdot\vec v_1)^2(\vec v_1\cdot\vec v_2)
+ 3(\vec n_{12}\cdot\vec v_1)(\vec n_{12}\cdot\vec v_2)(\vec v_1\cdot\vec v_2)
\right. \right. 
\nonumber \\
\mbox{} &&- \left. \left.
 \frac{55}{2}v_1^2(\vec v_1\cdot\vec v_2)
+ \frac{17}{2}(\vec v_1\cdot\vec v_2)^2
+ \frac{31}{4}v_1^2v_2^2
\right)
\right]
\nonumber  \\
\mbox{} &+& 
\epsilon^6
\left[
\frac{35}{128}m_1v_1^8 
+ \frac{3 m_1^4m_2}{8r_{12}^4} 
+ \frac{469 m_1^3m_2^2}{18r_{12}^4} 
\right. 
\nonumber \\
\mbox{} &&+ \left. 
\frac{m_1^2m_2^2}{2r_{12}^3}
\left(
\frac{547}{6}(\vec n_{12}\cdot\vec v_1)^2
- \frac{3115}{24}(\vec n_{12}\cdot\vec v_1)(\vec n_{12}\cdot\vec v_2)
-\frac{123\pi^2}{32}(\vec n_{12}\cdot\vec v_1)(\vec n_{12}\cdot\vec V)
\right. \right. 
\nonumber \\
\mbox{} &&- \left. \left. 
\frac{575}{9}v_1^2
+ \frac{41\pi^2}{32}(\vec V\cdot\vec v_2)
+ \frac{4429}{72}(\vec v_{1}\cdot\vec v_2)
\right)
\right. 
\nonumber \\
\mbox{} &&+ \left. 
\frac{m_1^3m_2}{2r_{12}^3}
\left(
- \frac{437}{4}(\vec n_{12}\cdot\vec v_1)^2
+ \frac{317}{4}(\vec n_{12}\cdot\vec v_1)(\vec n_{12}\cdot\vec v_2)
+ 3(\vec n_{12}\cdot\vec v_2)^2
+ \frac{301}{12}v_1^2
\right. \right. 
\nonumber \\
\mbox{} &&- \left. \left. 
\frac{337}{12}(\vec v_{1}\cdot\vec v_2)
+\frac{5}{2}v_2^2
\right)
\right. 
\nonumber \\
\mbox{} &&+ \left. 
\frac{m_1m_2}{r_{12}}
\left(
- \frac{5}{16}
(\vec n_{12}\cdot\vec v_1)^5(\vec n_{12}\cdot\vec v_2)
-\frac{5}{16}
(\vec n_{12}\cdot\vec v_1)^4(\vec n_{12}\cdot\vec v_2)^2
-\frac{5}{32}
(\vec n_{12}\cdot\vec v_1)^3(\vec n_{12}\cdot\vec v_2)^3
\right. \right. 
\nonumber \\
\mbox{} &&+ \left. \left. 
\frac{19}{16}
(\vec n_{12}\cdot\vec v_1)^3(\vec n_{12}\cdot\vec v_2)v_1^2
+\frac{15}{16}
(\vec n_{12}\cdot\vec v_1)^2(\vec n_{12}\cdot\vec v_2)^2v_1^2
+\frac{3}{4}(\vec n_{12}\cdot\vec v_1)(\vec n_{12}\cdot\vec v_2)^3v_1^2
\right. \right. 
\nonumber \\
\mbox{} &&+ \left. \left. 
\frac{19}{16}(\vec n_{12}\cdot\vec v_2)^4v_1^2
-\frac{21}{16}
(\vec n_{12}\cdot\vec v_1)(\vec n_{12}\cdot\vec v_2)v_1^4
-2(\vec n_{12}\cdot\vec v_2)^2v_1^4
+\frac{55}{16}v_1^6
-\frac{19}{16}
(\vec n_{12}\cdot\vec v_1)^4(\vec v_{1}\cdot\vec v_2)
\right. \right. 
\nonumber \\
\mbox{} &&- \left. \left. 
(\vec n_{12}\cdot\vec v_1)^3(\vec n_{12}\cdot\vec v_2)
(\vec v_{1}\cdot\vec v_2)
-\frac{15}{32}
(\vec n_{12}\cdot\vec v_1)^2(\vec n_{12}\cdot\vec v_2)^2
(\vec v_{1}\cdot\vec v_2)
\right. \right. 
\nonumber \\
\mbox{} &&+ \left. \left. 
\frac{45}{16}
(\vec n_{12}\cdot\vec v_1)^2v_1^2(\vec v_{1}\cdot\vec v_2)
+\frac{5}{4}
(\vec n_{12}\cdot\vec v_1)(\vec n_{12}\cdot\vec v_2)
v_1^2(\vec v_{1}\cdot\vec v_2)
+\frac{11}{4}
(\vec n_{12}\cdot\vec v_2)^2v_1^2(\vec v_{1}\cdot\vec v_2)
\right. \right. 
\nonumber \\
\mbox{} &&- \left. \left. 
\frac{139}{16}
v_1^4(\vec v_{1}\cdot\vec v_2)
-\frac{3}{4}(\vec n_{12}\cdot\vec v_1)^2(\vec v_{1}\cdot\vec v_2)^2
+ \frac{5}{16}
(\vec n_{12}\cdot\vec v_1)(\vec n_{12}\cdot\vec v_2)
(\vec v_{1}\cdot\vec v_2)^2
+\frac{41}{8}v_1^2(\vec v_{1}\cdot\vec v_2)^2
\right. \right. 
\nonumber \\
\mbox{} &&+ \left. \left. 
\frac{1}{16}(\vec v_{1}\cdot\vec v_2)^3
-\frac{45}{16}(\vec n_{12}\cdot\vec v_1)^2v_1^2v_2^2
-\frac{23}{32}(\vec n_{12}\cdot\vec v_1)(\vec n_{12}\cdot\vec v_2)
v_1^2v_2^2
+\frac{79}{16}v_1^4v_2^2
-\frac{161}{32}v_1^2v_2^2(\vec v_{1}\cdot\vec v_2)
\right)
\right.
\nonumber \\
\mbox{} &&+ \left.
\frac{m_1^2m_2}{r_{12}^2}
\left(
-\frac{49}{8}(\vec n_{12}\cdot\vec v_1)^4
+\frac{75}{8}(\vec n_{12}\cdot\vec v_1)^3(\vec n_{12}\cdot\vec v_2)
-\frac{187}{8}
(\vec n_{12}\cdot\vec v_1)^2(\vec n_{12}\cdot\vec v_2)^2
+\frac{11}{2}v_1^4
\right. \right. 
\nonumber \\
\mbox{} &&+ \left. \left. 
\frac{247}{24}
(\vec n_{12}\cdot\vec v_1)(\vec n_{12}\cdot\vec v_2)^3
+\frac{49}{8}(\vec n_{12}\cdot\vec v_1)^2v_1^2
+\frac{81}{8}
(\vec n_{12}\cdot\vec v_1)(\vec n_{12}\cdot\vec v_2)v_1^2
-\frac{21}{4}
(\vec n_{12}\cdot\vec v_2)^2v_1^2
\right. \right. 
\nonumber \\
\mbox{} &&- \left. \left. 
\frac{15}{2}(\vec n_{12}\cdot\vec v_1)^2(\vec v_{1}\cdot\vec v_2)
-\frac{3}{2}
(\vec n_{12}\cdot\vec v_1)(\vec n_{12}\cdot\vec v_2)
(\vec v_{1}\cdot\vec v_2)
+\frac{21}{4}
(\vec n_{12}\cdot\vec v_2)^2
(\vec v_{1}\cdot\vec v_2)
-27v_1^2(\vec v_{1}\cdot\vec v_2)
\right. \right. 
\nonumber \\
\mbox{} &&+ \left. \left. 
\frac{55}{2}(\vec v_{1}\cdot\vec v_2)^2
+\frac{49}{4}
(\vec n_{12}\cdot\vec v_1)^2v_2^2
-\frac{27}{2}
(\vec n_{12}\cdot\vec v_1)
(\vec n_{12}\cdot\vec v_2)v_2^2
+\frac{3}{4}(\vec n_{12}\cdot\vec v_2)^2v_2^2
+\frac{55}{4}v_1^2v_2^2
\right. \right. 
\nonumber \\
\mbox{} &&- \left. \left. 
28(\vec v_{1}\cdot\vec v_2)v_2^2
+\frac{135}{16}v_2^4
\right)
\right] + (1 \leftrightarrow 2) + O(\epsilon^7).
\label{eq:binaryconservedenergyat3PN}
\end{eqnarray}
This orbital energy 
of the binary is computed based on that found in Blanchet 
and Faye \cite{BF01a}, the relation between 
their 3PN equation of motion and 
our result described in Sec. \ref{subsec:comparison} 
below, and Eq. (\ref{eq5-45}). 
(After constructing $E$ given as 
Eq. (\ref{eq:binaryconservedenergyat3PN}),  
we have checked that our 3PN equations of motion make  
$E$ conserved.)

We note that Eq. (\ref{3PNEOM}) 
gives a correct geodesic equation in the test-particle limit, 
is Lorentz invariant, and admits the conserved energy.  
These facts can be seen by the form of $a_1^i|_{\delta_{A\ln}}$, 
Eq. (\ref{eq:f1dlog}); it is zero when $m_1 \rightarrow 0$,
is Lorentz invariant up to 3PN order, and is the effect 
of the mere redefinition 
of the dipole moments which does not break energy conservation.

Finally, we mention here two computational details. 
In the course of calculation, $\vec z_1$, $\vec z_2$ appears 
independently, that is, not in a form as $\vec r_{12}$.  This can 
be seen, for instance,  from Eq. (\ref{eq5-6}). As another 
example, the surface integral over the near zone boundary in 
Eq. (\ref{NBcontribution}) in general gives terms depending on 
$\vec z_A$ explicitly. All such $\vec z_A$ dependences, 
when collected in the equation of motion, are combined into 
$\vec r_{12}$.    
We have retained during our calculation 
${\cal R}$-dependent terms with 
positive powers of ${\cal R}$ or logarithms of ${\cal R}$. 
As stated below Eq. (\ref{RetardedEIREE}), it is a good computational 
check to show that our equation of motion does not depend on  
${\cal R}$ physically. 
In fact, we found that 
${\cal R}$-dependent terms canceled each other out in the final result. 
There was no need to employ a gauge transformation to remove 
such ${\cal R}$ dependences.

\section{Comparison and Summary}
\label{sec:summary}

\subsubsection{Comparison}
\label{subsec:comparison}

By comparing Eq. (\ref{3PNEOMFinal}) with the Blanchet and 
Faye 3PN equation of motion \cite{BF01a}, 
we find the following relationship:   
\begin{eqnarray}
m_1 \vec a_1^{{\rm this ~work}} &=& 
m_1 (\vec a_1^{{\rm BF}})_{\lambda = - \frac{1987}{3080}} 
+ m_1 \vec a_{1}|_{\delta_{A \ln}}
+ m_1 \vec a_{1}|_{\delta_{A,{\rm BF}}},  
\end{eqnarray}
where 
$m_1 \vec a_1^{{\rm this ~work}}$ is the 3PN acceleration 
given in Eq. (\ref{3PNEOMFinal}), 
$(\vec a_1^{{\rm BF}})_{\lambda = - 1987/3080}$ is 
the Blanchet and Faye 3PN acceleration with 
$\lambda = - 1987/3080$, and 
$m_1 \vec a_{1}|_{\delta_{A \ln}}$ is given in Eq. 
(\ref{eq:f1dlog}) with $\epsilon R_A$ replaced by 
$r_A'$ for notational consistency with 
the Blanchet and Faye 3PN equation of motion 
shown in \cite{BF01a}.
$m_1 \vec a_{1}|_{\delta_{A,{\rm BF}}}$ is an acceleration 
induced by the following 
dipole moments of the stars:    
\begin{eqnarray}
\delta_{A,{\rm BF}}^i &=& - \frac{3709}{1260}m_A^3a_A^i.
\end{eqnarray}
We can compute $m_1 a_{1}^i|_{\delta_{A,{\rm BF}}}$
by substituting $\delta^i_{A,{\rm BF}}$ 
instead of $\delta^i_{A\Theta}$
into Eq. (\ref{EffectOf3PNDtoEOMviaMVR}).
Thus, by choosing the dipole moments,  
\begin{eqnarray}
D_{A\Theta,{\rm BF}}^i = \epsilon^4 
\delta_{A\Theta}^i - 
\epsilon^4 \delta^i_{A,{\rm BF}},  
\end{eqnarray} 
we have the 3PN equation of motion in completely the same 
form as $(\vec a_1^{{\rm BF}})_{\lambda = - 1987/3080}$. 
In other words, 
our 3PN equation of motion physically agrees with 
$(\vec a_1^{{\rm BF}})_{\lambda = - 1987/3080}$ modulo 
the definition of the dipole moments (or equivalently, 
the coordinate transformation under the harmonic coordinate 
condition). 
In \cite{IF03}, we have shown 
some arguments that support this conclusion.

The value of $\lambda$ that we found, $\lambda = - 1987/3080$,
 is perfectly consistent with the relation  
(\ref{omegalambdarelation}) and the result of \cite{DJS01a} 
($\omega_{{\rm static}} = 0$).

\subsubsection{Summary}
\label{subsec:summary}

To deal with strongly self-gravitating objects
such as neutron stars, we have used the 
surface integral approach with 
the strong field point-particle limit. 
The surface integral approach is achieved by using 
the local conservation of the energy momentum, which led  
us to the general form of the
equation of motion which is expressed entirely in terms of 
surface integrals. 
The use of the strong field point-particle limit and the 
surface integral approach makes our 3PN equation of motion 
applicable to inspiraling compact binaries which consist  
of strongly self-gravitating regular stars (modulo the 
scalings imposed on the initial hypersurface). 
Our 3PN equation of motion depends only on masses 
of the stars and is independent of their internal structure such 
as their density profiles or radii.  
Thus our result supports the strong equivalence principle 
up to 3PN order.

The multipole moments previously defined in papers I and II 
are found to be unsatisfactory at 3PN order 
in the sense that 
these moments are not defined in a proper reference 
coordinate  
and consequently they contain monopole terms. 
By taking account of the effect of Lorentz contraction on these moments, 
we succeeded in 
extracting cleanly the monopole terms from 
these  multipole moments up to the required order.

At 3PN order, it does not seem possible  to derive the 
field in a closed form. This is because not all the  
superpotentials required are available, and thus we 
could not evaluate all the Poisson-type $N/B$ integrals. 
Some of the integrands allow us to derive  
superpotentials in series forms in the neighborhood 
of a star. 
For others,  
we have adopted an idea 
that Blanchet and Faye have used in \cite{BF00a,BF01a,BF00b}.
The idea is that while abandoning complete derivation of the 
3PN gravitational field throughout $N/B$, one exchanges  
the order of integration. We first evaluate 
the surface integrals in 
the evolution equation for the energy of a star and  
the general form of the equation of motion, and 
then we evaluate the remaining volume integrals.  
Using these methods, we first derived 
the 3PN mass-energy relation and the momentum-velocity 
relation. The 3PN mass-energy relation admits a natural 
interpretation. We then evaluated the surface integrals in the 
general form of the equation of motion, and obtained an equation 
of motion up to 3PN order of accuracy.

At 3PN order, 
our equation of motion contains logarithms of 
the body zone radii $R_A$. 
Practically, we cannot discard $R_A$ dependences if 
$R_A$ is in logarithms. We showed that we could remove 
the logarithms by suitable redefinition of the representative 
points of the stars.  Thus we could transform our 
3PN equation of motion into an unambiguous equation 
which does not contain any  
arbitrarily introduced free parameters.

Our so-obtained 3PN equation of motion  
agrees physically  
(modulo a definition of the representative 
points of the stars)  
with the result derived by 
Blanchet and Faye \cite{BF01a} 
with $\lambda = -1987/3080$,  
which is consistent 
with Eq. (\ref{omegalambdarelation}) and  
$\omega_{{\rm static}} = 0$ 
reported by Damour, Jaranowski, and Sch\"afer \cite{DJS01a}.
This result indirectly supports the validity of the 
dimensional regularization in the ADM canonical approach 
in the ADMTT gauge.

Blanchet and Faye \cite{BF00a,BF01a} introduced four arbitrary 
parameters. In Hadamard's partie finie regularization, 
one has to introduce a sphere around each singular 
point (representing a point-mass) whose radius is a 
free parameter. In their framework, 
regularizations are employed in the evaluation of both  
a gravitational field having two singular points 
and two equations of motion.   Since 
there is {\it a priori} no reason to expect that 
the spheres introduced for the evaluation 
of the field and the equations of motion coincide, there 
arise four arbitrary parameters. This is in contrast with 
our formalism where each body zone introduced in the 
evaluation of the field is inevitably the same as  
the body zone with which we defined the energy and 
the three-momentum of each star for which we derived 
our equation of motion.
 
Actually, the redefinition of the representative points 
in our formalism corresponds to the gauge transformation 
in \cite{BF01a}, and only two of the four parameters remain 
in \cite{BF01a}. Then they have used 
one of the
 remaining two free parameters to ensure the 
energy conservation and there remains only one arbitrary 
parameter $\lambda$ which they could not fix in their 
formalism.

On the other hand, our 3PN equation of motion has 
no ambiguous parameter, admits conservation of 
an orbital energy of the binary system (when we 
neglect the 2.5 PN radiation reaction effect), and 
respects Lorentz invariance in the post-Newtonian 
perturbative sense. We emphasize that we do not 
need to {\it a posteriori} adjust some parameters 
to make our 3PN equation of motion satisfy 
the above three physical features.

Finally, we note here that Blanchet {\it et al.} 
\cite{BDEF03}
have recently obtained the same 
value of $\lambda$ by computing   
a 3PN equation 
of motion in the harmonic gauge 
using the dimensional regularization.

\section*{Acknowledgments}
This work was partly done when the author was at the 
Astronomical Institute of Tohoku University, Japan. 
The author was partly supported 
by the Japan Society for the Promotion of Science 
Research for Young Scientists.
He is grateful to T. Futamase and H. Asada for valuable  
discussions and comments. H. Asada let the author know of 
the recent achievement of Blanchet {\it et al.}\cite{BDEF03}. 
Extensive use of the softwares 
MATHEMATICA, MATHTENSOR, and MAPLE has been made.

\appendix
\section{$Q_A^{K_li}$ and $R_A^{K_lij}$}
\label{RLijandQLi}

In this section we  evaluate $Q_A^{K_li}$ and $R_A^{K_lij}$ 
up to $O(\epsilon^4)$, the order required to compute the 
3PN field.
In the following, we shall omit the $\epsilon R_A$ dependence 
as explained in the appendix of paper II.
In the evaluation of the field up to 
3PN order, $\mbox{}_{\le {10}}h^{\tau \tau}$ and 
$\mbox{}_{\le 8}h^{\mu i}$, 
we use 
$\mbox{}_{\le 8}\Lambda_N^{\mu \nu}$ 
as the integrand in the 
surface integrals of $Q_A^{K_li}$ and $R_A^{K_lij}$. 
Then  straightforward calculation gives
 \begin{eqnarray} 
Q_A^{K_li} &=& \epsilon^{-4}
\sum_{n=4}^{8}\epsilon^{n}\oint_{\pa B_A}dS_m
\left[\mbox{}_n\Lambda^{\tau m}_N - 
v_A^m\mbox{}_n\Lambda^{\tau \tau}_N  
\right]y_A^{K_li} = O(\epsilon^5) 
\end{eqnarray}
for ${}^{\forall}$ l and    
\begin{eqnarray} 
R_1^{K_lij} &=& \epsilon^{-4}
\sum_{n=4}^{8}\epsilon^{n}\oint_{\pa B_1}dS_m
\left[\mbox{}_n\Lambda^{j m}_N - 
v_1^m\mbox{}_n\Lambda^{j \tau}_N  
\right]y_1^{K_li}  \nonumber \\ 
\mbox{} &=& \left\{
\begin{array}{ll} 
- \epsilon^4\frac{3 m_1^3 m_2}{5 r_{12}^5}r_{12}^{<ij>} 
+ O(\epsilon^5) 
& {\rm for~} l = 0, \\ 
\epsilon^4 \frac{3 m_1^3 m_2}{5r_{12}^3}
\left(4 \delta^{ki}r_{12}^j - \delta^{jk}r_{12}^i - \delta^{ij}r_{12}^k   
\right) 
+ O(\epsilon^5) 
& {\rm for~} l = 1, \\ 
O(\epsilon^5) 
& \mbox{otherwise}.  
\end{array}
\right.   
\end{eqnarray}

$R_A^{K_lij}$ contribute to the field 
at  3PN order through the moments $Z_A^{K_lij}$. 
See Eqs. 
(\ref{StrVelRelation})-(\ref{ZLijToJLij}).
Combining the above results, the $Q_A^{K_li}$ and $R_A^{K_lij}$ 
contributions to the 3PN field $h_{QR}^{\mu\nu}$ become  
\begin{eqnarray}
\mbox{}_{10}h_{QR}^{\tau\tau} +   
\mbox{}_8h_{QR}^{k}\mbox{}_{k} &=& 
4 \sum_{A=1,2}
\frac{r^k_A}{r_A^3}\left(
\mbox{}_4R_A^{kll} 
-\frac{1}{2}\mbox{}_4R_A^{llk}
 \right)
= 
C_{QR} \sum_{A=1,2}\frac{m_A^3 (\vec a_A \cdot \vec r_A)}{r_A^3}, 
\end{eqnarray}
with $C_{QR} = 12$ and $h^{\tau i}_{QR} = O(\epsilon^9)$,   
where we used Eqs. (\ref{hBttInAp}), (\ref{hBtiInAp}), and 
(\ref{hBijInAp}) below.

The $h^{\mu\nu}_{QR}$ field affects the equation of motion,  
\begin{eqnarray}
&&
m_1 a_{1QR}^i = 
- \epsilon^6 
\frac{C_{QR}}{2}\frac{m_1^3m_2^2}{r_{12}^6}r_{12}^i
- \epsilon^6 
\frac{C_{QR}}{2}\frac{m_1^2m_2^3}{r_{12}^6}r_{12}^i, 
\label{QREffecttoEOMinAp}
\end{eqnarray}
where $a_{1 QR}^i$ is the $QR$ field contribution to the 
acceleration of the star 1. Note that in the above equation 
we have not yet taken into account the effect of 
the 3PN $Q_A^i$ integral ($\mbox{}_6Q_A^i$). 
$\mbox{}_6Q_A^i$ does not affect the 3PN field, but the 3PN 
equation of motion through the 3PN momentum-velocity relation. 
We evaluate $\mbox{}_6Q_A^i$ in 
Sec. \ref{sec:3PNMomVelRelation} and Appendix 
\ref{chipart}.

For convenience, we list $Q_A^{K_li}$ integrals, 
$R_A^{K_lij}$ integrals, and the body zone contribution including 
multipole moments up to 3PN order. We retain here (and only here) 
the dipole moments of order $O(\epsilon^2)$, since it is appropriate 
to define the representative points of the stars by 
$D_A^i = \epsilon^2 M_A^{ik}v_A^k + O(\epsilon^3)$ 
when we are concerned with the spin effects (see paper I).     
\begin{eqnarray}
Q_1^{i} &=&  
 \epsilon^4 \frac{2m_2M_1^{ik}n_{12}^k}{3r_{12}^2} 
+ O(\epsilon^5), 
\label{eqB5}   
\end{eqnarray}
and $Q_1^{K_li} =O(\epsilon^5)$ for $l\ne0$,  
\begin{eqnarray}
R_1^{ji} &=& 
\epsilon^4\frac{m_2}{3r_{12}^2}
\left(
\mbox{}_2D_1^kn_{12}^k\delta^{ij} + 2\mbox{}_2D_1^jn_{12}^i-
\mbox{}_2D_1^in_{12}^j
\right)
\nonumber \\
&&+ 
\mbox{}
\epsilon^4\frac{m_2}{3 r_{12}^2}
\left(
 3 \delta^{ij}M_1^{kl}n_{12}^kv_1^l 
-3 M_1^{ik}v_1^kn_{12}^j 
+4 M_1^{ik}n_{12}^k v_1^j
-4 \delta^{ij}M_1^{kl}n_{12}^kv_2^l 
\right.
\nonumber \\
\mbox{} &&- \left. 
2 M_1^{jk}v_2^kn_{12}^i
+2 M_1^{jk}n_{12}^kv_2^i
+4 M_1^{ik}v_2^kn_{12}^j
-4 M_1^{ik}n_{12}^kv_2^j
\right)
\nonumber \\
\mbox{} &&+
\epsilon^4\frac{m_2}{5 r_{12}^3}
\left(3 I_1^{ij} 
- 3 \delta^{ij}I_1^{kl}n_{12}^kn_{12}^l 
- 12 I_1^{jk}n_{12}^kn_{12}^i 
+ 3 I_1^{ik}n_{12}^kn_{12}^j 
+ 3I_1^{k}\mbox{}_kn_{12}^in_{12}^j
\right) 
\nonumber \\
\mbox{} &&+
\epsilon^4\frac{2m_2}{5 r_{12}^3}
\left(
-2 n_{12}^in_{12}^jZ_1^{k[lk]l}
- n_{12}^kn_{12}^jZ_1^{k[li]l}
- n_{12}^kn_{12}^jZ_1^{i[lk]l}
\right.
\nonumber \\
\mbox{} &&+ \left. 
 4n_{12}^kn_{12}^iZ_1^{k[lj]l}
+ 4n_{12}^kn_{12}^iZ_1^{j[lk]l}
+2\delta^{ij}n_{12}^kn_{12}^lZ_1^{k[ml]m}
\right.
\nonumber \\
\mbox{} &&- \left. 
Z_1^{i[kj]k}-Z_1^{j[ki]k}
\right) 
+ ({\rm monopole~part}) + O(\epsilon^5), 
\label{eqB6}   \\
R_1^{jki} &=& \epsilon^4\frac{m_2}{5r_{12}^2}\left(
 2 \delta^{i(j}I_1^{k)l}n_{12}^l 
- 2I_1^{i(j}n_{12}^{k)}
- \delta^{jk}I_1^{l}\mbox{}_ln_{12}^i 
+ 3 I_1^{jk}n_{12}^i
\right)
\nonumber \\
\mbox{} &&+ 
\epsilon^4\frac{2 m_2}{15r_{12}^2}\left(
- 2 \delta^{jk}Z_1^{l[lm]m}n_{12}^i 
-  \delta^{ij}Z_1^{l[mk]m}n_{12}^l
-  \delta^{ik}Z_1^{l[mj]m}n_{12}^l
-  \delta^{ij}Z_1^{k[ml]m}n_{12}^l
\right. 
\nonumber \\
\mbox{} &&- \left. 
  \delta^{ik}Z_1^{j[ml]m}n_{12}^l
+ Z_1^{i[lj]l}n_{12}^k
+ Z_1^{i[lk]l}n_{12}^j
+ Z_1^{j[li]l}n_{12}^k
+ Z_1^{k[li]l}n_{12}^j
\right. 
\nonumber \\
\mbox{} &&- \left. 
 3Z_1^{j[lk]l}n_{12}^i
- 3Z_1^{k[lj]l}n_{12}^i
\right) 
+ ({\rm monopole~part}) 
+ O(\epsilon^5), 
\label{eqB7}  
\end{eqnarray}
and $R_1^{K_lij} =O(\epsilon^5)$ for $l \ne 0,1$. 

In the following, we have used 
Eqs. (\ref{MomVelRelation})-(\ref{ZLijToJLij}): 
\begin{eqnarray}
h^{\tau\tau}_B &=& 4 \epsilon^4 \sum_{A=1,2}
\left[
\frac{P_A^{\tau}}{r_A} + \epsilon^2 \frac{D_A^k r^k_A}{r_A^3} +
 \epsilon^4 \frac{3 I_A^{kl} r^{<kl>}_A }{2 r_A^5} +
 \epsilon^6 \frac{5 I_A^{klm} r^{<klm>}_A}{2 r_A^7} 
\right] 
-4 \epsilon^5 \frac{\pa }{\pa\tau}\sum_{A=1,2}P_A^{\tau}
\nonumber \\
\mbox{} &&+ 2 
\epsilon^6 \frac{\pa^2}{\pa \tau^2}\sum_{A=1,2}
\left[P_A^{\tau}r_A - \epsilon^2 \frac{r_A^kD_A^k}{r_A} + 
\epsilon^4 \frac{\delta^{kl} -n_A^kn_A^l}{r_A}I_A^{kl}
\right] \nonumber \\
\mbox{} && - \frac{2}{3} 
\epsilon^7 \frac{\pa^3}{\pa \tau^3}\sum_{A=1,2}
\left[
P_A^{\tau}r_A^2 - 2 \epsilon^2 r_A^k D_A^k 
\right] 
+ \frac{1}{6} 
\epsilon^8 \frac{\pa^4}{\pa \tau^4}\sum_{A=1,2}
\left[
P_A^{\tau}r_A^3 - 3 \epsilon^2 r_A r_A^k D_A^k 
\right] 
\nonumber \\
\mbox{} &&- \frac{1}{30} 
\epsilon^9 \frac{\pa^5}{\pa \tau^5}\sum_{A=1,2}
\left[
P_A^{\tau}r_A^4 \right] 
+ \frac{1}{180} 
\epsilon^{10} \frac{\pa^6}{\pa \tau^6}\sum_{A=1,2}
\left[
P_A^{\tau}r_A^5 \right] 
 + O(\epsilon^{11}), 
\label{hBttInAp} 
\end{eqnarray} 
\begin{eqnarray}
h^{\tau i}_B &=& 4 \epsilon^4 \sum_{A=1,2}
\left[
\frac{P_A^{\tau}v_A^i }{r_A}
+ \epsilon^2 \left(
\frac{1}{r_A}\frac{dD_A^{i}}{d\tau} 
+  \frac{M_A^{ki} r^k_A}{2 r_A^3} 
+ \frac{v_A^{(k}D_A^{i)} r^k_A}{r_A^3} 
\right)
\right.
\nonumber \\
\mbox{}&&+ \left.
\epsilon^4 \left(\frac{\mbox{}_4Q_A^i}{r_A}
+ \frac{r^k_A}{2 r_A^3}\frac{dI_A^{ki}}{d\tau}
+ \frac{3 v_A^{(i}I_A^{kl)} r_A^{<kl>}}{2r_A^5}  
+ \frac{2 J_A^{k[li]} r_A^{<kl>}}{r_A^5}
\right)
\right]
\nonumber \\
\mbox{}&& 
-4 \epsilon^5 \frac{\pa }{\pa\tau}\sum_{A=1,2}
\left[
P_A^{\tau}v_A^i + \epsilon^2\frac{dD_A^i}{d\tau}
\right]
\nonumber \\
\mbox{}&&+ 
 2 
\epsilon^6 \frac{\pa^2}{\pa \tau^2}\sum_{A=1,2}
\left[P_A^{\tau}r_A v_A^i -  
\epsilon^2 \left(
\frac{r_A^kM_A^{ki}}{2 r_A}
+ \frac{r_A^kv_A^{(k}D_A^{i)}}{r_A}
\right) 
\right] \nonumber \\
\mbox{} && - \frac{2}{3} 
\epsilon^7 \frac{\pa^3}{\pa \tau^3}\sum_{A=1,2}
\left[
P_A^{\tau}r_A^2 v_A^i 
\right] 
+ \frac{1}{6} 
\epsilon^8 \frac{\pa^4}{\pa \tau^4}\sum_{A=1,2}
\left[
P_A^{\tau}r_A^3v_A^i \right] 
 + O(\epsilon^9),
\label{hBtiInAp} 
\end{eqnarray} 
\begin{eqnarray}
h^{ij}_B &=& 4 \epsilon^4 \sum_{A=1,2}
\left[
\frac{P_A^{\tau}v_A^{i}v_A^j}{r_A}
\right.
\nonumber \\
\mbox{} &&+ \left. 
 \epsilon^2 
\left(
 \frac{2v_A^{(i}}{r_A}\frac{dD_A^{j)}}{d\tau}
+ \frac{D_A^{(i}}{r_A}\frac{dv_A^{j)}}{d\tau}
+ \frac{M_A^{k(i}v_A^{j)} r^k_A}{r_A^3} 
+ \frac{D_A^{(i}v_A^{j)} v_A^kr^k_A}{r_A^3} 
+ \frac{Z_A^{k[li]j}r_A^{<kl>}}{r_A^5}
+ \frac{Z_A^{k[lj]i}r_A^{<kl>}}{r_A^5}
\right)
\right.
\nonumber \\
\mbox{} &&+ \left. 
\epsilon^4 
\left(
\frac{1}{2r_A}\frac{d^2I_A^{ij}}{d\tau^2} 
+ \frac{v_A^k r^k_A}{2 r_A^3}\frac{dI_A^{ij}}{d\tau}
+ \frac{r_A^k}{2r_A^3}\frac{d}{d\tau}\left(v_A^{(k}I_A^{ij)}\right)
- \frac{2r_A^k}{3r_A^3}\left(
\frac{d J_A^{i[jk]}}{d\tau} + \frac{d J_A^{j[ik]}}{d\tau}
\right)
\right.\right.
\nonumber \\
\mbox{} &&+ \left.\left. 
 \frac{r_A^{<kl>}}{6r_A^5}
\left( 2 v_A^kv_A^lI_A^{ij}+ v_A^iv_A^jI_A^{kl}+ 
6v_A^kv_A^{(i}I_A^{j)l}
\right)
\right.\right.
\nonumber \\
\mbox{} &&+ \left.\left. 
 \frac{r_A^{<kl>}}{6r_A^5}
\left(v_A^kJ_A^{i[lj]}+v_A^kJ_A^{j[li]}+2v_A^iJ_A^{k[lj]}
+2v_A^jJ_A^{k[li]}
\right)
\right.\right.
\nonumber \\
\mbox{} &&+ \left.\left. 
 \frac{15 Z_A^{kl[mi]j}r_A^{<klm>}}{8r_A^7}
+ \frac{15 Z_A^{kl[mj]i}r_A^{<klm>}}{8r_A^7}
+ \frac{\mbox{}_4Q_A^{(i}v_A^{j)}}{r_A}
\right.\right.
\nonumber \\
\mbox{} &&+ \left.\left. 
\frac{5\mbox{}_4R_A^{klm(ij)}r_A^{<klm>}}{8r_A^7}
+ \frac{\mbox{}_4R_A^{(ij)}}{r_A}
+ \frac{r^k_A}{2 r_A^3}(\mbox{}_4R_A^{kji} + 
\mbox{}_4R_A^{kij} - 
\mbox{}_4R_A^{ijk} ) 
 \right)
 \right]
\nonumber \\
\mbox{} &&- 
4 \epsilon^5 \frac{\pa}{\pa \tau}
\sum_{A=1,2}\left[
P_A^{\tau}v_A^iv_A^j 
+ \epsilon^2\left(
2 v_A^{(i}\frac{dD_A^{j)}}{d\tau}
+ D_A^{(i}\frac{dv_A^{j)}}{d\tau}
\right)
\right]
\nonumber \\
\mbox{} &&+ 
2 \epsilon^6 \frac{\pa^2}{\pa \tau^2}
\sum_{A=1,2}\left[
P_A^{\tau}v_A^iv_A^j r_A 
+\epsilon^2\left(
2 r_A v_A^{(i}\frac{dD_A^{j)}}{d\tau}
+ r_A D_A^{(i}\frac{dv_A^{j)}}{d\tau} 
- \frac{M_A^{k(i}v_A^{j)} r^k_A}{r_A} 
\right.\right.
\nonumber \\
\mbox{} && \left.\left.
- \frac{D_A^{(i}v_A^{j)} v_A^kr^k_A}{r_A} 
+ \frac{\delta^{kl}-n_A^{kl}}{3 r_A}Z_A^{k[li]j}
+ \frac{\delta^{kl}-n_A^{kl}}{3 r_A}Z_A^{k[lj]i}
\right) 
\right] 
\nonumber \\
\mbox{} &&- 
 \frac{2}{3}\epsilon^7
\frac{\pa^3}{\pa \tau^3}
\sum_{A=1,2}\left[P_A^{\tau}v_A^iv_A^jr_A^2\right]
+ \frac{1}{6}\epsilon^8\frac{\pa^4}{\pa \tau^4}
\sum_{A=1,2}\left[
P_A^{\tau}v_A^iv_A^jr_A^3
\right]
\nonumber \\
\mbox{} &&+  O(\epsilon^9).
 \label{hBijInAp}
\end{eqnarray}
We restrict our attention to 
the equation of motion for two spherical compact stars in  
this paper, however our formulation can be extended to  
an extended body with higher multipole moments, as shown in 
paper I.

\section{Derivation of Eq. (3.31)}
\label{sec:proofeq331}

In this section, we show a derivation of Eq. (\ref{NBcontribution}) 
without using the Dirac delta distribution. The following proof 
is essentially due to \cite{Terasawa}.

Suppose that a two-dimensional surface $S$ surrounds a 
three-dimensional volume $V$. Suppose that there is a point $P(\vec x)$ 
and a point $Q(\vec y)$ both inside of $V$, and the distance 
between the two points is $r$. Thus,
$$
r=|\vec x- \vec y|.
$$
Note that $\nabla^2(1/r)=0$ except for $r=0$.

Define a sphere $V'$ which is centered at $P(\vec x)$ 
and has a radius $d$ that is sufficiently small so that 
$V'$ is enclosed completely by $V$. The surface $\Sigma$ 
of $V'$ divides $V$ into two regions. We call the outer region $V''$.

The Green's theorem (e.g., \cite{Morse53}) 
states for a certain function $v=v(\vec x)$ that  
\begin{eqnarray}
\int_{V''}\frac{\nabla^2 v(\vec y)}{r}d^3y
&+& \oint_{S}\left\{
\frac{1}{r}\frac{\partial v(\vec y)}{\partial y^i}
-v(\vec y)\frac{\partial}{\partial y^i}\left(\frac{1}{r}\right)
\right\}dS_i
\nonumber \\
\mbox{} 
&+& 
\oint_{\Sigma}\left\{
\frac{1}{r}\frac{\partial v(\vec y)}{\partial y^i}
-v(\vec y)\frac{\partial}{\partial y^i}\left(\frac{1}{r}\right)
\right\}dS_i = 0.
\end{eqnarray}
Note that $V''$ does not include the point  $P(\vec x)$. 
We assume here that for $v(\vec x)$ the second derivative with respect 
to $\vec x$ exists and is continuous in $V''$.  In the end of 
this section, we mention  
whether this assumption holds for our particular example, 
Eq. (\ref{NBcontribution}).

The third term can be evaluated as 
\begin{eqnarray}
\oint_{\Sigma}\left\{
\frac{1}{r}\frac{\partial v(\vec y)}{\partial y^i}
-v(\vec y)\frac{\partial}{\partial y^i}\left(\frac{1}{r}\right)
\right\}dS_i &=& 
d \oint_{\Sigma}\frac{\partial v(\vec y)}{\partial y^i}n_id\Omega 
+ \oint_{\Sigma} v(\vec y)d\Omega, 
\end{eqnarray} 
where $n_i$ is the outward normal of the surface $\Sigma$.
If $\partial v(\vec y)/\partial y^i$ is finite in the 
$d \rightarrow 0$ limit, then the first term is zero in this 
limit. Thus, 
\begin{eqnarray}
\oint_{\Sigma}\left\{
\frac{1}{r}\frac{\partial v(\vec y)}{\partial y^i}
-v(\vec y)\frac{\partial}{\partial y^i}\left(\frac{1}{r}\right)
\right\}dS_i &=& 
4\pi v(\vec x), 
\end{eqnarray} 
in the $d \rightarrow 0$ limit. 

In the same manner, 
\begin{eqnarray}
\lim_{d  \rightarrow 0}
\int_{V'}\frac{\nabla^2 v(\vec y)}{r}d^3y 
&=&
\lim_{d \rightarrow 0}
\int_{V'}\nabla^2 v(\vec y)rdrd\Omega = 0.
\end{eqnarray}
Then we have  
\begin{eqnarray}
\int_V\frac{\nabla^2 v(\vec y)}{r}d^3y 
+ \oint_{S}\left\{
\frac{1}{r}\frac{\partial v(\vec y)}{\partial y^i}
-v(\vec y)\frac{\partial}{\partial y^i}\left(\frac{1}{r}\right)
\right\}dS_i 
+ 4\pi v(\vec x)=0.
\end{eqnarray}
This is essentially Eq. (\ref{NBcontribution}).

In the particular case of Eq. (\ref{NBcontribution}), $V=N/B$. 
We recall that the stars are completely 
enclosed by the body zones $B_A$ and thus there exists no matter 
in $N/B$. The integrand $v(\vec x)$  
or $f(x)$ and $g(x)$ in Eq. (\ref{NBcontribution}) have thus no singularity 
in $N/B$ even in the point-particle limit. 
These functions have a smooth second derivative with 
respect to $\vec x$ in $N/B$.

\section{Correction to the multipole moments}
\label{Appendix:CorrectionToMoments}

A natural reference coordinate where we would define  
multipole moments of a star may be a coordinate in which effects 
of its orbital motion and the companion star 
are removed (modulo, namely, the tidal effect). 
In other words, such a natural reference coordinate may be 
the generalized Fermi coordinate \cite{AB86}, 
in which the metric is Lorentzian at $z_A^i(\tau)$.  
We define the two stars to be spherical in such a coordinate
in this paper \cite{PW02}.

In paper II, we defined a multipole moment, which  
we call the NZC moment in this section, as a volume integral  
over the body zone which is spherical in the near zone 
coordinate (NZC).
Then a question specific to our formalism is whether  
the NZC moments affect the 
orbital motion of the stars which are spherical in the  
generalized Fermi coordinate (GFC).

At lowest order ($O(\epsilon^0)$), 
the NZC multipole moments and the GFC multipole 
moments defined as volume integrals over 
a sphere in GFC 
must be the same. 
For a spherically symmetric star, the NZC moments 
and the GFC moments with trace-free operation or 
antisymmetrization on their indexes vanish 
at the lowest order.
We further assume that the GFC moments and 
 the NZC moments are zero for a spherical compact 
star at the lowest order because of its compactness.

For the mass monopole, 
we have already obtained the relation between the 
NZC monopole (the energy) and the mass 
via the evolution equation of the energy. 
Thus, we seek the 1PN correction to the NZC $L$th 
multipole moments with $L \ge 2$ 
since physically relevant multipole moments appear at 
2PN order and we are concerned with the 3PN equation 
of motion in this paper. 
In the following, we compute the 
corrections which do not include higher order multipole moments 
than the energy monopole. 

As for construction of GFC, 
we assume, 
up to the relevant order here, that the transformation from 
NZC to GFC or vice versa takes apparently the same form for 
a strongly self-gravitating star in which we are interested here 
and for a weakly self-gravitating star for which GFC has been 
constructed in \cite{AB86}. This is because 
the construction of the generalized Fermi coordinate depends 
on how the star moves, and because the equations of motion for  
binary stars take the same form regardless of 
the strength of 
the stars' internal gravity up to 2.5PN order (paper II). 
An important difference 
is that 
the mass parameter in the transformation from NZC to GFC 
for the strongly self-gravitating star includes the strong field 
effect, as explained below Eq. (\ref{DefOfMass}).

We now assume the following coordinate transformations 
from GFC  $(\hat \tau,\hat  x^k)$ to NZC $(\tau, x^k)$: 
\begin{eqnarray}
y^i &=& 
z_A^i(\tau) + 
\hat y^i_A  + 
\delta y^i(\tau,\hat y_A^k),  
\label{LC:NCtoGFofY} 
\\
\tau - \tau_P &=& \hat \tau - \hat \tau_P + 
\delta \tau(\tau,\hat y_A^i), 
\label{LC:NCtoGFofT}
\end{eqnarray}
with 
\begin{eqnarray}
\delta y^i(\tau,\hat y_A^k) 
&=& 
\sum_{n=1}A^{iJ_n}(\tau)\hat y_A^{J_n},
\\
\delta \tau(\tau,\hat y_A^k) 
&=& 
\sum_{n=0}B^{J_n}(\tau)\hat y_A^{J_n},
\end{eqnarray}
where the capital index denotes a set of collective indexes, 
e.g.,  $A^{iJ_n} = A^{i j_1\cdot\cdot\cdot j_n}$.
$\hat y_A^i = \hat y^i - \hat z_A^i$.
In Eq. (\ref{LC:NCtoGFofY}), $\hat y_A^i$ is on the 
$\hat \tau = $ const surface. $\tau_P$ and $\hat \tau_P$ 
are fiducial time coordinates. 
We assume that $A^{iJ_n}, B^{J_n} = O(\epsilon^2)$.

The NZC moments are defined 
on $B_A(\tau_P)$  which is 
a sphere centered at ($\tau_P$,$z_A^i(\tau_P))$ 
with radius $\epsilon R_A$ in NZC,   
while the GFC moments are defined 
on $\hat B_A(\hat \tau_P)$ which 
is centered at the same event 
($\hat \tau_P$,$\hat z_A^i$) in GFC 
(($\tau_P$,$z_A^i(\tau_P))$  in NZC)  
with the radius $\epsilon R_A$ in GFC.
The corrections we shall compute are then 
\begin{eqnarray}
\epsilon^{2 l + 4 - s} \delta I_A^{J_l\mu\nu}
&\equiv& 
\epsilon^{2 l + 4 - s}  I_{A,{\rm NZC}}^{J_l\mu\nu}
- \epsilon^{2 l + 4 - s}  I_{A,{\rm GFC}}^{J_l\mu\nu}, 
\\ 
\epsilon^{2 l + 4 - s}  I_{A,{\rm NZC}}^{J_l\mu\nu} 
&\equiv& \int_{B_A(\tau = \tau_P)} 
d^3y_A y_A^{J_l}(\tau)
\Lambda^{\mu\nu}_N(\tau,y^i),
\label{def:NZCmmoments}
\\
\epsilon^{2 l + 4 - s}  I_{A,{\rm GFC}}^{J_l\mu\nu} 
&\equiv& \int_{\hat B_A (\hat \tau = \hat \tau_P)} d^3\hat y_A 
\hat y_A^{J_l}
\hat \Lambda^{\mu\nu}_{G'}(\hat \tau,\hat y^i),
\end{eqnarray}
where $s = 2$ for $(\mu,\nu) = (i,j)$ or $0$ otherwise.
$\hat \Lambda^{\mu\nu}_{G'}(\hat \tau,\hat y^i) = 
\hat \Lambda^{\mu\nu}_G(\hat \tau,\hat y_A^i) 
$.

We now express  $I_{A,{\rm NZC}}^{J_l\mu\nu}$ using the generalized 
Fermi coordinate. Note that $y_A^{J_l}(\tau)$ 
in Eq. (\ref{def:NZCmmoments}) is on the $\tau=\tau_P = $ const  
surface. A necessary coordinate transformation that  
relates $y_A^i(\tau)$ 
on the $\tau =$ const surface 
with $\hat y_A^i$ 
on the $\hat \tau =$ const surface can be obtained  
by modifying Eq. (\ref{LC:NCtoGFofY}) using retardation 
expansion. Up to 1PN order, the result is  
\begin{eqnarray}
y^i &=& z_A^i(\tau) + 
\hat {y}^i_A
- \tilde \delta y^i(\tau_P,\hat y_A^k), 
\label{LC:NCtoGFPrimeofY} 
\end{eqnarray}
with 
\begin{eqnarray}
\tilde \delta y^i(\tau_P,\hat y_A^k) 
&=& 
\sum_{n=1}
\left(
B^{J_n}(\tau_P)v_A^i(\tau_P)
- A^{iJ_n}(\tau_P)
\right)
\hat {y}_A^{J_n}.
\end{eqnarray}

Using tetrad $e^{\mu}\mbox{}_{\hat \alpha}(\hat \tau,\hat y^i) = 
\delta^{\mu}\mbox{}_{\hat \alpha} + O(\epsilon^2)$,  
we have 
\begin{eqnarray}
\epsilon^{2 l + 4 - s}  I_{A,{\rm NZC}}^{J_l\mu\nu} 
&=& 
\int_{
\tilde B_A\left(
\hat \tau = \hat \tau(\hat \tau_P,\hat y_A^i)
\right)
} 
d^3\hat y_A 
\left|\frac{\pa (y_A^i)}{\pa (\hat y_A^j)}\right|
\nonumber \\ 
\mbox{} &&\times 
\prod^l_{k=1}
\left\{
\hat y_A^{j_k} - 
\delta \tilde y^{j_k}(\tau,\hat y_A^i) 
\right\}
\nonumber \\ 
\mbox{} &&\times
e^{\mu}\mbox{}_{\hat \alpha}(\hat \tau,\hat y_A^i) 
e^{\nu}\mbox{}_{\hat \beta}(\hat \tau,\hat y_A^i) 
\hat \Lambda^{\alpha\beta}_{G'}(\hat \tau,\hat y^i) 
\nonumber \\
\mbox{} &=&
\int_{
\tilde B_A\left(
\hat \tau = \hat \tau(\hat \tau_P,\hat y_A^i)
\right)
} 
d^3\hat y_A 
\hat y_A^{J_l}
\hat \Lambda^{\mu\nu}_{G'}(\hat \tau,\hat y^i),
\label{LC:NZCmom1}
\end{eqnarray}
where in the last equality 
we neglected all the corrections that result in   
the multipole moments of order $L' \ge L$.
The integral region 
$\tilde B_A\left(
\hat \tau = \hat \tau(\hat \tau_P,\hat y_A^i)
\right)$ 
which corresponds to $B_A(\tau = \tau_P = {\rm const})$
is not a $\hat \tau = \hat \tau_P$ const 3-surface 
nor spherical in GFC. In fact, from Eq. (\ref{LC:NCtoGFofT}),  
\begin{eqnarray}
\hat \tau(\hat \tau_P,\hat y_A^i) &=& \hat \tau_P 
- 
\delta \tau(\tau_P,\hat y_A^i). 
\label{LC:tauhattotauP}
\end{eqnarray}
Thus we make a retardation expansion around $\hat \tau = \hat
\tau_P$ in the integrand. Then to make the slightly 
perturbed sphere 
$\tilde B_A\left(\hat \tau = \hat \tau_P\right)$ 
into the sphere 
$\hat B_A\left(\hat \tau = \hat \tau_P\right)$ 
with radius $\epsilon R_A$, 
we change the integration variable 
$\hat y_A^i$ into $\check y_A^i$. 
Up to 1PN order, the transformation may be obtained by 
inverting Eq. (\ref{LC:NCtoGFPrimeofY}),  
\begin{eqnarray}
\hat y^i &=& \hat z_A^i(\hat \tau_P) + 
\check{y}^i_A
+ \tilde \delta y^i(\tau_P,\check y_A^k).  
\label{LC:GFtoNCPrimeofY} 
\end{eqnarray}

Using Eq. (\ref{LC:tauhattotauP}) and 
Eq. (\ref{LC:GFtoNCPrimeofY}),  
we simplify Eq. (\ref{LC:NZCmom1}) up to 1PN order as 
\begin{eqnarray}
\lefteqn{\epsilon^{2 l + 4 - s}  I_{A,{\rm NZC}}^{J_l\mu\nu}} \nonumber \\
&=& 
\int_{
\tilde B_A\left(\hat \tau_P
\right)
} 
d^3\hat y_A 
\hat y_A^{J_l}
\left\{
\hat \Lambda^{\mu\nu}_{G'}
\left(
\hat \tau_P - 
\delta \tau(\tau_P,\hat y_A^i) 
,\hat y^i
\right) 
\right\}
\nonumber \\
&=& 
\int_{
\tilde B_A\left(\hat \tau_P
\right)
} 
d^3\hat y_A 
\hat y_A^{J_l}
\left\{
\hat \Lambda^{\mu\nu}_{G'}
(\hat \tau_P,\hat y^i) 
- \delta \tau(\tau_P,\hat y_A^i) 
\frac{\pa}{\pa \hat \tau_P}
\hat \Lambda^{\mu\nu}_{G'}
(\hat \tau_P,\hat y^i) 
\right\}
\nonumber \\
&=& 
\int_{\hat B_A(\hat \tau_P)}  d^3\check y_A 
\left|\frac{\pa (\hat y_A^i)}{\pa (\check y_A^j)}\right|
\prod^l_{k=1}
\left\{
\check y_A^{j_k} + 
\tilde \delta y^{j_k}(\tau_P,\check y_A^i) 
\right\}
\nonumber \\
&&\times
\left\{
\hat \Lambda^{\mu\nu}_{G}
\left(
\hat \tau_P,\check y_A^i  
+ \tilde \delta y^i(\tau_P,\check y_A^i) 
\right) 
- \delta \tau(\tau_P,\check y_A^i) 
\frac{\pa}{\pa \hat \tau_P}
\hat \Lambda^{\mu\nu}_{G'}
(\hat \tau_P,\check y^i) 
\right\}
\nonumber \\
&=& 
\int_{\hat B_A(\hat \tau_P)} 
d^3\check y_A 
\left\{
\check y_A^{J_l}
\hat \Lambda^{\mu\nu}_{G'}
\left(
\hat \tau_P,\check y^i 
\right) 
+ \tilde \delta y^m(\tau_P,\check y_A^i) 
\frac{\pa}{\pa \check y^m}
\left(
\check y_A^{J_l}
\hat \Lambda^{\mu\nu}_{G'}
\left(\hat \tau_P,\check y^i \right) 
\right)
\right.
\nonumber \\
&&- \left.
\check y_A^{J_l}
\delta \tau(\tau_P,\check y_A^i) 
\frac{\pa}{\pa \hat \tau_P}
\hat \Lambda^{\mu\nu}_{G'}
(\hat \tau_P,\check y^i) 
\right\}.
\end{eqnarray}
As before, we have discarded terms which end up with multipole 
moments of order $L' \ge L$. 
Integrating by parts and rewriting $\check y^i$ by $\hat y^i$, we 
finally obtain a formula for  $\delta I_A^{J_l\mu\nu}$, 
\begin{eqnarray}
\lefteqn{\epsilon^{2 l + 4 - s}  
\delta I_{A}^{J_l\mu\nu}} 
\nonumber \\ 
&=& 
\sum_{n=1}
\left(
B^{K_n}(\tau_P)v_A^m(\tau_P) -
A^{mK_n}(\tau_P)
\right)
\oint_{\pa \hat B_A(\hat \tau_P)}
d\hat S_m 
\hat y_A^{J_l}
\hat y_A^{K_n}
\hat \Lambda^{\mu\nu}_{G'}(\hat \tau_P,\hat y^i)
\nonumber \\
&&- 
\sum_{n=1}
B^{K_n}(\tau_P)
\frac{d}{d \hat \tau_P}
\int_{\hat B_A(\hat \tau_P)} d^3\hat y_A
\hat y_A^{J_l}
\hat y_A^{K_n}
\hat \Lambda^{\mu\nu}_{G'}(\hat \tau_P,\hat y^i)
\nonumber \\
&&-
\sum_{n=1}
\left(
B^{K_n}(\tau_P)v_A^m(\tau_P) -
A^{mK_n}(\tau_P)
\right)
\nonumber \\
&&\times
\int_{\hat B_A(\hat \tau_P)} d^3\hat y_A
\frac{\pa \hat y_A^{K_n}
}{\pa \hat y^m}
\hat y_A^{J_l}
\hat \Lambda^{\mu\nu}_{G'}(\hat \tau_P,\hat y^i).
\label{LC:CorrectionToMultipole}
\end{eqnarray}
Notice that $d \hat z_A^i(\hat \tau_P)/d\hat \tau_P = 0$.

The last two volume integrals in the above equation 
result in the 
multipole moments and we neglect them here. As for the first term, 
we can evaluate the surface integral to 1PN order 
explicitly by 
replacing all the hatted quantities by those without a hat 
(and $G'$ by $N$). 
We found that $\mbox{}_n\Lambda^{\mu\nu}_N$ ($n \le 7$) do not 
contribute for any $L$. 
On the other hand, 
$\mbox{}_8\Lambda^{\tau\tau}_N$ for $L=2$ and $n=1$ in 
the summation in the first term of 
Eq. (\ref{LC:CorrectionToMultipole})  
gives a monopole correction 
to the 3PN gravitational field 
(through 
the 1PN correction to the quadrupole 
moment which itself appears at 2PN order in the field),     
\begin{eqnarray}
\oint_{\pa \hat B_A(\hat \tau_P)}
d\hat S_m 
\hat y_A^{i}\hat y_A^{j}
\hat y_A^{k}
\hat \Lambda^{\tau\tau}_{G'}(\hat \tau_P,\hat y^i)
&=& 
\oint_{\pa B_A(\tau_P)}
dS_m 
y_A^{i}y_A^j
y_A^{k}
\Lambda^{\tau\tau}_{N}(\tau_P,y^i)(1 + O(\epsilon^2))
\nonumber \\
\mbox{} &=&
\epsilon^8 \oint_{\pa B_A(\tau_P)}
dS_m 
y_A^{i}y_A^j
y_A^{k}
\mbox{}_8\Lambda^{\tau\tau}_{N}(\tau_P,y^i) 
+ O(\epsilon^{10})
\nonumber \\
\mbox{} &=&
- \epsilon^8\frac{4m_A^3}{5}\left(
\delta^{ij}\delta^{k}_m + 
\delta^{kj}\delta^{i}_m + 
\delta^{ik}\delta^{j}_m
\right).
\label{LC:GeneralCorrection}
\end{eqnarray} 
Neither 
$\mbox{}_8\Lambda_N^{\mu\nu}$ for $L\ge3$ 
nor $n \ge 2$ contribute to the 3PN field.

The coefficients $A^{mk}(\tau_P)$ and $B^{k}(\tau_P)$ 
may be read off 
from the results in \cite{AB86,BK89}, which are 
\begin{eqnarray}
B^{i}(\tau_P)v_1^j(\tau_P) 
&=& \epsilon^2 v_1^iv_1^j
+ O(\epsilon^3)  
\\
A^{ij}(\tau_P) &=& \epsilon^2\left(
\frac{1}{2}v_1^iv_1^j - \frac{m_2}{r_{12}}\delta^{ij}
\right) 
+ O(\epsilon^3)  
\end{eqnarray}
for the star $A = 1$. Thus, the coefficient in front 
of the surface integral, Eq. (\ref{LC:GeneralCorrection}),  
becomes $\epsilon^2(v_1^iv_1^j/2 + \delta^{ij}m_2/r_{12})$.

Finally using the coefficients above, 
we obtain the 1PN corrections of the multipole moments 
that affect a 3PN equation of motion for spherical stars 
(we rename $\delta I_A^{ij\tau\tau}$ by
$\delta I_A^{ij}$), 
\begin{eqnarray}
\delta I_1^{ij} 
&=&   
\epsilon^2\left(
-  \frac{4 m_1^3}{5}v_1^{i}v_1^j
-  \frac{2 m_1^3 v_1^2}{5}\delta^{ij}
-  \frac{4 m_1^3 m_2}{r_{12}} \delta^{ij}
\right)
+ O(\epsilon^3).
\label{LC:CorrectionToQuadrupole}
\end{eqnarray}
A similar equation holds for the star $A = 2$ by exchanging 
$1 \leftrightarrow 2$ in the above equation.
Notice that since the quadrupole 
moments at 2PN order is symmetric-trace-free, 
the last two terms 
do not contribute to the 3PN field. 

The correction $\delta I_1^{<ij>}$ 
should appear 
even when $m_2 \rightarrow 0$. Not surprisingly, 
with the correction $\delta I_1^{<ij>}$, the 3PN gravitational 
field for a single star moving at a constant velocity 
derived by solving the harmonically relaxed 
Einstein equations iteratively agrees   
with the boosted Schwarzschild metric in the harmonic coordinate 
up to 3PN order.

It is possible to use Eq. (\ref{LC:CorrectionToMultipole}) 
for the $L=1$ case and the result becomes again the 3PN correction 
to the field
(we again rename $\delta I_1^{i\tau\tau}$ as $\delta D_1^i$), 
\begin{eqnarray}
 \delta D_1^i = \epsilon^4 \frac{2 m_1^3 m_2}{r_{12}^3}r_{12}^i.  
\label{truedipolemoments}
\end{eqnarray}
(In the computation of $\delta D_1^i$, 
the surface integral with 
$\mbox{}_8\chi^{\tau\tau\alpha\beta}_N\mbox{}_{,\alpha\beta}$ 
as the integrand is found to vanish.)
However, any change of the dipole moment amounts merely 
to a redefinition of the representative point of the star and 
it causes no physical effect. 
In fact, we have not taken into account $\delta D_1^i$ to 
derive our 3PN equation of motion, Eq. (\ref{3PNEOMFinal}).

\section{Renormalization of the multipole moments}
\label{sec:renommultipole}

This section explains the ``renormalization'' 
of the multipole moments and that the field does not 
depend on $\epsilon R_A$. This is rather trivial, 
however we show this section for clarity.

We first define a symbol 
$\mathop{{\rm part}}_{\epsilon R_A} F$ which 
is the $\epsilon R_A$-dependent part in an expression $F$ 
except for the logarithmic dependence of $\epsilon R_A$. 
Correspondingly, we define 
$\mathop{{\rm disc}}_{\epsilon R_A} F$,   
which means to discard all the $\epsilon R_A$-dependent terms 
in $F$ other than $\ln \epsilon R_A$.
By construction, 
\begin{eqnarray}
F &=& 
\mathop{{\rm disc}}_{\epsilon R_A}F +
\mathop{{\rm part}}_{\epsilon R_A}F.
\end{eqnarray} 
We give an example of $\mathop{{\rm disc}}_{\epsilon R_A}$ in 
Sec. \ref{ExplanationforCompStarInt}.

Now, to derive the field, we study the following Poisson 
integral for a certain function $f(\vec x)$. 
$f(\vec x)$ is some combination of $\Lambda^{\mu\nu}(\tau,\vec x)$ 
and is assumed to be nonsingular in $N$. 
For notational simplicity, we do not write time dependence explicitly 
in $f(\vec x)$, 
\begin{eqnarray}
\int_{N}\frac{d^3y}{|\vec x - \vec y|}f(\vec y)
&=& 
\sum_{A=1,2}
\int_{B_A}\frac{d^3y}{|\vec x - \vec y|}f(\vec y)
+
\int_{N/B}\frac{d^3y}{|\vec x - \vec y|}f(\vec y).
\label{eq:splittheintegral}
\end{eqnarray}

The second volume integral is evaluated with the help of 
the superpotential $g(\vec x)$ 
that satisfies $\Delta g(\vec x) = f(\vec y)$ in $N/B$. 
Using Eq. (\ref{NBcontribution}) 
and expanding the kernel $1/|\vec x - \vec y|$ 
around $\vec y_A=\vec y-\vec z_A$, we obtain  
\begin{eqnarray}
\int_{N/B}\frac{d^3y}{|\vec x - \vec y|}f(\vec y)
&=& 
-
\sum_{A=1,2}\sum_{n=0}\frac{(2n-1)!!r_A^{<K_n>}}{n!r_A^{2n+1}}
\oint_{B_A}dS_k y_A^{K_n}
\frac{\partial g(\vec y_A+\vec z_A)}{\partial y_A^k} 
\nonumber \\
\mbox{} &+&
\sum_{A=1,2}\sum_{n=0}\frac{(2n-1)!!r_A^{<K_n>}}{n!r_A^{2n+1}}
\oint_{B_A}dS_k g(\vec y_A+\vec z_A)
\frac{\partial y_A^{K_n}}{\partial y_A^k} 
\nonumber \\
\mbox{} &+&
 \cdots.
\label{eq:RAdependenceinNB}
\end{eqnarray} 
Here we only show explicitly the terms which possibly depend on 
$\epsilon R_A$, and 
``$\cdots$'' denotes $\epsilon R_A$-independent terms.
Thus, using the symbols introduced above, we have 
\begin{eqnarray}
\int_{N/B}\frac{d^3y}{|\vec x-\vec y|}f(\vec y)
&=&
\mathop{{\rm disc}}_{\epsilon R_A}
\int_{N/B}\frac{d^3y}{|\vec x-\vec y|}f(\vec y)
+
\mathop{{\rm part}}_{\epsilon R_A}
\int_{N/B}\frac{d^3y}{|\vec x-\vec y|}f(\vec y),
\end{eqnarray}
with 
\begin{eqnarray}
\lefteqn{\mathop{{\rm part}}_{\epsilon R_A}
\int_{N/B}\frac{d^3y}{|\vec x-\vec y|}f(\vec y)}\nonumber \\
&=& 
\mathop{{\rm part}}_{\epsilon R_A}\left[
-\sum_{A=1,2}\sum_{n=0}\frac{(2n-1)!!r_A^{K_n}}{n!r_A^{2n+1}}
\oint_{B_A}dS_k y_A^{K_n}
\frac{\partial g(\vec y_A+\vec z_A)}{\partial y_A^k} 
\right.
\nonumber \\
\mbox{} &+&\left.
\sum_{A=1,2}\sum_{n=0}\frac{(2n-1)!!r_A^{K_n}}{n!r_A^{n+1}}
\oint_{B_A}dS_k g(\vec y_A+\vec z_A)
\frac{\partial y_A^{K_n}}{\partial y_A^k} 
\right].
\end{eqnarray}

For the first volume integral in Eq. (\ref{eq:splittheintegral}), 
we use the multipole expansion. 
\begin{eqnarray}
\sum_{A=1,2}
\int_{B_A}\frac{d^3y}{|\vec x - \vec y|}f(\vec y)
&=&
\sum_{A=1,2}
\sum_{n=0}
\frac{(2n-1)!! r_A^{<K_n>} I_A^{K_n}}{n! r_A^{2n+1}},
\end{eqnarray}
with 
\begin{eqnarray}
I_A^{K_n} &\equiv&
\int_{B_A}d^3y_Af(\vec y_A+\vec z_A)y_A^{K_n}.
\label{eq:definemultipole}
\end{eqnarray}
$I_A^{K_n}$ is the multipole moment of the star $A$.
We simplify here the definition of the multipole moments 
so that we omit the scalings on the integrand 
$\Lambda^{\mu\nu}(\tau,\vec x)$ and hence $f(\vec x)$. Obviously, 
this definition is enough to study the $\epsilon R_A$ dependence 
in the multipole moments and the field.

$I_A^{K_n}$ in general depends on $\epsilon R_A$ since 
the integrand $f(\vec y)$ is noncompact support. A possible 
$\epsilon R_A$ dependence in $I_A^{K_n}$ may be examined by 
the following. First, since we are studying the nonsingular 
sources, we expect that $f(\vec x)$ is smooth so that we 
can assume $\Delta g(\vec x) = f(\vec x)$ just inside of 
$\partial B_A$. Thus, the $\epsilon R_A$ dependence in 
$I_A^{K_n}$ can be examined via 
\begin{eqnarray}
I_A^{K_n} &=& 
\int_{{\rm in~ the~ neighborhood~ of~ }\partial B_A}d^3y_A
\Delta g(\vec y_A+\vec z_A)
y_A^{K_n} + \cdots 
\nonumber \\
\mbox{} &=& 
\oint_{\partial B_A}dS_k
\left(
y_A^{K_n}
\frac{\partial g(\vec y_A+\vec z_A)}{\partial y_A^k}
-\frac{\partial y_A^{K_n}}{\partial y_A^k}
g(\vec y_A+\vec z_A)
\right) 
\nonumber \\
\mbox{} &+& 
\cdots.
\label{eq:RAdependenceinmoments}
\end{eqnarray}
Here ``$\cdots$'' again denotes terms that do not depend on 
$\epsilon R_A$.

Using the symbols introduced in this section, we have 
\begin{eqnarray}
I_A^{K_n} &=& 
\mathop{{\rm disc}}_{\epsilon R_A}I_A^{K_n}
+ 
\mathop{{\rm part}}_{\epsilon R_A}I_A^{K_n},
\\
\mathop{{\rm disc}}_{\epsilon R_A}I_A^{K_n}
&=& 
\mathop{{\rm disc}}_{\epsilon R_A}
\int_{B_A}d^3y_Af(\vec y_A+\vec z_A)y_A^{K_n},
\\
\mathop{{\rm part}}_{\epsilon R_A}I_A^{K_n}
\mbox{}&=& 
\mathop{{\rm part}}_{\epsilon R_A}
\int_{B_A}d^3y_Af(\vec y_A+\vec z_A)y_A^{K_n}
\nonumber \\
\mbox{} &=&
\mathop{{\rm part}}_{\epsilon R_A}\left[
\oint_{\partial B_A}dS_k
\left(
y_A^{K_n}
\frac{\partial g(\vec y_A+\vec z_A)}{\partial y_A^k}
-\frac{\partial y_A^{K_n}}{\partial y_A^k}
g(\vec y_A+\vec z_A)
\right)\right].
\label{eRAdependenceinMPM} 
\end{eqnarray}
We call $I_{A,r}^{K_n}\equiv\mathop{{\rm disc}}_{\epsilon R_A} I_A^{K_n}$ the 
``renormalized'' multipole moments that are independent of powers of 
$\epsilon R_A$ by definition of $\mathop{{\rm disc}}_{\epsilon R_A}$.

It is clear that the $\epsilon R_A$ dependences in 
the multipole moments Eq. (\ref{eRAdependenceinMPM})    
cancel out those of the $N/B$ contribution shown as the first 
and the second terms in Eq. (\ref{eq:RAdependenceinNB}).   
In conclusion,  we find that we can discard (or neglect) all the  
$\epsilon R_A$-independent terms other than $\ln \epsilon R_A$ terms 
when we compute the field, 
\begin{eqnarray}
\int_{N}\frac{d^3y }{|\vec x - \vec y|}f(\vec y)
&=& 
\sum_{A=1,2}
\int_{B_A}\frac{d^3y}{|\vec x - \vec y|}f(\vec y)
+
\int_{N/B}\frac{d^3y}{|\vec x - \vec y|}f(\vec y)
\nonumber \\
\mbox{} &=&
\sum_{A=1,2}
\sum_{n=0}
\frac{(2n-1)!!r_A^{<K_n>}I_{A}^{K_n}}{n!r_A^{2n+1}}+ 
\int_{N/B}\frac{d^3y f(\vec y)}{|\vec x-\vec y|}
\nonumber \\
\mbox{} &=&
\sum_{A=1,2}
\sum_{n=0}
\frac{(2n-1)!!r_A^{<K_n>}I_{A,r}^{K_n}}{n!r_A^{2n+1}}+ 
\mathop{{\rm disc}}_{\epsilon R_A}
\int_{N/B}\frac{d^3y f(\vec y)}{|\vec x-\vec y|}
\nonumber \\
\mbox{} &+& 
\sum_{A=1,2}
\sum_{n=0}
\frac{(2n-1)!!r_A^{<K_n>}}{n!r_A^{2n+1}}
\mathop{{\rm part}}_{\epsilon R_A} I_A^{K_n}
+ 
\mathop{{\rm part}}_{\epsilon R_A}
\int_{N/B}\frac{d^3y f(\vec y)}{|\vec x-\vec y|}
\nonumber \\
\mbox{} &=&
\sum_{A=1,2}
\sum_{n=0}
\frac{(2n-1)!!r_A^{<K_n>}I_{A,r}^{K_n}}{n!r_A^{2n+1}}+ 
\mathop{{\rm disc}}_{\epsilon R_A}
\int_{N/B}\frac{d^3y f(\vec y)}{|\vec x-\vec y|}.
\end{eqnarray} 
In the main body of this paper other than this section, 
we write $I_{A,r}^{K_n}$ 
as $I_{A}^{K_n}$ and omit the symbol 
$\mathop{{\rm disc}}_{\epsilon R_A}$ in front 
of the Poisson integral over $N/B$ for notational 
simplicity and from triviality of the fact that the 
total field is independent of $\epsilon R_A$.
As an exception, we write 
$\mathop{{\rm disc}}_{\epsilon R_A}$ in Secs.  
\ref{sec:derivatoinOfh8ti}
and \ref{TPNGravitationalField} 
to make our discarding $\epsilon R_A$ procedure clear.

Finally, we mention that it is straightforward to extend 
the arguments here to show that the cancellation of 
$\epsilon R_A$ terms between the body zone contribution 
and the $N/B$ contribution occurs for any retarded 
field, that is, $n\ge 1$ terms in Eq. (\ref{TotFieldRetExpandAbst}).

\section{$\chi$ Part}
\label{chipart}

We derive the functional expressions 
of $P_{A \chi}^{\mu}$ 
on $m_A, v_A^i$, and $r_{12}^i$. Here 
we defined $P_{A \chi}^{\mu}$ as 
\begin{eqnarray}
&&
P_{A \chi}^{\mu} \equiv \epsilon^{-4}\int_{B_A}d^3y
\chi^{\mu\tau\alpha\beta}_N\mbox{}_{,\alpha\beta}.
\end{eqnarray}
By the definition of $\chi^{\mu\nu\alpha\beta}_N\mbox{}_{,\alpha\beta}$,
$$
16 \pi \chi^{\tau\tau\alpha\beta}_N\mbox{}_{,\alpha\beta}
= (h^{\tau k}h^{\tau l} - h^{\tau\tau}h^{kl})_{,kl},
$$
$$
16 \pi \chi^{\tau i\alpha\beta}_N\mbox{}_{,\alpha\beta}
= (h^{\tau \tau}h^{i k} - h^{\tau i}h^{\tau k})_{,\tau k}
+ (h^{\tau k}h^{i l} - h^{\tau i}h^{kl})_{,kl},
$$
thus we can obtain the functional expressions of $P_{A \chi}^{\mu}$
using Gauss' law. In fact, up to 3PN order, the definition of 
$P_{A\chi}^{\tau}$ gives
\begin{eqnarray}
P_{1 \chi}^{\tau} &=& \epsilon^4\frac{m_1 m_2}{3 r_{12}}
\left[
4 V^2 +\frac{m_2}{r_{12}} - \frac{2 m_1}{r_{12}}\right]
\nonumber \\ 
\mbox{} &-& \epsilon^5
 \frac{2}{3}m_1\mbox{}^{(3)}I_{{\rm orb}}^k\mbox{}_k
\nonumber \\ 
\mbox{} &+& \epsilon^6 \frac{m_1m_2}{r_{12}}
\left[
- \frac{2m_1^2}{3r_{12}^2}  
- \frac{5m_1m_2}{r_{12}^2}
+ \frac{m_2^2}{r_{12}^2}
+  \frac{m_1}{r_{12}}
\left(
\frac{14}{5}v_1^2 
+ \frac{11}{3}v_2^2 
\right.
\right. 
\nonumber \\
\mbox{} &-& \left. \left. 
\frac{22}{3}(\vec v_{1}\cdot\vec v_2)
- \frac{2}{5}(\vec n_{12}\cdot\vec v_1)^2
+ \frac{20}{3}(\vec n_{12}\cdot\vec v_1)
(\vec n_{12}\cdot\vec v_2)
-4(\vec n_{12}\cdot\vec v_2)^2
\right)
\right.
\nonumber \\
\mbox{} &+& \left.
\frac{m_2}{r_{12}}
\left(
\frac{197}{30}v_1^2 + \frac{19}{3}v_2^2 
- \frac{38}{3}(\vec v_{1}\cdot\vec v_2)
+ \frac{2}{15}(\vec n_{12}\cdot\vec v_1)^2
-\frac{2}{3}(\vec n_{12}\cdot\vec v_1)
(\vec n_{12}\cdot\vec v_2)
\right)
\right.
\nonumber \\
\mbox{} &+& \left.
\frac{22}{15}v_1^4 + \frac{34}{15}v_1^2v_2^2 + 
\frac{4}{3}v_2^4 - \frac{44}{15}v_1^2 (\vec v_1\cdot \vec v_2)
-\frac{8}{3}v_2^2(\vec v_1\cdot\vec v_2) 
+ \frac{8}{15}(\vec v_{1}\cdot\vec v_2)^2 
\right.
\nonumber \\
\mbox{} &-& \left.  
 \frac{2}{3}v_1^2 (\vec n_{12}\cdot\vec v_2)^2 
-\frac{2}{3}v_2^2 (\vec n_{12}\cdot\vec v_2)^2  
+ \frac{4}{3}(\vec v_{1}\cdot\vec v_2)
(\vec n_{12}\cdot\vec v_2)^2  
\right]
\nonumber \\ 
\mbox{} &+& O(\epsilon^7).
\label{Ptchi3PN}
\end{eqnarray}
$P_{A\chi}^{\tau}$ of $O(\epsilon^6)$ affects a 3PN 
equation of motion only through the field $\mbox{}_{10}h^{\tau\tau}$.

The dipole moment and $Q_A^i$ integral of the 
$\chi$ part give a nonzero contribution 
starting from 3PN order in the momentum-velocity relation,   
\begin{eqnarray}
D_{1\chi}^i &=& \epsilon^4\frac{175m_1^3m_2}{18r_{12}^3}r_{12}^i
+ O(\epsilon^5), 
\label{eq:Dichi3PNinApChi} \\
Q_{1\chi}^i &=& \epsilon^6\frac{m_1^3m_2}{6 r_{12}^3}
\left(- \frac{73 n_{12}^{<ij>}v_1^j}{5} +  
11 n_{12}^{<ij>}v_2^j 
\right) + O(\epsilon^7). 
\label{eq:Q1ichi3PNinApChi}
\end{eqnarray}
Here again, we used Gauss law to derive $D_{1\chi}^i$.
On the other hand, we evaluate $P_{A\chi}^i$ directly 
from the definition of the $P_{A\chi}^i$ using  
Gauss law 
and found that 
\begin{eqnarray}
P_{A\chi}^i &=& P_{A\chi}^{\tau}v_A^i + Q_{A\chi}^i 
+ \epsilon^2\frac{d D_{A\chi}^i}{d\tau}+O(\epsilon^7)    
\label{eq:P1ichi3PNinApChi}
\end{eqnarray}
is an identity up to 3PN order.  
Thus, the representative point of the star is defined 
with the $\Theta$ part of 
the momentum-velocity relation, not with the 
$\chi$ part.

Finally, by evaluating the surface integrals in the 
evolution equation,  
\begin{eqnarray}
\frac{dP_{A \chi}^{\mu}}{d\tau} &=&
 - \epsilon^{-4} \oint_{\partial B_A}dS_k
 \chi^{\mu k\alpha\beta}_N\mbox{}_{,\alpha\beta}
 + \epsilon^{-4} v_A^k \oint_{\partial B_A}dS_k
 \chi^{\mu \tau\alpha\beta}_N\mbox{}_{,\alpha\beta},  
\label{eqC36}
\end{eqnarray}
we found that the resulting equations  
for $d P_{A \chi}^{\mu}/d\tau$ are consistent with the 
explicit expressions of  $P_{A \chi}^{\mu}$ 
directly obtained from their definitions,  
Eqs. (\ref{Ptchi3PN})
and (\ref{eq:P1ichi3PNinApChi}), as expected.

\section{Landau-Lifshitz Pseudotensor Expanded in Epsilon}
\label{pNELLPT}

The Landau-Lifshitz pseudotensor \cite{LL1975}
in terms of $h^{\mu\nu}$ which
satisfies the harmonic condition is as follows.
\begin{eqnarray}
(-16 \pi g)t_{LL}^{\mu\nu} &=&
 g_{\alpha\beta}g^{\gamma\delta}
 h^{\mu\alpha}\mbox{}_{,\gamma}
 h^{\nu\beta}\mbox{}_{,\delta}
+ \frac{1}{2}g^{\mu\nu}g_{\alpha\beta}
 h^{\alpha\gamma}\mbox{} _{,\delta}
 h^{\beta\delta}\mbox{}_{,\gamma}
 -2 g_{\alpha\beta}g^{\gamma {\scriptscriptstyle (} \mu}
 h^{\nu {\scriptscriptstyle )}\alpha}\mbox{}_{,\delta}
 h^{\delta\beta}\mbox{}_{,\gamma}
\nonumber \\
\mbox{} &+& 
\frac{1}{2}\left(g^{\mu\alpha}g^{\nu\beta}
			- \frac{1}{2}g^{\mu\nu}g^{\alpha\beta}
		   \right)
           \left(g_{\gamma\delta}g_{\epsilon\zeta}
			- \frac{1}{2}g_{\gamma\epsilon}g_{\delta\zeta}
		   \right)
			h^{\gamma\epsilon}\mbox{} _{,\alpha}
            h^{\delta\zeta}\mbox{}_{,\beta}.
\end{eqnarray}

We expand 
the deviation field $h^{\mu\nu}$ in a power series 
of $\epsilon$; 
$$
h^{\mu\nu} = \sum_{n=0}\epsilon^{4+n} \mbox{}_{n+4}h^{\mu\nu}.
$$
Using this equation, we expand $t_{LL}^{\mu\nu}$ 
in $\epsilon$.

Here we only show $\mbox{}_{10}[-16 \pi g t_{LL}^{\mu \nu}]$. 
See paper II for    
$\mbox{}_{\le 9}[-16 \pi g t_{LL}^{\mu \nu}]$. 
Note that all the divergence such as $h^{\mu k}\mbox{}_{,k}$ 
in paper II should be replaced by $- h^{\mu \tau}\mbox{}_{,\tau}$ 
consistent with the following results. This is simply 
because it is practically much 
easier to use $h^{\mu \tau}\mbox{}_{,\tau}$ 
than $- h^{\mu k}\mbox{}_{,k}$,

\noindent
%
%
%
%
%
%
%
%
%
\begin{eqnarray}
\lefteqn{
\mbox{}_{10}[-16 \pi g t_{LL}^{\tau\tau}]
}\nonumber \\ 
&=& 
- \frac{7}{4}
\mbox{}_4h^{\tau\tau}\mbox{}_{,k}
\mbox{}_8h^{\tau\tau,k} 
+\frac{1}{4}
\mbox{}_4h^{\tau\tau,l}
\mbox{}_6h^{k}\mbox{}_{k,l} 
- \mbox{}_4h^{\tau\tau}\mbox{}_{,k}
\mbox{}_6h^{\tau k}\mbox{}_{,\tau}
+ 2\mbox{}_4h^{\tau k,l}\mbox{}_6h^{\tau}\mbox{}_{(k,l)}
\nonumber \\
\mbox{} &&- 
\frac{3}{4}
\mbox{}_4h^{\tau\tau}\mbox{}_{,\tau}
\mbox{}_6h^{\tau\tau}\mbox{}_{,\tau} 
+ \frac{7}{4}
\mbox{}_4h^{\tau\tau}
\mbox{}_4h^{\tau\tau}\mbox{}_{,k}
\mbox{}_6h^{\tau\tau,k} 
- \mbox{}_4h^{\tau k}\mbox{}_{,\tau}
\mbox{}_6h^{\tau\tau}\mbox{}_{,k} 
+ \frac{1}{4}
\mbox{}_4h^{k}\mbox{}_{k,l}
\mbox{}_6h^{\tau\tau,l} 
\nonumber \\
\mbox{} &&-
\frac{7}{8}
\mbox{}_6h^{\tau\tau}\mbox{}_{,k}
\mbox{}_6h^{\tau\tau,k} 
+ \frac{1}{4}
\mbox{}_4h^{\tau\tau}\mbox{}_{,\tau}
\mbox{}_4h^{k}\mbox{}_{k,\tau} 
+\mbox{}_4h^{\tau k,l}
\mbox{}_4h_{kl,\tau} 
\nonumber \\ 
&&+ 
\frac{1}{4}
\mbox{}_4h^{kl,m}\mbox{}_4h_{kl,m}
- \frac{1}{2}
\mbox{}_4h^{kl,m}\mbox{}_4h_{km,l}
- \frac{1}{8}
\mbox{}_4h^{k}\mbox{}_{k,m}
\mbox{}_4h^{l}\mbox{}_{l}\mbox{}^{,m}
\nonumber \\ 
&&+ 
\frac{1}{4}
\mbox{}_4h^{\tau k}
\mbox{}_4h^{\tau\tau}\mbox{}_{,\tau}
\mbox{}_4h^{\tau\tau}\mbox{}_{,k}
+\frac{7}{8}
\mbox{}_4h^{kl}
\mbox{}_4h^{\tau\tau}\mbox{}_{,k}
\mbox{}_4h^{\tau\tau}\mbox{}_{,l}
-\frac{7}{8}
(\mbox{}_4h^{\tau\tau})^2
\mbox{}_4h^{\tau\tau}\mbox{}_{,k}
\mbox{}_4h^{\tau\tau,k}
+\frac{7}{8}
\mbox{}_6h^{\tau\tau}
\mbox{}_4h^{\tau\tau}\mbox{}_{,k}
\mbox{}_4h^{\tau\tau,k}
\nonumber \\ 
&&- 
 2 
\mbox{}_4h^{\tau k}
\mbox{}_4h^{\tau\tau}\mbox{}_{,l}
\mbox{}_4h^{\tau l}\mbox{}_{,k}
- \frac{3}{2}
\mbox{}_4h^{\tau k}
\mbox{}_4h^{\tau\tau,l}
\mbox{}_4h^{\tau}\mbox{}_{k,l}, 
\label{tLLtt10}
\end{eqnarray}
\noindent
\begin{eqnarray}
\lefteqn{
\mbox{}_{10}[-16 \pi g t_{LL}^{\tau i}]
}\nonumber \\ 
&=& 
2 
\mbox{}_4h^{\tau\tau}\mbox{}_{,k}
\mbox{}_8h^{\tau [k,i]}
+ \frac{3}{4}
\mbox{}_4h^{\tau\tau,i}
\mbox{}_8h^{\tau\tau}\mbox{}_{,\tau}
+ 2
\mbox{}_4h^{\tau [k,i]}
\mbox{}_8h^{\tau\tau}\mbox{}_{,k}
+ \frac{3}{4}
\mbox{}_4h^{\tau\tau}\mbox{}_{,\tau}
\mbox{}_8h^{\tau\tau,i}
\nonumber \\
\mbox{} &&- 
\frac{1}{4}
\mbox{}_4h^{\tau\tau,i}
\mbox{}_6h^{k}\mbox{}_{k,\tau}
- \frac{1}{4}
\mbox{}_4h^{\tau\tau}\mbox{}_{,\tau}
\mbox{}_6h^{k}\mbox{}_{k}\mbox{}^{,i}
+ 2 
\mbox{}_4h^{\tau}\mbox{}_{k,l}
\mbox{}_6h^{k[i,l]}
\nonumber \\ 
\mbox{} &&-
\mbox{}_4h^{\tau i}\mbox{}_{,k}
\mbox{}_6h^{\tau k}\mbox{}_{,\tau}
- \mbox{}_4h^{\tau k}\mbox{}_{,\tau}
 \mbox{}_6h^{\tau i}\mbox{}_{,k}
+ 2\mbox{}_4h^{\tau\tau}
\mbox{}_4h^{\tau\tau}\mbox{}_{,k}
\mbox{}_6h^{\tau [i,k]} 
\nonumber \\ 
\mbox{} &&+
2 
\mbox{}_4h^{k [i,l]}
\mbox{}_6h^{\tau}\mbox{}_{k,l} 
+ 2
\mbox{}_6h^{\tau\tau}\mbox{}_{,k}\mbox{}_6h^{\tau [k,i]} 
- \frac{3}{4}
\mbox{}_4h^{\tau\tau}
\mbox{}_4h^{\tau\tau,i}
\mbox{}_6h^{\tau\tau}\mbox{}_{,\tau}
- \frac{1}{4}
\mbox{}_4h^{k}\mbox{}_{k}\mbox{}^{,i}
\mbox{}_6h^{\tau\tau}\mbox{}_{,\tau}
\nonumber \\
&&+ \frac{3}{4}
\mbox{}_6h^{\tau\tau,i}
\mbox{}_6h^{\tau\tau}\mbox{}_{,\tau}
\nonumber \\
&&- \frac{3}{4}
\mbox{}_4h^{\tau\tau}
\mbox{}_4h^{\tau\tau}\mbox{}_{,\tau}
\mbox{}_6h^{\tau\tau,i}
+ \frac{1}{4}
\mbox{}_4h^{\tau i}
\mbox{}_4h^{\tau\tau,k}
\mbox{}_6h^{\tau\tau}\mbox{}_{,k}
- \frac{1}{2}
\mbox{}_4h^{\tau}\mbox{}_k
\mbox{}_4h^{\tau\tau,(i}
\mbox{}_6h^{|\tau\tau|,k)}
+ 2 \mbox{}_4h^{\tau\tau}
\mbox{}_4h^{\tau [i,k]}
\mbox{}_6h^{\tau\tau}\mbox{}_{,k}
\nonumber \\
\mbox{} &&-
\frac{1}{4}
\mbox{}_4h^{k}\mbox{}_{k,\tau}
\mbox{}_6h^{\tau\tau,i}
- \frac{1}{2}
\mbox{}_4h^{kl,i}
\mbox{}_4h_{kl,\tau}
+ \mbox{}_4h^{ki,l}
\mbox{}_4h_{kl,\tau} 
+ \frac{1}{4}
\mbox{}_4h^{k}\mbox{}_{k}\mbox{}^{,i}
\mbox{}_4h^{l}\mbox{}_{l,\tau}
\nonumber \\
\mbox{} &&-
\frac{3}{8}
\mbox{}_4h^{\tau i}
(\mbox{}_4h^{\tau\tau}\mbox{}_{,\tau})^2
- \frac{3}{4}
\mbox{}_4h^{ik}
\mbox{}_4h^{\tau\tau}\mbox{}_{,\tau}
\mbox{}_4h^{\tau\tau}\mbox{}_{,k}
+ \frac{3}{4}
(\mbox{}_4h^{\tau \tau})^2
\mbox{}_4h^{\tau\tau}\mbox{}_{,\tau}
\mbox{}_4h^{\tau\tau,i}
- \frac{3}{4}
\mbox{}_6h^{\tau \tau}
\mbox{}_4h^{\tau\tau}\mbox{}_{,\tau}
\mbox{}_4h^{\tau\tau,i}
\nonumber \\
&&- \frac{1}{4}
\mbox{}_4h^{\tau i}
\mbox{}_4h^{\tau \tau}
\mbox{}_4h^{\tau\tau}\mbox{}_{,k}
\mbox{}_4h^{\tau\tau,k}
+ \frac{1}{8}
\mbox{}_6h^{\tau i}
\mbox{}_4h^{\tau\tau}\mbox{}_{,k}
\mbox{}_4h^{\tau\tau,k}
+ \frac{1}{2}
\mbox{}_4h^{\tau \tau}
\mbox{}_4h^{\tau k}
\mbox{}_4h^{\tau\tau}\mbox{}_{,k}
\mbox{}_4h^{\tau\tau,i}
\nonumber \\
&&- \frac{1}{4}
\mbox{}_6h^{\tau k}
\mbox{}_4h^{\tau\tau}\mbox{}_{,k}
\mbox{}_4h^{\tau\tau,i}
+ \frac{1}{2}
\mbox{}_4h^{\tau k}
\mbox{}_4h^{\tau}\mbox{}_{k,\tau}
\mbox{}_4h^{\tau\tau,i}
- \mbox{}_4h^{ik}
\mbox{}_4h^{\tau\tau}\mbox{}_{,l}
\mbox{}_4h^{\tau l}\mbox{}_{,k}
+ 2 (\mbox{}_4h^{\tau\tau})^2
\mbox{}_4h^{\tau\tau}\mbox{}_{,k}
\mbox{}_4h^{\tau [k,i]}
\nonumber \\
&&
- 
2
\mbox{}_6h^{\tau\tau}
\mbox{}_4h^{\tau\tau}\mbox{}_{,k}
\mbox{}_4h^{\tau [k,i]}
+ 
\mbox{}_4h^{\tau i}
\mbox{}_4h^{\tau k,l}
\mbox{}_4h^{\tau}\mbox{}_{[l,k]}
+ \frac{1}{2}
\mbox{}_4h^{\tau}\mbox{}_{k}
\mbox{}_4h^{\tau\tau}\mbox{}_{,\tau}
\mbox{}_4h^{\tau k,i}
\nonumber \\
\mbox{} &&+ 
2 \mbox{}_4h^{\tau k}
\mbox{}_4h^{\tau}\mbox{}_{k,l}
\mbox{}_4h^{\tau [l,i]} 
+ 
2 \mbox{}_4h^{\tau k}
\mbox{}_4h^{\tau}\mbox{}_{l,k}
\mbox{}_4h^{\tau [l,i]} 
+ \mbox{}_4h^{\tau k}
\mbox{}_4h^{\tau\tau}\mbox{}_{,\tau}
\mbox{}_4h^{\tau i}\mbox{}_{,k}
+ \mbox{}_4h^{k l}
\mbox{}_4h^{\tau\tau}\mbox{}_{,k}
\mbox{}_4h^{\tau i}\mbox{}_{,l} 
\nonumber \\
\mbox{} &&+ 
\frac{1}{4}
\mbox{}_4h^{\tau i}
\mbox{}_4h^{\tau\tau,k}
\mbox{}_4h^{l}\mbox{}_{l,k}
- \frac{1}{2}
\mbox{}_4h^{\tau}\mbox{}_{k}
\mbox{}_4h^{l}\mbox{}_{l}\mbox{}^{(,k}
\mbox{}_4h^{|\tau\tau|,i)} 
+ 2 \mbox{}_4h^{\tau}\mbox{}_{k}
\mbox{}_4h^{\tau\tau}\mbox{}_{,l}
\mbox{}_4h^{k [l,i]},  
\label{tLLti10}
\end{eqnarray}
%
%
%
%
%
%
%
\noindent 
\begin{eqnarray}
\lefteqn{
\mbox{}_{10}[-16 \pi g t_{LL}^{ij}]
}\nonumber \\ 
&=& \frac{1}{4}
(\delta^i\mbox{}_k\delta^j\mbox{}_l + 
\delta^j\mbox{}_k\delta^i\mbox{}_l - 
\delta^{ij}\delta_{kl})
\left\{
\mbox{}_4h^{\tau\tau,k}
(\mbox{}_{10}h^{\tau\tau,l} + \mbox{}_{8}h^m\mbox{}_m\mbox{}^{,l} 
+ 4 \mbox{}_8h^{\tau l}\mbox{}_{,\tau}) 
+ 8 \mbox{}_4h^{\tau}\mbox{}_{m}\mbox{}^{,k}\mbox{}_8h^{\tau [l,m]}
\right\}
\nonumber \\ 
&&+ 
2 \mbox{}_4h^{\tau i}\mbox{}_{,k}\mbox{}_8h^{\tau [k,j]}
+ 2 \mbox{}_4h^{\tau j}\mbox{}_{,k}\mbox{}_8h^{\tau [k,i]}
-\frac{3}{4}\delta^{ij}
\mbox{}_4h^{\tau\tau}\mbox{}_{,\tau}
\mbox{}_8h^{\tau\tau}\mbox{}_{,\tau} 
\nonumber \\ 
&&+  
\frac{1}{4}
(\delta^i\mbox{}_k\delta^j\mbox{}_l + 
\delta^j\mbox{}_k\delta^i\mbox{}_l - 
\delta^{ij}\delta_{kl})
\left(
\mbox{}_6h^{\tau\tau,k}  
+ \mbox{}_4h^m\mbox{}_m\mbox{}^{,k} 
- 2 \mbox{}_4h^{\tau\tau}\mbox{}_4h^{\tau\tau,k}
+ 4 \mbox{}_4h^{\tau k}\mbox{}_{,\tau}
\right)\mbox{}_8h^{\tau\tau,l}
\nonumber \\ 
&&+  
\frac{1}{4}\delta^{ij}
\mbox{}_4h^{\tau\tau}\mbox{}_{,\tau}
\mbox{}_6h^k\mbox{}_{k,\tau} -  
(\delta^i\mbox{}_k\delta^j\mbox{}_l + 
\delta^j\mbox{}_k\delta^i\mbox{}_l - 
\delta^{ij}\delta_{kl})
\mbox{}_4h^{\tau m,k}
\mbox{}_6h^{l}\mbox{}_{m,\tau}  
%
%
\nonumber \\ 
&&+  
\frac{1}{4}
(\delta^i\mbox{}_k\delta^j\mbox{}_l + 
\delta^j\mbox{}_k\delta^i\mbox{}_l - 
\delta^{ij}\delta_{kl})
\left(
\mbox{}_6h^{\tau\tau,k} - 
\mbox{}_4h^m\mbox{}_m\mbox{}^{,k}
- \mbox{}_4h^{\tau\tau}\mbox{}_4h^{\tau\tau,k}
\right)
\mbox{}_6h^n\mbox{}_n\mbox{}^{,l}
\nonumber \\
&&+
(\delta^i\mbox{}_k\delta^j\mbox{}_l + 
\delta^j\mbox{}_k\delta^i\mbox{}_l - 
\delta^{ij}\delta_{kl})
\left(\frac{1}{2}
\mbox{}_4h^{mn,k}\mbox{}_6h_{mn}\mbox{}^{,l} - 
\mbox{}_4h^{mk,n}\mbox{}_6h_{mn}\mbox{}^{,l}  
\right)
\nonumber \\
&&+ 
 2 \mbox{}_4h^{k [i,l]}\mbox{}_6h^j\mbox{}_{k,l}
+ 2 \mbox{}_4h^{k [j,l]}\mbox{}_6h^i\mbox{}_{k,l}
%
%
\nonumber \\
&&+
(\delta^i\mbox{}_k\delta^j\mbox{}_l + 
\delta^j\mbox{}_k\delta^i\mbox{}_l - 
\delta^{ij}\delta_{kl})
(
\mbox{}_6h^{\tau\tau,k}\mbox{}_6h^{\tau l}\mbox{}_{,\tau}
- \mbox{}_4h^{\tau\tau}\mbox{}_4h^{\tau\tau,k}
\mbox{}_6h^{\tau l}\mbox{}_{,\tau}) 
+2 \mbox{}_4h^{\tau (i}\mbox{}_{,|\tau|} 
\mbox{}_6h^{j)\tau}\mbox{}_{,\tau} 
\nonumber \\
&&-  (\delta^i\mbox{}_k\delta^j\mbox{}_l + 
\delta^j\mbox{}_k\delta^i\mbox{}_l - 
\delta^{ij}\delta_{kl})
\mbox{}_4h^{k}\mbox{}_{m,\tau}\mbox{}_6h^{\tau m,l}
%
%
\nonumber \\
&&+
(\delta^i\mbox{}_k\delta^j\mbox{}_l + 
\delta^j\mbox{}_k\delta^i\mbox{}_l - 
\delta^{ij}\delta_{kl})
\left(\mbox{}_4h^{\tau\tau}
\mbox{}_4h^{\tau}\mbox{}_{m}\mbox{}^{,k}
\mbox{}_6h^{\tau m,l} 
- \mbox{}_4h^{\tau\tau}
\mbox{}_4h^{\tau k}\mbox{}_{,m}
\mbox{}_6h^{\tau m,l}
+ \frac{1}{2}
\mbox{}_4h^{\tau}\mbox{}_m
\mbox{}_4h^{\tau\tau,k}\mbox{}_6h^{\tau m,l} 
\right) 
\nonumber \\
&&+
 2 \mbox{}_4h^{\tau\tau}
\mbox{}_4h^{\tau [i,k]}\mbox{}_6h^{\tau j}\mbox{}_{,k}
+ 2 \mbox{}_4h^{\tau\tau}
\mbox{}_4h^{\tau [j,k]}\mbox{}_6h^{\tau i}\mbox{}_{,k}
%
%
\nonumber \\
&&+
\frac{1}{8}\delta^{ij}\mbox{}_6h^{kl}
\mbox{}_4h^{\tau\tau}\mbox{}_{,k}
\mbox{}_4h^{\tau\tau}\mbox{}_{,l} + 
\frac{1}{8}\mbox{}_6h^{ij}
\mbox{}_4h^{\tau\tau}\mbox{}_{,k}
\mbox{}_4h^{\tau\tau,k} - 
\frac{1}{2}\mbox{}_4h^{\tau\tau}\mbox{}_{,k}
\mbox{}_4h^{\tau\tau,(i}\mbox{}_6h^{j) k} 
%
%
\nonumber \\
&&+ \frac{1}{4}(\delta^i\mbox{}_k\delta^j\mbox{}_l + 
\delta^j\mbox{}_k\delta^i\mbox{}_l - 
\delta^{ij}\delta_{kl})
(2 \mbox{}_6h^{\tau}\mbox{}_m
\mbox{}_4h^{\tau\tau,k}
\mbox{}_4h^{\tau m,l}
- \mbox{}_6h^{\tau k}
\mbox{}_4h^{\tau\tau}\mbox{}_{,\tau}
\mbox{}_4h^{\tau\tau,l}) 
\nonumber \\
&&+ 
\delta^{ij}\mbox{}_6h^{\tau k,l}
\mbox{}_6h^{\tau}\mbox{}_{[k,l]} 
+ 2 \mbox{}_6h^{\tau k,(i}\mbox{}_6h^{j)\tau}\mbox{}_{,k}
- \mbox{}_6h^{\tau k,i}\mbox{}_6h^{\tau}\mbox{}_{k}\mbox{}^{,j}
- \mbox{}_6h^{\tau i,k}\mbox{}_6h^{\tau j}\mbox{}_{,k}
%
%
\nonumber \\
&&+\frac{1}{4}
\delta^{ij}
\left(
2
\mbox{}_4h_{kl}
\mbox{}_4h^{\tau k,m}
\mbox{}_4h^{\tau l}\mbox{}_{,m}- 
2
\mbox{}_4h^{kl}
\mbox{}_4h^{\tau m}\mbox{}_{,k}
\mbox{}_4h^{\tau}\mbox{}_{m,l} +
\mbox{}_4h^{kl}
\mbox{}_4h^{\tau \tau}\mbox{}_{,k}
\mbox{}_6h^{\tau\tau}\mbox{}_{,l} - 
\mbox{}_4h^{\tau\tau}\mbox{}_4h^{kl}
\mbox{}_4h^{\tau \tau}\mbox{}_{,k}
\mbox{}_4h^{\tau\tau}\mbox{}_{,l}  
\right)
%
%
%
%
\nonumber \\
&&+
\frac{3}{8}\mbox{}_4h^{ij}
(\mbox{}_4h^{\tau\tau}\mbox{}_{,\tau})^2 
- \frac{1}{4}\mbox{}_4h^{\tau\tau}
\mbox{}_4h^{ij}
\mbox{}_4h^{\tau\tau}\mbox{}_{,k}\mbox{}_4h^{\tau\tau,k} 
+ \mbox{}_4h^{\tau\tau}
\mbox{}_4h^{\tau\tau}\mbox{}_{,k}\mbox{}_4h^{\tau\tau,(i}
\mbox{}_4h^{j)k} 
+ \mbox{}_4h^{ij}\mbox{}_4h^{\tau\tau}\mbox{}_{,k}
\mbox{}_4h^{\tau k}\mbox{}_{,\tau} 
\nonumber \\
&&- \mbox{}_4h^{ij}\mbox{}_4h^{\tau k,l}
\mbox{}_4h^{\tau}\mbox{}_{[k,l]} 
+ \mbox{}_4h^{kl}
\mbox{}_4h^{\tau i}\mbox{}_{,k}
\mbox{}_4h^{\tau j}\mbox{}_{,l}
- \mbox{}_4h_{kl}
\mbox{}_4h^{\tau k,i}
\mbox{}_4h^{\tau l,j} 
\nonumber \\
&&+  
 2\mbox{}_4h^{ki}\mbox{}_4h^{\tau [l,j]}\mbox{}_4h^{\tau}\mbox{}_{l,k}
+ 2\mbox{}_4h^{kj}\mbox{}_4h^{\tau [l,i]}\mbox{}_4h^{\tau}\mbox{}_{l,k}
- 2\mbox{}_4h^{k(i}\mbox{}_4h^{j)\tau}\mbox{}_{,\tau}
\mbox{}_4h^{\tau\tau}\mbox{}_{,k}
\nonumber \\
&&+  
 \frac{1}{4}
\mbox{}_4h^{ij}
\mbox{}_4h^{\tau\tau}\mbox{}_{,k}
\mbox{}_6h^{\tau\tau,k}  
- \frac{1}{2}
\mbox{}_4h^{\tau\tau}\mbox{}_{,k}
\mbox{}_6h^{\tau\tau,(i} 
\mbox{}_4h^{j)k}
- \frac{1}{2}
\mbox{}_6h^{\tau\tau}\mbox{}_{,k}
\mbox{}_4h^{\tau\tau,(i}  
\mbox{}_4h^{j)k}
%
%
%
%
\nonumber \\
&&+  
\delta^{ij}
\left(
2 \mbox{}_4h^{\tau}\mbox{}_k
\mbox{}_4h^{\tau}\mbox{}_{l,m}
\mbox{}_4h^{k[l,m]} 
- \frac{1}{2}
\mbox{}_4h^{\tau}\mbox{}_{k}
\mbox{}_4h^{\tau k,l}
\mbox{}_4h^m\mbox{}_{m,l}
-\frac{1}{4} (\mbox{}_4h^{\tau\tau})^2
\mbox{}_4h^{\tau\tau,k}\mbox{}_4h^l\mbox{}_{l,k}
\right.
\nonumber \\
\mbox{} &&+\left.
 \frac{1}{4}\mbox{}_4h^{\tau k}
\mbox{}_4h^{\tau\tau}\mbox{}_{,\tau}
\mbox{}_4h^l\mbox{}_{l,k}
\right)
\nonumber \\
&&+   
\delta^{ij}\left(
\frac{1}{4}\mbox{}_4h^{\tau\tau}\mbox{}_4h^l\mbox{}_{l,k}
\mbox{}_6h^{\tau\tau,k} 
+ \frac{1}{4}\mbox{}_6h^{\tau\tau}\mbox{}_4h^l\mbox{}_{l,k}
\mbox{}_4h^{\tau\tau,k} 
\right) \nonumber
\end{eqnarray}
%
%
%
\begin{eqnarray}
\mbox{} 
&&-   
\frac{1}{2}
\mbox{}_4h^{\tau\tau}\mbox{}_{,\tau}
\mbox{}_4h^k\mbox{}_{k}\mbox{}^{,(i}
\mbox{}_4h^{j)\tau} 
+ \frac{1}{2}(\mbox{}_4h^{\tau\tau})^2
\mbox{}_4h^k\mbox{}_{k}\mbox{}^{,(i}
\mbox{}_4h^{|\tau\tau|,j)} 
- \frac{1}{2}\mbox{}_6h^{\tau\tau}
\mbox{}_4h^k\mbox{}_{k}\mbox{}^{,(i}
\mbox{}_4h^{|\tau\tau|,j)} 
\nonumber \\
&&+ 
\mbox{}_4h^{\tau}\mbox{}_k
\mbox{}_4h^l\mbox{}_{l}\mbox{}^{,(i}
\mbox{}_4h^{|\tau k|,j)}
+ 2\mbox{}_4h^{\tau k}
\mbox{}_4h^{\tau [i,l]}
\mbox{}_4h_{kl}\mbox{}^{,j}
+ 2\mbox{}_4h^{\tau k}
\mbox{}_4h^{\tau [j,l]}
\mbox{}_4h_{kl}\mbox{}^{,i}
\nonumber \\
&&+ 
2\mbox{}_4h^{\tau k}
\mbox{}_4h^{\tau [l,i]}
\mbox{}_4h^{j}\mbox{}_{k,l}
+ 2\mbox{}_4h^{\tau k}
\mbox{}_4h^{\tau [l,j]}
\mbox{}_4h^{i}\mbox{}_{k,l}
- \frac{1}{2}\mbox{}_4h^{\tau\tau}
\mbox{}_4h^{k}\mbox{}_k\mbox{}^{,(i}
\mbox{}_6h^{|\tau\tau|,j)}
%
%
%
%
\nonumber \\
&&+ 
(\delta^i\mbox{}_k\delta^j\mbox{}_l + 
\delta^j\mbox{}_k\delta^i\mbox{}_l - 
\delta^{ij}\delta_{kl})(
\mbox{}_4h^{\tau}\mbox{}_m
\mbox{}_4h^{\tau\tau,k}\mbox{}_4h^{ml}\mbox{}_{,\tau}
-\frac{1}{4}\mbox{}_4h^{\tau k}
\mbox{}_4h^{\tau\tau,l}
\mbox{}_4h^m\mbox{}_{m,\tau}
)
%
%
%
%
\nonumber \\
&&+ 
\delta^{ij}\left(
- \frac{3}{8}(\mbox{}_4h^{\tau\tau})^2 
(\mbox{}_4h^{\tau\tau}\mbox{}_{,\tau})^2
+  \frac{3}{8}\mbox{}_6h^{\tau\tau} 
(\mbox{}_4h^{\tau\tau}\mbox{}_{,\tau})^2  
+\frac{1}{2}(\mbox{}_4h^{\tau\tau})^3 
\mbox{}_4h^{\tau\tau}\mbox{}_{,k}
\mbox{}_4h^{\tau\tau,k} 
-\frac{1}{2}
\mbox{}_4h^{\tau\tau}
\mbox{}_4h^{\tau k}
\mbox{}_4h^{\tau\tau}\mbox{}_{,\tau}
\mbox{}_4h^{\tau\tau}\mbox{}_{,k} 
\right.
\nonumber \\
&&-
\left.
 \frac{3}{4}
\mbox{}_4h^{\tau\tau}
\mbox{}_6h^{\tau\tau}
\mbox{}_4h^{\tau\tau}\mbox{}_{,k}
\mbox{}_4h^{\tau\tau,k} 
+ \frac{1}{4}
\mbox{}_8h^{\tau\tau}
\mbox{}_4h^{\tau\tau}\mbox{}_{,k}
\mbox{}_4h^{\tau\tau,k}  
+ \frac{1}{4}
\mbox{}_4h^{\tau}\mbox{}_k\mbox{}_4h^{\tau k}
\mbox{}_4h^{\tau\tau}\mbox{}_{,l}
\mbox{}_4h^{\tau\tau,l} 
-  \frac{1}{2}\mbox{}_4h^{\tau k}
\mbox{}_4h^{\tau\tau}\mbox{}_{,\tau}
\mbox{}_4h^{\tau}\mbox{}_{k,\tau} 
\right.
\nonumber \\
&&-
\left.
(\mbox{}_4h^{\tau\tau})^2
\mbox{}_4h^{\tau\tau,k}
\mbox{}_4h^{\tau}\mbox{}_{k,\tau}
+ \mbox{}_6h^{\tau\tau}
\mbox{}_4h^{\tau\tau,k}
\mbox{}_4h^{\tau}\mbox{}_{k,\tau} 
+ \mbox{}_4h^{\tau\tau}
\mbox{}_4h^{\tau k}
\mbox{}_4h^{\tau\tau,l}
\mbox{}_4h^{\tau}\mbox{}_{k,l} 
-\mbox{}_4h^{\tau}\mbox{}_{k}
\mbox{}_4h^{\tau k}\mbox{}_{,l}
\mbox{}_4h^{\tau l}\mbox{}_{,\tau}
\right.
\nonumber \\
&&+
\left.
(\mbox{}_4h^{\tau\tau})^2
\mbox{}_4h^{\tau k,l}\mbox{}_4h^{\tau}\mbox{}_{[k,l]}
- \mbox{}_6h^{\tau\tau}
\mbox{}_4h^{\tau k,l}\mbox{}_4h^{\tau}\mbox{}_{[k,l]}
- \mbox{}_4h^{\tau k}
\mbox{}_4h^{\tau}\mbox{}_{l,k}
\mbox{}_4h^{\tau l}\mbox{}_{,\tau}
+ \frac{1}{4}\mbox{}_4h^k\mbox{}_{l,\tau}
\mbox{}_4h^l\mbox{}_{k,\tau}
\right.
\nonumber \\
&&+
\left.
 \frac{1}{4}\mbox{}_4h^{kl}
\mbox{}_4h^{\tau\tau}\mbox{}_{,k}
\mbox{}_4h^m\mbox{}_{m,l}
- \frac{1}{8}\mbox{}_4h^k\mbox{}_{k,\tau}
\mbox{}_4h^l\mbox{}_{l,\tau}
+ \frac{1}{4}
\mbox{}_4h^k\mbox{}_{k,\tau}
\mbox{}_6h^{\tau\tau}\mbox{}_{,\tau}
- \frac{1}{4}
\mbox{}_4h^{kl}
\mbox{}_4h^{\tau\tau,m}
\mbox{}_4h_{kl,m} 
+ \frac{3}{4}\mbox{}_4h^{\tau\tau}
\mbox{}_4h^{\tau\tau}\mbox{}_{,\tau}
\mbox{}_6h^{\tau\tau}\mbox{}_{,\tau}
\right.
\nonumber \\
&&+
\left.
\frac{1}{4}
\mbox{}_4h^{\tau k}
\mbox{}_4h^{\tau\tau}\mbox{}_{,k}
\mbox{}_6h^{\tau\tau}\mbox{}_{,\tau}
- \frac{3}{8}(\mbox{}_6h^{\tau\tau}\mbox{}_{,\tau})^2
+ \frac{1}{4}\mbox{}_4h^{\tau k}
\mbox{}_4h^{\tau\tau}\mbox{}_{,\tau}
\mbox{}_6h^{\tau\tau}\mbox{}_{,k}
+ \mbox{}_4h^{\tau\tau}
\mbox{}_4h^{\tau k}\mbox{}_{,\tau}
\mbox{}_6h^{\tau\tau}\mbox{}_{,k}
\right.
\nonumber \\
&&-
\left.
\frac{1}{2}\mbox{}_4h^{\tau k}
\mbox{}_4h^{\tau}\mbox{}_{k,l}\mbox{}_6h^{\tau\tau,l}
- \frac{3}{4}(\mbox{}_4h^{\tau\tau})^2
\mbox{}_4h^{\tau\tau}\mbox{}_{,k}
\mbox{}_6h^{\tau\tau,k}  
+ \frac{1}{2}\mbox{}_6h^{\tau\tau}
\mbox{}_4h^{\tau\tau}\mbox{}_{,k}
\mbox{}_6h^{\tau\tau,k}  
+ \frac{1}{4}
\mbox{}_4h^{\tau\tau}
\mbox{}_6h^{\tau\tau}\mbox{}_{,k}
\mbox{}_6h^{\tau\tau,k}  
\right)
%
%
%
%
\nonumber \\
&&+ 
\mbox{}_4h^{\tau\tau}
\mbox{}_4h^{\tau\tau}\mbox{}_{,\tau}
\mbox{}_4h^{\tau\tau,(i}
\mbox{}_4h^{j)\tau} 
- (\mbox{}_4h^{\tau\tau})^3
\mbox{}_4h^{\tau\tau,i}
\mbox{}_4h^{\tau\tau,j}
- \frac{1}{2}
\mbox{}_4h^{\tau k}
\mbox{}_4h^{\tau}\mbox{}_k
\mbox{}_4h^{\tau\tau,i}
\mbox{}_4h^{\tau\tau,j}
\nonumber \\
&&+ \frac{3}{2}
\mbox{}_4h^{\tau\tau}
\mbox{}_6h^{\tau\tau}
\mbox{}_4h^{\tau\tau,i}
\mbox{}_4h^{\tau\tau,j}
- \frac{1}{2}\mbox{}_8h^{\tau\tau}
\mbox{}_4h^{\tau\tau,i}
\mbox{}_4h^{\tau\tau,j}
\nonumber \\
\mbox{} &&
- 2 \mbox{}_4h^{\tau\tau}
\mbox{}_4h^{\tau}\mbox{}_{k}
\mbox{}_4h^{\tau\tau,(i}
\mbox{}_4h^{|\tau k|,j)} 
+ 2\mbox{}_4h^{\tau}\mbox{}_{k,\tau}
\mbox{}_4h^{\tau k,(i}
\mbox{}_4h^{j)\tau}
\nonumber \\
&&+ 
2 (\mbox{}_4h^{\tau\tau})^2
\mbox{}_4h^{\tau}
\mbox{}_k\mbox{}^{,j}
\mbox{}_4h^{\tau [i,k]}
- 2 (\mbox{}_4h^{\tau\tau})^2
\mbox{}_4h^{\tau j}\mbox{}_{,k}
\mbox{}_4h^{\tau [i,k]}
+ 2 (\mbox{}_4h^{\tau\tau})^2
\mbox{}_4h^{\tau\tau,(i}
\mbox{}_4h^{j)\tau}\mbox{}_{,\tau}
\nonumber \\
&&- 
2 \mbox{}_6h^{\tau\tau}
\mbox{}_4h^{\tau}\mbox{}_k\mbox{}^{,j}
\mbox{}_4h^{\tau [i,k]}
+ 2 \mbox{}_6h^{\tau\tau}
\mbox{}_4h^{\tau j}\mbox{}_{,k}
\mbox{}_4h^{\tau [i,k]} 
- 2 \mbox{}_6h^{\tau\tau}
\mbox{}_4h^{\tau\tau,(i}
\mbox{}_4h^{j)\tau}\mbox{}_{,\tau}  
\nonumber \\
&&-  
2 \mbox{}_4h^{\tau\tau}\mbox{}_{,\tau}
\mbox{}_4h^{\tau (i}
\mbox{}_4h^{j)\tau}\mbox{}_{,\tau} 
+ 2 \mbox{}_4h^{\tau}\mbox{}_k
\mbox{}_4h^{\tau k,(i}
\mbox{}_4h^{j)\tau}\mbox{}_{,\tau}
- 2 \mbox{}_4h^{\tau k}\mbox{}_{,\tau}
\mbox{}_4h^{\tau (i}\mbox{}_4h^{j)\tau}\mbox{}_{,k}
+ 2 \mbox{}_4h^{\tau}\mbox{}_{k}
\mbox{}_4h^{\tau (i,|k|}
\mbox{}_4h^{j)\tau}\mbox{}_{,\tau}
\nonumber \\
&&+  
 \frac{1}{4}\mbox{}_4h^{ij}
\mbox{}_4h^{\tau\tau,k}
\mbox{}_4h^l\mbox{}_{l,k}
- \frac{1}{2}
\mbox{}_4h^l\mbox{}_{l,k}
\mbox{}_4h^{\tau\tau,(i}
\mbox{}_4h^{j)k}
- \frac{1}{2}
\mbox{}_4h^{\tau\tau}\mbox{}_{,k}
\mbox{}_4h^l\mbox{}_{l}\mbox{}^{,(i}
\mbox{}_4h^{j)k} 
+ \frac{1}{2}
\mbox{}_4h_{kl}
\mbox{}_4h^{\tau\tau,(i}
\mbox{}_4h^{|kl|,j)}
\nonumber \\
&&- 
\mbox{}_4h^{ik}\mbox{}_{,\tau}
\mbox{}_4h^{j}\mbox{}_{k,\tau}
- \frac{1}{2}
\mbox{}_6h^{\tau\tau}\mbox{}_{,\tau}
\mbox{}_4h^{\tau\tau,(i}\mbox{}_4h^{j)\tau}
- \frac{1}{2}
\mbox{}_4h^{\tau\tau}\mbox{}_{,\tau}
\mbox{}_6h^{\tau\tau,(i}\mbox{}_4h^{j)\tau} 
+ \frac{3}{2}(\mbox{}_4h^{\tau\tau})^2
\mbox{}_4h^{\tau\tau,(i}
\mbox{}_6h^{|\tau\tau|,j)}
\nonumber \\
&&-
 \mbox{}_6h^{\tau\tau}\mbox{}_6h^{\tau\tau,(i}
\mbox{}_4h^{|\tau\tau|,j)}
+ \mbox{}_4h^{\tau}\mbox{}_k
\mbox{}_4h^{\tau k,(i}
\mbox{}_6h^{|\tau\tau|,j)}
- 2 \mbox{}_4h^{\tau\tau}
\mbox{}_6h^{\tau\tau,(i}\mbox{}_4h^{j)\tau}\mbox{}_{,\tau}
- \frac{1}{2}\mbox{}_4h^{\tau\tau}
\mbox{}_6h^{\tau\tau,i} 
\mbox{}_6h^{\tau\tau,j}. 
\label{tLLij10}
\end{eqnarray}

\section{Superpotentials}
\label{splist}
Here we list useful superpotentials. 
Below, $f^{(m,n)}$ satisfies $\Delta f^{(m,n)} = r_1^m r_2^n$. 
Similarly, $\Delta 
f^{(m;\ln,n;\ln)} = r_1^m r_2^n \ln r_1 \ln r_2$.
The appendix in \cite{JS98} gives a greatly useful list 
of superpotentials including 
$f^{(\ln S)}$, $f^{(1,1)}$, $f^{(-1,1)}$, and  
$f^{(3,-1)}$. Other useful superpotentials are 
given in \cite{BFP98,BF01a},

\begin{eqnarray*}
f^{(-2,2)} &=& 
\frac{r_1^2}{6} + 
\frac{1}{3}(-r_1^2 + 2 r_{12}^2 + r_2^2)\ln r_1,
\nonumber \\
f^{(-3,-3)} &=&
\frac{1}{r_{12}^3 r_1}
\ln\left(\frac{S}{r_1}\right) + 
\frac{1}{r_{12}^3 r_2}
\ln\left(\frac{S}{r_2}\right) -
\frac{1}{r_{12}^2r_1r_2},
\nonumber \\ 
f^{(-3,-2)} &=& 
\frac{1}{r_1r_{12}^2}\ln\left(\frac{r_2}{r_1}\right), 
\nonumber \\ 
f^{(-3,-1)} &=& 
\frac{1}{r_1r_{12}}\ln\left(\frac{S}{r_1}\right),
\nonumber   \\ 
f^{(-3,0)} &=& \frac{-\ln r_1}{r_1}, \nonumber   \\ 
f^{(-3,1)} &=&
- \frac{r_2}{r_1} + \ln S + \frac{r_{12}}{r_1}\ln 
\left(\frac{S}{r_1}\right), \nonumber \\
f^{(-4,-3)} &=& \frac{1}{2}\Delta_{11}f^{(-2,-3)},
\nonumber  \\
f^{(-4,-1)} &=&  
\frac{r_2}{2r_1^2r_{12}^2}, \nonumber \\
f^{(0;\ln,0)} &=& \frac{1}{6}r_1^2\ln r_1 - \frac{5}{36}r_1^2.
\nonumber  
\end{eqnarray*}

An example below shows how we construct superpotentials   
from simpler ones, 
\begin{eqnarray}
f^{(6,-2)} &=& -6 f^{(0;0,4;\ln)} - 3 f^{(2,2)} 
+ \frac{24r_{12}^2}{7}f^{(0;0,2;\ln)}  
-\frac{24r_{12}^2}{7}f^{(2;0,0;\ln)} +  \frac{6r_{12}^2}{7}f^{(4,-2)}
\nonumber \\ 
\mbox{} &&+ \frac{r_1^6}{14} - \frac{3}{35}r_{12}^2r_1^4 + 
\frac{3}{70}r_{12}^2r_2^4 - \frac{r_2^6}{49} + \frac{r_1^6}{7}\ln r_2 
- \frac{3}{7}r_1^4r_2^2\ln r_2 + \frac{3}{7}r_1^2r_2^4\ln r_2,
\nonumber
\end{eqnarray}
which is computed from $
f^{(0;0,4;\ln)}, f^{(2,2)}, 
f^{(0;0,2;\ln)}, f^{(2;0,0;\ln)}
$, 
and $f^{(4,-2)}$.
They are 
\begin{eqnarray}
f^{(0;0,4;\ln)} &=& \frac{1}{42}r_2^6\ln r_2 - \frac{13}{1764}r_2^6,
\nonumber \\ 
f^{(2,2)} &=& - \frac{r_1^6}{84} + \frac{1}{70} r_1^4r_{12}^2
+ \frac{1}{28}r_1^4r_2^2
,\nonumber  \\
f^{(0;0,2;\ln)} &=& \frac{1}{20}r_2^4\ln r_2 - \frac{9}{400}r_2^4,
\nonumber \\
f^{(2;0,0;\ln)} &=& 
- \frac{7}{200}r_1^4 
+ \frac{1}{60}r_1^2r_{12}^2
- \frac{7}{180}r_{12}^2r_2^2 
+ \frac{1}{80}r_2^4 
+ \frac{1}{10}r_1^2r_2^2\ln r_2
+ \frac{1}{15}r_{12}^2r_2^2\ln r_2 
- \frac{1}{20}r_2^4\ln r_2 
,\nonumber \\
f^{(4,-2)} &=& 
\frac{1}{5}r_1^4 \ln r_2 - \frac{1}{25}r_1^4 + 
\frac{4}{5}r_{12}^2 f^{(2,-2)} - 4 f^{(2;0,0;\ln)}
.\nonumber  
\end{eqnarray}

\end{document}